\documentclass[aps,prd,twocolumn,nofootinbib,showpacs,10pt,superscriptaddress]{revtex4-2}

\usepackage{textcomp}
\usepackage{amsmath}
\usepackage{amsfonts}
\usepackage[bbgreekl]{mathbbol}
\usepackage{calc}
\usepackage{mathrsfs}
\usepackage{amssymb}
\usepackage{array}
\usepackage{color}
\usepackage{bbm}
\usepackage{graphicx}
\usepackage[pdftex,breaklinks,colorlinks,
linkcolor=Blue,
citecolor=teal,
anchorcolor=red,
urlcolor=cyan]{hyperref}
\usepackage{color}
\usepackage[usenames,dvipsnames,svgnames,table]{xcolor}
\usepackage{tabulary}
\usepackage{booktabs}
\usepackage{orcidlink}
\usepackage{enumitem}

\newcommand{\beq}{\begin{equation}}
\newcommand{\eeq}{\end{equation}}
\renewcommand{\P}{{\cal P}}
\newcommand{\res}{{\cal R}}
\newcommand{\R}{{\cal R}}

\newcommand{\psifour} {\psi_{4}}

\newcommand{\rstar}{r^{*}}
\newcommand{\lm}  {\ell m}
\newcommand{\lmw}  {\ell m}
\newcommand{\Y}[2] {\,{_{#1}Y_{#2}}}

\newcommand{\teukR}[3] {\,_{#1}R^{#2}_{#3}}
\newcommand{\teukpot}[2] {\,_{#1}V_{#2}}

\newcommand{\Tgensource}[2] {\,_{#1}S_{#2}}
\newcommand{\gexact}{{\sf g}}
\newcommand{\hexact}{{\sf h}}
\newcommand{\placeholder}{a}

\definecolor{colour1}{HTML}{0571b0} %--- Blue
\hypersetup{colorlinks=true, linkcolor=colour1, citecolor=colour1,
filecolor=colour1, urlcolor=colour1}

\newcommand{\e}{\varepsilon}

\newcommand{\diff}[2]  {\frac{d #1}{d #2}}
\newcommand{\sdiff}[2]  {\frac{d^2 #1}{d #2^2}}
\newcommand{\nn}{\nonumber}

\font\ec=ecrm0800 at 11pt
\def\th{\hbox{\ec\char'336}}
\def\edth{\hbox{\ec\char'360}}

\begin{document}
\title{Worldtube puncture scheme for first- and second-order self-force calculations in the Fourier domain} 
\author{Jeremy Miller}
\affiliation{Shamoon College of Engineering, Jabotinsky 84, Ashdod, 77245, Israel}
\author{Benjamin Leather\,\orcidlink{0000-0001-6186-7271}}
\affiliation{Max Planck Institute for Gravitational Physics (Albert Einstein Institute), Am M{\"u}hlenberg 1, Potsdam 14476, Germany}
\author{Adam Pound\,\orcidlink{0000-0001-9446-0638}}
\affiliation{School of Mathematical Sciences and STAG Research Centre, University of 
Southampton, Southampton, SO17 1BJ, United Kingdom}
\author{Niels Warburton\,\orcidlink{0000-0003-0914-8645}}
\affiliation{School of Mathematics \& Statistics, University College Dublin, Belfield, Dublin 4, Ireland, D04 V1W8}
\date{\today}

\begin{abstract}

Second-order gravitational self-force theory has recently led to the breakthrough calculation of ``first post-adiabatic'' (1PA) compact-binary waveforms~[Phys. Rev. Lett. 130, 241402 (2023)]. The computations underlying those waveforms depend on a method of solving the perturbative second-order Einstein equation on a Schwarzschild background in the Fourier domain. In this paper we present that method, which involves dividing the domain into several regions. Different regions utilize different time slicings and allow for the use of ``punctures'' to tame sources and enforce physical boundary conditions. We demonstrate the method for Lorenz-gauge and Teukolsky equations in the relatively simple case of calculating parametric derivatives (``slow time derivatives'') of first-order fields, which are an essential input at second order.
\end{abstract}
\maketitle

%--------------------------------------------------------------------------------------------------------------------------------------%
\section{Introduction}

\subsection{Waveform generation and second-order self-force theory}

In recent years, gravitational self-force theory~\cite{Barack:2018yvs,Pound:2021qin} has reached a mature stage of producing practical models of compact-binary waveforms~\cite{Katz:2021yft,Hughes:2021exa,Isoyama:2021jjd,McCart:2021upc,Wardell:2021fyy}. These models, targeted at asymmetric binaries in which one body is much more massive than the other, have traditionally been motivated by the need to model waveforms from extreme-mass-ratio inspirals (EMRIs) with mass ratios $\e:=\mu/M\sim 10^{-5}$~\cite{Mino:97}, where $M$ is the mass of the larger body, and $\mu$ is the mass of the companion. However, the resulting waveforms have proved to be quite accurate even for mass ratios $\sim 10^{-1}$~\cite{Wardell:2021fyy}.

The method underlying these models is an expansion of the spacetime metric in powers of $\e$, with the assumption that the zeroth-order spacetime is a stationary black hole. From that starting point, a combination of perturbative techniques are used, including broad strategies adapted from singular perturbation theory (matched asymptotic expansions, multiscale expansions, and related methods) as well as the specific tools of black hole hole perturbation theory~\cite{Barack:2018yvs,Pound:2021qin}.

Most of this progress in waveform modeling has been driven by calculations in the Fourier domain~\cite{Miller:2020bft,Chua:2020stf,Katz:2021yft,Hughes:2021exa,Isoyama:2021jjd,Pound:2021qin,McCart:2021upc,Wardell:2021fyy}. While there has been continued progress in time-domain calculations~\cite{Thornburg:2016msc,Long:2021ufh,OBoyle:2022yhp,Markakis:2023pfh}, and while it is possible to construct practical surrogate models~\cite{Rifat:2019ltp} from a bank of time-domain waveforms, most development has been on Fourier methods that leverage the disparate time scales in small-mass-ratio binaries: the fast orbital time scale $\sim M$ and the slow time scale $\sim M/\e$ over which the system evolves. This separation of scales allows one to divide waveform generation into two steps: an expensive offline step in which one solves Fourier-domain field equations on a grid of slowly evolving parameter values (e.g., eccentricity, semi-latus rectum, the mass and spin of the primary black hole, etc.); and a fast, cheap online step of solving simple ordinary differential equations (ODEs) to evolve through the parameter space. The flexibility and efficiency of such a framework is exemplified by the \texttt{Fast EMRI Waveforms} package~\cite{Katz:2021yft}. 

This method can be carried to any order in $\e$ by using a multiscale expansion of the Einstein equations~\cite{Miller:2020bft,Pound:2021qin}, which builds on the multiscale form of the companion's orbital motion around the primary~\cite{Hinderer-Flanagan:08}. Orbits around a Kerr black hole generically have three slowly evolving frequencies $\Omega_A=(\Omega_r,\Omega_\theta,\Omega_\phi)$ corresponding to azimuthal motion ($\Omega_\phi$), orbital precession associated with eccentricity ($\Omega_r$), and precession of the orbital plane around the primary's spin axis ($\Omega_\theta$). Any given $(\ell, m)$ multipole of the resulting waveform then takes the simple form~\cite{Pound:2021qin}
\begin{multline}
h_{\ell m} = \sum_{k^i}\left[\e h^{(1,k^i)}_{\ell m}({\cal J}_I) \right.\\
\left.+ \e^2 h^{(2,k^i)}_{\ell m}({\cal J}_I) + \ldots\right] e^{-i(m\varphi_\phi+k^i\varphi_i)},\label{generic waveform}
\end{multline}
where $k^i=(k^r,k^\theta)$ are integers running from $-\infty$ to $+\infty$, ${\cal J}_I$ are the binary's slowly evolving parameters, and $\varphi_A=(\varphi_r,\varphi_\theta,\varphi_\phi)$ are the orbital phases associated with the three frequencies $\Omega_A$. The time dependence of the waveform is governed by simple ODEs of the form
\begin{align}
\frac{d\varphi_A}{du} &= \Omega_A({\cal J}_I),\label{phiAdot}\\
\frac{d{\cal J}_I}{du} &= \e[F^{(0)}_I({\cal J}_K) +\e F^{(1)}_I({\cal J}_K)+\ldots],\label{JIdot}
\end{align}
where $u$ denotes retarded time at future null infinity. The slowly evolving amplitudes $h^{(n,k^i)}_{\ell m}$, frequencies $\Omega_A$, and driving forces $F^{(n)}_I$ are pre-computed in the offline step, and the waveform is then rapidly generated by solving the ODEs~\eqref{phiAdot} and \eqref{JIdot}.

A model that uses only the leading forcing term $F^{(0)}_I$ is referred to as adiabatic (``0PA''); this requires solving the linearized Einstein or Teukolsky equation in the offline step. A model that includes terms up to and including $F^{(n)}_I$ is referred to as $n$th post-adiabatic ($n$PA); this requires solving the Einstein equations through order $\e^{n+1}$. In the offline step, the Einstein equations are formulated in a discrete Fourier domain based on mode expansions in the orbital phases $\varphi_A$, as displayed in Eq.~\eqref{generic waveform}. 

The most advanced self-force calculations use this multiscale method to solve the Einstein equations through second order in $\e$~\cite{Pound-etal:19,Warburton:2021kwk,Wardell:2021fyy}, yielding 1PA waveforms~\cite{Wardell:2021fyy}. Those calculations are currently restricted to the simplest scenario of quasicircular inspirals into Schwarzschild black holes, in which case the problem simplifies because (i) there is only one orbital frequency ($\Omega_\phi$) and its associated azimuthal phase, and (ii) the perturbative Einstein equations on the Schwarzschild background are fully separable.

In Ref.~\cite{Miller:2020bft}, we presented the multiscale Einstein equations for this special case of quasicircular, nonspining binaries. In this paper, we present a method of solving such equations: a worldtube puncture scheme in the Fourier domain. This scheme, which builds on earlier work by Warburton and Wardell~\cite{Warburton-Wardell:14, Wardell:2015ada}, was a key tool in the second-order calculations in Refs.~\cite{Pound-etal:19,Warburton:2021kwk,Wardell:2021fyy}. It extends Refs.~\cite{Warburton-Wardell:14, Wardell:2015ada} by allowing for noncompact sources, irregular boundary conditions, and arbitrary choices of time variable. Its main new ingredients are subtractions of punctures in multiple regions and differing choices of time slicing in different regions. 

Punctures have traditionally been used because self-force calculations work by ``skeletonizing'' the small companion, reducing it to a point-particle singularity. In that context, puncture schemes subtract off the dominant, singular part of the particle's gravitational field in a worldtube surrounding the particle and then solve a field equation for the regular residual piece. First presented in practical forms in Refs.~\cite{Barack-Golbourn-Sago:07,Vega-Detweiler:07}, these schemes are now a standard method in self-force theory~\cite{Wardell:2015kea,Barack:2018yvs,Pound:2021qin}; most pertinently, they have underpinned most descriptions of second-order self-force theory~\cite{Rosenthal:2006nh,Detweiler:12,Pound:12a,Pound:2012dk,Gralla:2012db}. Our scheme leans even more heavily on this puncture method by introducing additional punctures at the black hole horizon and at infinity.

\begin{figure}[t]
\centering
\includegraphics[width=\columnwidth]{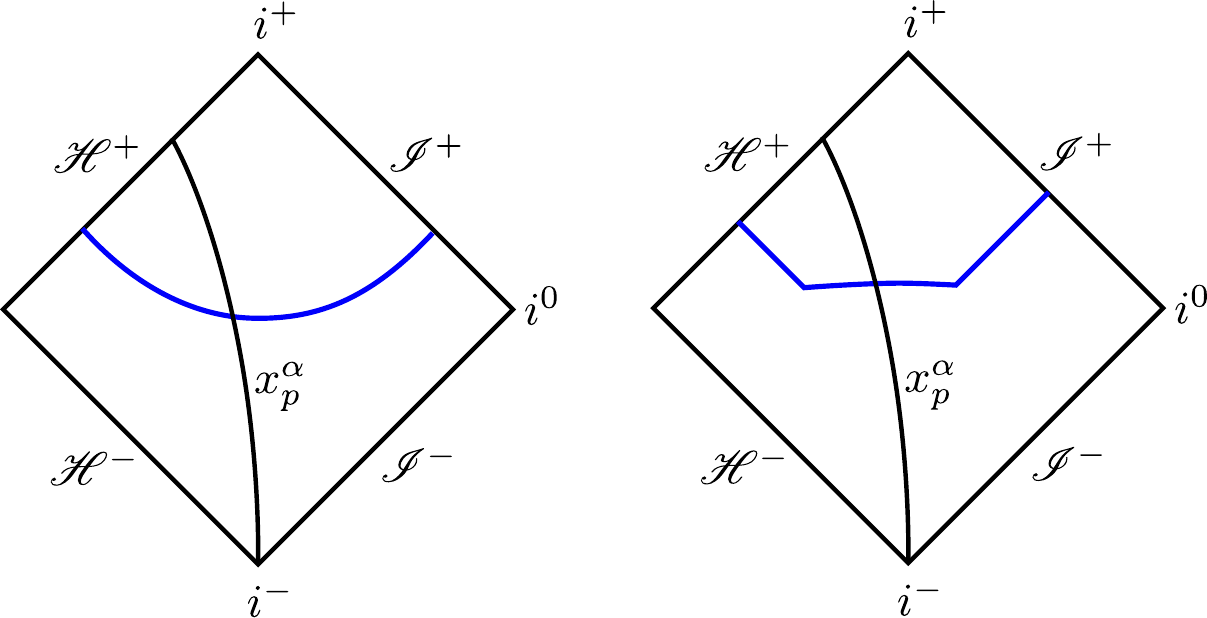}
\caption{Penrose diagrams of the Schwarzschild exterior illustrating a particle's trajectory $x_p^\alpha$ along with slices (blue curves) of constant hyperboloidal time $s=t-k(r^*)$. $s$ transitions from advanced time $v=t+r^*$ near the future horizon to retarded time $u=t-r^*$ near future null infinity. Left: a smooth choice of slicing (which may or may not be everywhere spacelike in the Schwarzschild exterior). Right: sharp slicing in which $s=v$ in a region to the left of the particle, $s=t$ in a region containing the particle, and $s=u$ in a region to the right of the particle.}
\label{fig:penrose_diagram}
\end{figure}

Use of alternative slicings in Fourier-domain self-force calculations is a more recent development. The multiscale field equations in Ref.~\cite{Miller:2020bft}, which were the basis for the calculations in Refs.~\cite{Pound-etal:19,Warburton:2021kwk,Wardell:2021fyy}, were formulated using a hyperboloidal time variable\footnote{We use the term ``hyperboloidal'' loosely. In the standard definition, hyperboloidal slices are required to be everywhere spacelike in the black hole's exterior, while we allow for slices containing null segments.}. Slices of constant hyperboloidal time penetrate the future horizon of the primary black hole and extend to future null infinity rather than to spatial infinity, as illustrated in Fig.~\ref{fig:penrose_diagram}. This has key advantages in a multiscale expansion, significantly improving the behaviour of the source terms in the second-order field equations, reducing the need to derive punctures, and simplifying waveform extraction. More recently, Ref.~\cite{PanossoMacedo:2022fdi} extended this approach by compactifying the hyperboloidal surfaces and working with a spectral method. Those modifications bring additional advantages and are part of a longer-term introduction of hyperboloidal methods into black hole perturbation theory~\cite{Zenginoglu:2007jw,Zenginoglu:2011jz,PanossoMacedo:2019npm}. Here, for historical reasons, we do not adopt these additional tools, but we delineate the relative merits of each method. We are also careful to note that compactified hyperboloidal slices do not evade the fundamental breakdown of the multiscale expansion at large distances; this breakdown, which was explored in Ref.~\cite{Pound:2015wva}, has necessitated the use of an alternative, post-Minkowski expansion at large distances in current second-order self-force calculations. Details of that expansion will be presented elsewhere.

\subsection{Outline and conventions}

We begin in Sec.~\ref{sec_field_equations} by reviewing the multiscale expansion of the Einstein field equations. This review broadly follows Ref.~\cite{Miller:2020bft}'s treatment of quasicircular inspirals into Schwarschild black holes, but we take the opportunity to present that treatment in a more geometrical form that is not tied to the Lorenz gauge or to a tensor-harmonic decomposition. We also discuss how it straightforwardly extends to the case of eccentric orbits.

In Secs.~\ref{sec_twotimescale_expansion} and \ref{sec_teukolsky_hyperboloidal_slicing} we summarize two specific forms of the multiscale equations. Section~\ref{sec_twotimescale_expansion} summarizes the Lorenz-gauge field equations, again following Ref.~\cite{Miller:2020bft}. Here we decompose the multiscale metric perturbation into tensor spherical harmonics, reducing the field equations to a radial ODE for each mode. Since our worldtube scheme is quite generic, in Sec.~\ref{sec_teukolsky_hyperboloidal_slicing} we also present the Teukolsky equation in this multiscale framework, building on recent work in Refs.~\cite{coupling-paper} and~\cite{Spiers:2023cip}.

Sections \ref{sec_worldtube_scheme}--\ref{sec_worldtube_G11} then present our worldtube puncture scheme in a generic form applicable to both the Lorenz-gauge and Teukolsky equations. The method of solving the equations is based on variation of parameters. We consider various formulations of that method and its application to the various types of field equations that arise.

In Secs.~\ref{sec_demonstration_Lorenz} and \ref{sec_demonstration_Teukolsky}, we demonstrate the method in the Lorenz-gauge and Teukolsky versions. The demonstrations consist of solving a field equation for a parametric derivative (a derivative with respect to orbital radius) of the first-order-in-$\e$ field (the metric perturbation in the Lorenz-gauge case and the Weyl scalar in the Teukolsky case). Such parametric derivatives are important in the multiscale expansion because they enter into the source terms in the second-order field equations. In the case of Lorenz-gauge perturbations, we find agreement with results for the same quantity as calculated using a different method in Ref.~\cite{Durkan:2022fvm}.

Finally, in the concluding section, Sec.~\ref{sec_conclusion}, we discuss the relative merits of our variation-of-parameters approach versus the more recent alternatives in Refs.~\cite{Durkan:2022fvm} and~\cite{PanossoMacedo:2022fdi}.

Throughout the paper we use a mostly positive metric signature, $(-,+,+,+)$, and geometrical units with $G=c=1$. Indices are raised and lowered with the background Schwarzschild metric $g_{\alpha\beta}$, and $\nabla$ and a semicolon both denote the covariant derivative compatible with $g_{\alpha\beta}$. $(t,r,\theta,\phi)$ denote Schwarzschild coordinates, in which $g_{\alpha\beta}={\rm diag}\left(-f,f^{-1},r^2,r^2\sin^2\theta\right)$, where $f:=1-2M/r$.

%-------------------------------------------------------------------------------------------------------------------------------------------------------%
%
\section{Einstein field equations in multiscale form}\label{sec_field_equations}
%
%-------------------------------------------------------------------------------------------------------------------------------------------------------%

In this section we review the perturbative Einstein field equations for a binary with a small mass ratio $\e$. We first explain the expansion as formulated in regular perturbation theory and then explain how we formulate it in our multiscale form. We particularly highlight (i) the role of spacetime foliations, (ii) the discrete Fourier expansion of the field equations, and  (iii) the appearance of parametric derivatives as source terms. We refer to Refs.~\cite{Miller:2020bft,Pound:2021qin} for more details. Our formulation here is a more geometrical form of the expansion described in Appendix~A of Ref.~\cite{Miller:2020bft}.

\subsection{Regular perturbation theory}

In regular perturbation theory, we expand the exact metric $\gexact_{\mu\nu}$ and stress-energy tensor $T_{\mu\nu}$ as
\beq
\gexact_{\mu\nu}(y,\e) = g_{\mu\nu}(y) + \e h^{(1)}_{\mu\nu}(y) + \e^2 h^{(2)}_{\mu\nu}(y) + O(\e^3)\label{g expansion}
\eeq
and 
\beq
T_{\mu\nu}(y,\e) = \e T^{(1)}_{\mu\nu}(y) + \e^2 T^{(2)}_{\mu\nu}(y) + O(\e^3),\label{T expansion}
\eeq
where $y$ stands for some coordinates $y^\mu$. At least through second order in $\e$, $T_{\mu\nu}$ can be taken to be the Detweiler stress-energy tensor~\cite{Detweiler:12,Upton:2021oxf},\footnote{We note that because of the strongly divergent fields at second order, the stress-energy tensor~\eqref{Detweiler T} and field equation~\eqref{EFE2} have only been derived in a class of ``highly regular'' gauges and in gauges smoothly related to Lorenz~\cite{Upton:2021oxf}. In the Lorenz-gauge case, the derivation requires adopting a certain canonical distributional definition of $G^{(2)}_{\mu\nu}[h^{(1)},h^{(1)}]$. We will not concretely require that definition here, but we return to this point in the Conclusion, Sec.~\ref{sec_conclusion}. However, we also note that no concrete second-order calculations have directly used Eq.~\eqref{EFE2} but instead have used a puncture scheme, which is a more primitive formulation that does not involve these subtleties. We write the field equations here in terms of a stress-energy tensor only to simplify the presentation. Like the stress-energy tensor, concrete punctures are only available in highly regular and smoothly deformed Lorenz gauges~\cite{Pound:2014xva,Upton:2023tcv}, but this does not restrict the choice of gauge when solving the field equations: independent of the puncture's gauge, the numerical variable (the residual field) can be in any convenient gauge; see, for example, Refs.~\cite{Toomani:2021jlo, Bourg:2024vre}.}
\beq\label{Detweiler T}
T_{\mu\nu} = \mu\int \tilde u_\mu \tilde u_\nu \frac{\delta^4(y-y_p)}{\sqrt{-\tilde g}}d\tilde\tau, 
\eeq
where $\mu$ is the mass of the particle, $y_p^\mu$ is its worldline, $\tilde u_\mu :=\tilde g_{\mu\alpha}\frac{dy_p^\alpha}{d\tilde\tau}$, $\tilde g_{\mu\nu}$ is a certain effective metric of the form
\beq
\tilde g_{\mu\nu}(y,\e) = g_{\mu\nu}(y) + \e h^{{\rm R}(1)}_{\mu\nu}(y) + \e^2 h^{{\rm R}(2)}_{\mu\nu}(y) + O(\e^3),\label{g eff}
\eeq
$\tilde\tau$ is the proper time in that metric, and $h^{{\rm R}(n)}_{\mu\nu}$ are certain smooth vacuum perturbations extracted from $h^{(n)}_{\mu\nu}$. The worldline $y_p^\mu$ obeys the geodesic equation in  $\tilde g_{\mu\nu}$ or an equivalent self-forced equation of motion in $g_{\mu\nu}$~\cite{Pound:12a,Pound:2017psq}.

Given these expansions, the Einstein equation $G_{\mu\nu}[\gexact]=8\pi T_{\mu\nu}$ becomes a hierarchical sequence of equations, the first three of which are
\begin{align}
 G_{\mu\nu}[g] &= 0,\label{EFE0}\\
 G^{(1)}_{\mu\nu}[h^{(1)}] &= 8\pi T^{(1)}_{\mu\nu},\label{EFE1}\\
 G^{(1)}_{\mu\nu}[h^{(2)}] &= 8\pi T^{(2)}_{\mu\nu} - G^{(2)}_{\mu\nu}[h^{(1)},h^{(1)}].\label{EFE2}
\end{align}
Here $G^{(1)}_{\mu\nu}[h]$ is the linearized Einstein tensor constructed from a perturbation $h_{\mu\nu}$, and $G^{(2)}_{\mu\nu}[h,h]$ is the quadratic term in the expansion of the Einstein tensor. We write $G^{(1)}_{\mu\nu}[h]$ as
\beq
G^{(1)}_{\mu\nu}[h] = -\frac{1}{2}E_{\mu\nu}[\bar h] + \nabla_{(\mu}Z_{\nu)}[\bar h] -\frac{1}{2}g_{\mu\nu} \nabla_{\alpha}Z^{\alpha}[\bar h],
\eeq
in terms of the trace-reversed field  
\beq
\bar h_{\mu\nu}:=h_{\mu\nu}-\tfrac{1}{2}g_{\mu\nu}g^{\alpha\beta}h_{\alpha\beta}
\eeq
and the linear operators
\begin{align}
E_{\alpha\beta}[\bar h] &:= \Box \bar h_{\alpha\beta}+2R_\alpha{}^\mu{}_\beta{}^\nu \bar h_{\mu\nu},\\
Z_\alpha[\bar h] &:= g^{\beta\gamma}\nabla_\gamma\bar h_{\alpha\beta},
\end{align}
with  $\Box:=g^{\mu\nu}\nabla_{\!\mu}\nabla_{\!\nu}$. The quadratic Einstein tensor $G^{(2)}_{\mu\nu}[h,h]$ is written explicitly in Eq.~(4) of Ref.~\cite{Pound:2021qin}, but we will not need its explicit form here.

Given the background metric $g_{\mu\nu}$, regular perturbation theory reduces to solving the sequence of linear field equations~\eqref{EFE1}, \eqref{EFE2}, and so on to higher order.

\subsection{Multiscale expansion}\label{sec_multiscale_expansion}

The multiscale expansion of the field equations differs in important ways from a regular expansion. It is designed to maintain uniform accuracy while capturing the binary's ``fast'' evolution on the time scale $\sim M$, due to the orbiting particle, and the ``slow'' evolution on the time scale $\sim M/\e$, due to dissipation.

The method begins with a choice of time function, which we write as% 
\beq%
s=s(t,r) := t - k(r^*).
\eeq
Here $r^*$ is the standard tortoise coordinate, $\rstar = r + 2M \ln \left[r/(2M) - 1 \right]$, and $k(r^*)$ is a height function. We choose $k(r^*)$ such that
\begin{align}
k(r^*) &\to +r^*  \quad\text{for } r^*\to\infty,\\
k(r^*) &\to -r^* \quad\text{for } r^*\to-\infty. 
\end{align}
This ensures that slices of constant $s$ are hyperboloidal, by which we mean they reduce to surfaces of constant advanced time $v=t+r^*$ at the future horizon ($\mathscr{H}^+$) and to surfaces of constant retarded time $u=t-r^*$ at future null infinity ($\mathscr{I}^+$), as illustrated in Fig.~\ref{fig:penrose_diagram}. With our loose definition of the term ``hyperboloidal'', we do not require these slices to be everywhere spacelike in the Schwarzschild exterior, nor do we require them to be smooth.

For most of this paper, we leave $s$ unspecified. However, we mention here our preferred, ``sharp'' foliation used in our numerical calculations. This slicing, illustrated in the right panel of Fig.~\ref{fig:penrose_diagram}, uses a piecewise height function,
\begin{equation}
k(r^*)=\begin{cases}-r^* & \text{for }r <r_1,\\ 
0 & \text{for }r_1 < r < r_2,\\ 
+r^* & \text{for }r > r_2,\end{cases}\label{sharp_slicing}
\end{equation}
where $r_1$ ($r_2$) is a radius smaller (larger) than the particle's orbital radius. With this choice, $s=v$ in a region extending to $\mathscr{H}^+$; $s=t$ in a region containing the particle; and $s=u$ in a region extending to $\mathscr{I}^+$. We will refer to this as $v$-$t$-$u$ slicing. It provides a simple way of avoiding steep gradients within a numerical domain. The sharp change in slicing will cause finite differentiability or jumps in various fields when crossing $r_1$ and $r_2$, but these effects are easily controlled. The switching points are placed at boundaries between domains, and the sharp change in slicing is accounted for using easily derived junction conditions at these domain boundaries (a familiar procedure from dealing with point-particle sources in radial differential equations). 

Given a choice of time function $s$, we assume that the metric's dependence on $s$ is fully encoded in a dependence on the binary's mechanical variables.\footnote{This means that we exclude a variety of effects, including gauges that do not conform to our assumed multiscale time dependence. For example, we exclude gauges that blow up with time and incoming gauge modes that could be sent into the system with arbitrary frequencies. In the case of eccentric orbits, our assumptions also fail at $r-\phi$ resonant kicks~\cite{vandeMeent:2014raa}, though that effect is higher order than we consider here. Hereditary effects associated with gravitational-wave memory (which enter into the multiscale solution at 2PA order through physical boundary conditions at large $r$~\cite{Pound:2015wva,Cunningham:2024}) also introduce a type of integrated dependence on the past history of the mechanical variables, as in post-Newtonian theory~\cite{Blanchet-Damour:88}. However, these hereditary effects are determined by the ``current'' state of the system's slowly evolving variables, meaning they do not spoil our assumptions~\cite{Cunningham:2024}.} For the quasicircular inspirals we focus on here, the time-dependent mechanical variables are (i) the particle's orbital phase $\phi_p$, (ii) its orbital frequency $\Omega:=d\phi_p/ds$, and (iii) corrections to the central black hole's mass and spin, $\e\delta M_A = (\e\delta M, \e\delta J)$ (with the overall factor of $\e$ pulled out to make $\delta M_A$ order unity). The spacetime's slow evolution is encoded in the metric's dependence on the parameters ${\cal J}_I = (\Omega, \delta M_A)$, which evolve due to the dissipative self-force (in the case of $\Omega$) or due to fluxes through the horizon (in the case of $\delta M_A$). The evolution on the fast orbital time scale $\sim 1/\Omega$ is encoded in a periodic dependence on the orbital phase $\phi_p$.
We comment below on how this extends to the case of eccentric orbits.

More concretely, for quasicircular orbits the system's evolution is governed by the rates of change of $\phi_p$ and ${\cal J}_I$. We expand those rates of change in powers of $\e$ at fixed phase-space coordinate values $(\phi_p,{\cal J}_I)$:
\begin{align}
    \dot\phi_p &= \Omega,\label{phidot}\\
    \dot\Omega &= \e\left[ F_{\Omega}^{(0)}(\Omega) + \e F_\Omega^{(1)}(\Omega,\delta M_A) + {\cal O}(\e^2)\right],\\
    \dot{\delta M_A} &= \e\left[ F_A^{(0)}(\Omega) + \e F_A^{(1)}(\Omega,\delta M_B) + {\cal O}(\e^2)\right],\label{Mdot}
\end{align}
where a dot denotes $d/ds$ [cf. Eqs.~\eqref{phiAdot}--\eqref{JIdot}]. In these expansions, the numerical labels denote the post-adiabatic order at which the quantity enters\footnote{This statement assumes that we calculate $F^{(n)}_{\Omega}$ from energy fluxes to infinity and into the horizon, using a balance law, meaning the leading-order horizon flux $F_A^{(0)}(\Omega)$ enters at 0PA order, and the first subleading horizon flux enters at 1PA order. If  $F^{(n)}_{\Omega}$ is instead calculated using the local self-force, then $F_A^{(0)}(\Omega)$ does not enter until 1PA order. $\delta M_A$ itself only enters at 1PA order in either approach.}.  The driving forces $F_{\Omega}^{(n)}$, whose explicit forms are not needed here, are given in terms of the gravitational self-force in Eqs.~(A9)--(A10) of Ref.~\cite{Miller:2020bft}, and $F_A^{(0)}(\Omega)$ are the standard leading-order fluxes of energy and angular momentum into the black hole due to an orbiting particle~\cite{Miller:2020bft,Pound:2021qin}. The particle's orbital trajectory $x_p^i$ in Schwarzschild spatial coordinates $x^i=(r,\theta,\phi)$ takes the simple form
\beq
x^i_p(\phi_p,\Omega,\e) = [r_p(\Omega,\e),\pi/2,\phi_p],
\eeq
where 
\beq
r_p(\Omega,\e) = r_0(\Omega) + \e r_1(\Omega) + O(\e^2).
\eeq
The leading coefficient $r_0=M(M\Omega)^{-2/3}$ is the standard geodesic relationship; the subleading coefficient, which will not be explicitly needed here, is given in Eq.~(A8) of Ref.~\cite{Miller:2020bft}.

In line with our assumption that the spacetime only depends on $s$ through a dependence on $(\phi_p,{\cal J}_I)$, we now treat the metric as a function on an extended manifold that includes the binary's mechanical phase space. Instead of using the regular expansions~\eqref{g expansion} and \eqref{T expansion}, in which we would expand in powers of $\e$ at fixed values of spacetime coordinates $(s,x^i)$, we now expand the metric and the stress-energy tensor in powers of $\e$ at fixed $(x^i,\phi_p,{\cal J}_I)$:
\begin{multline}
\gexact_{\mu\nu}(x^i,\phi_p,{\cal J}_I,\e) = g_{\mu\nu}(x^i) + \e h^{(1)}_{\mu\nu}(x^i,\phi_p,{\cal J}_I) \\+ \e^2 h^{(2)}_{\mu\nu}(x^i,\phi_p,{\cal J}_I) + O(\e^3),\label{g tt expansion}
\end{multline}
and 
\begin{multline}
T_{\mu\nu}(x^i,\phi_p,{\cal J}_I,\e) = \e T^{(1)}_{\mu\nu}(x^i,\phi_p,\Omega) \\+ \e^2 T^{(2)}_{\mu\nu}(x^i,\phi_p,{\cal J}_I) + O(\e^3).\label{T tt expansion}
\end{multline}
The expansion of $T_{\mu\nu}$ is obtained by substituting the multiscale expansions of $x^i_p$ and $\tilde g_{\mu\nu}$ into Eq.~\eqref{Detweiler T};\footnote{We note that Eq.~\eqref{Detweiler T} is derived within a self-consistent expansion~\cite{Pound:10a} that is more general than either regular perturbation theory or our multiscale expansion.} in turn, $\tilde g_{\mu\nu}$ (along with the punctures and residual fields we introduce later) inherits the multiscale structure of $\gexact_{\mu\nu}$ as described in Ref.~\cite{Miller:2020bft}. Each term in these expansions is assumed to be a $2\pi$-periodic function of $\phi_p$. The physical metric on spacetime is obtained once the $s$ dependence of $\phi_p$ and ${\cal J}_I$ is determined via Eqs.~\eqref{phidot}--\eqref{Mdot}. Prior to that determination, we treat $(x^i,\phi_p,{\cal J}_I)$ as independent variables.

Substituting the expansions~\eqref{g tt expansion} and \eqref{T tt expansion} into the Einstein equations, we obtain a modified version of Eqs.~\eqref{EFE1}--\eqref{EFE2}, 
\begin{align}
 G^{(1,0)}_{\mu\nu}[h^{(1)}] &= 8\pi T^{(1)}_{\mu\nu},\label{tt EFE1}\\
 G^{(1,0)}_{\mu\nu}[h^{(2)}] &= 8\pi T^{(2)}_{\mu\nu} - G^{(2,0)}_{\mu\nu}[h^{(1)},h^{(1)}]\nonumber\\&\quad-G^{(1,1)}_{\mu\nu}[h^{(1)}].\label{tt EFE2}
\end{align}
The operators $G^{(n,j)}_{\mu\nu}$ act on functions of $(x^i,\phi_p,{\cal J}_I)$. They are derived from $G^{(n)}_{\mu\nu}$ using the chain rule
\beq\label{chain rule}
\frac{\partial}{\partial y^\alpha} = e^i_\alpha \frac{\partial}{\partial x^i} + s_\alpha\left(\frac{d\phi_p}{ds}\frac{\partial}{\partial\phi_p} +\frac{d{\cal J}_I}{ds}\frac{\partial}{\partial {\cal J}_I}\right),
\eeq
where 
\beq
e^i_\alpha := \frac{\partial x^i}{\partial y^\alpha} \quad \text{and} \quad s_\alpha := \frac{\partial s}{\partial y^\alpha}
\eeq
are a basis of one-forms. $s^\alpha$ is normal to surfaces of constant $s$. If $x^i=(r,\theta,\phi)$, then  $e^r_\alpha s^\alpha = -dk/dr^*$ and $e^i_\alpha s^\alpha=0$ for $i=\theta$ or $\phi$. Note that in the simple case $y^\alpha=(s,x^i)$, $e^i_\alpha$ and $s_\alpha$ reduce to $\delta^i_\alpha$ and $\delta^s_\alpha$, but here we leave $y^\alpha$ generic.
 
Given Eqs.~\eqref{phidot}--\eqref{Mdot}, the chain rule~\eqref{chain rule} implies the expansion
\beq
\nabla_\alpha = \nabla^{(0)}_\alpha + \e s_\alpha \vec{\partial}_{\cal V} + {\cal O}(\e^2),
\eeq
where the zeroth-order covariant derivative is 
\beq
\nabla^{(0)}_\alpha = e^i_\alpha\frac{\partial}{\partial x^i} + s_\alpha \Omega\frac{\partial}{\partial\phi_p} + \text{Christoffel terms},\label{nabla0}
\eeq
${\cal V}_I=(F^{(0)}_\Omega,F^{(0)}_A)$ is the leading-order velocity through parameter space, and 
\beq\label{param D quasicircular}
\vec{\partial}_{\cal V}:={\cal V}_I\frac{\partial}{\partial{\cal J}_I} = F_\Omega^{(0)}\frac{\partial}{\partial\Omega} + F_A^{(0)}\frac{\partial}{\partial\delta M_A}
\eeq
is a directional derivative in the parameter space.

Using the above expansion of the covariant derivative, we see that $G^{(n,0)}_{\mu\nu}$ is given by $G^{(n)}_{\mu\nu}$  with $\nabla_\alpha\to \nabla^{(0)}_\alpha$. $G^{(n,1)}_{\mu\nu}$ is given by the terms in $G^{(n)}_{\mu\nu}$ that are linear in the velocity ${\cal V}_I$. Explicitly, the operator that appears at 1PA order is
\begin{multline}
G^{(1,1)}_{\mu\nu}[h] = -\frac{1}{2}E^{(1)}_{\mu\nu}[\bar h] + \nabla^{(0)}_{(\mu}Z^{(1)}_{\nu)}[\bar h] + \nabla^{(1)}_{(\mu}Z^{(0)}_{\nu)}[\bar h]\\ 
-\frac{1}{2}g_{\mu\nu}\left(\nabla^{(0)}_{\alpha}Z_{(1)}^{\alpha}[\bar h] +\nabla^{(1)}_{\alpha}Z_{(0)}^{\alpha}[\bar h]\right).\label{G11}
\end{multline}
The individual terms in this expression are
\begin{align}
E^{(1)}_{\mu\nu}[\bar h] &= s_\alpha s^\alpha F_\Omega^{(0)}\partial_{\phi_p}\bar h_{\mu\nu} + 2 s^\alpha\nabla^{(0)}_\alpha\!\left(\vec{\partial}_{\cal V}\bar h_{\mu\nu}\right) \nonumber\\
&\quad +\left(\nabla^{(0)}_\alpha s^\alpha\right)\vec{\partial}_{\cal V} \bar h_{\mu\nu},\label{E1}\\
Z^{(1)}_\mu[\bar h] &= s^\alpha \vec{\partial}_{\cal V}\bar h_{\mu\alpha},\label{Z1}\\
\nabla^{(1)}_{\mu}Z^{(0)}_{\nu}[\bar h] &= s_\mu \vec{\partial}_{\cal V} Z^{(0)}_\nu[\bar h].
\end{align}
The contractions involving $s_\alpha$ evaluate to
\begin{align}
    s_\alpha s^\alpha &= -f^{-1}(1-H^2),\\
    \nabla^{(0)}_\alpha s^\alpha &= -\frac{dH}{dr} -\frac{2H}{r},\\
    s^\alpha\nabla^{(0)}_\alpha &= -f^{-1}(1-H^2)\Omega\partial_{\phi_p} - H \partial_r \nonumber\\
    &\quad +\text{Christoffel terms},\label{s.del}
\end{align}
where 
\beq
H:=\frac{dk}{dr^*}. 
\eeq
These equations simplify significantly if $s=v$ (meaning $H=-1$), $s=t$ (meaning $H=0$), or $s=u$ (meaning $H=1$). We repeatedly return to those special cases in later sections.

For any choice of $s$, the field equations~\eqref{tt EFE1}--\eqref{tt EFE2} reduce to partial differential equations in $(x^i,\phi_p)$. These can be solved at fixed values of ${\cal J}_I$ because derivatives with respect to ${\cal J}_I$ only appear as sources, in the term $G^{(1,1)}_{\mu\nu}[h^{(1)}]$.

\subsection{Fourier expansion}

Since all functions of $\phi_p$ are periodic, we can expand them in Fourier series:
\begin{align}
h^{(n)}_{\alpha\beta}(x^i,\phi_p,{\cal J}_I) &= \sum_{m=-\infty}^\infty h^{(n,m)}_{\alpha\beta}(x^i,{\cal J}_I)e^{-im\phi_p},\label{hn Fourier}\\
T^{(n)}_{\alpha\beta}(x^i,\phi_p,{\cal J}_I) &= \sum_{m=-\infty}^\infty T^{(n,m)}_{\alpha\beta}(x^i,{\cal J}_I)e^{-im\phi_p}.
\end{align}
We then have $\frac{\partial}{\partial\phi_p} \to -im$ when acting on individual modes, implying
\beq\label{del0 modes}
\nabla^{(0)}_\alpha \to e^i_\alpha\frac{\partial}{\partial x^i} -i s_\alpha \omega_m + \text{Christoffel terms},
\eeq
where $\omega_m := m\Omega$. 

By substituting this Fourier expansion into the field equations~\eqref{tt EFE1}--\eqref{tt EFE2}, we obtain decoupled differential equations in $x^i$ for each mode $h^{(n,m)}_{\alpha\beta}(x^i,{\cal J}_I)$. Again, these can be solved at fixed values of ${\cal J}_I$. As summarized around Eqs.~\eqref{generic waveform}--\eqref{JIdot}, the waveform-generation scheme used in Ref.~\cite{Wardell:2021fyy} then consists of (i) computing and storing the waveform amplitudes $\lim_{r\to\infty}rh^{(n,m)}_{\alpha\beta}$ and the driving forces $F^{(n)}_I$ on a grid of $\Omega$ values, (ii) solving Eqs.~\eqref{phidot}--\eqref{Mdot} to generate a trajectory through phase space, and (iii) substituting the trajectory into $\lim_{r\to\infty}\sum_m r h^{(n,m)}_{\alpha\beta}[{\cal J}_I(s)]e^{-im\phi_p(s)}$ to obtain the waveform.

Before proceeding, we stress that the discrete Fourier series~\eqref{hn Fourier} is \emph{not} a Fourier transform in time. It would only be a Fourier transform if ${\cal J}_I$ were independent of $s$ and if $\phi_p(s)$ were equal to $\Omega s$ (with constant $\Omega$). Neither of these conditions hold true for the inspiraling system. Therefore, we are careful to refer to the resulting field equations as being in the Fourier domain but not being in the frequency domain. However, readers familiar with frequency-domain equations can apply virtually all their knowledge directly to our Fourier-domain equations: the left-hand side of the field equation for a mode coefficient $h^{(n,m)}_{\alpha\beta}(x^i,{\cal J}_I)$ is identical to the left-hand side of the field equation for the mode coefficient in a frequency-domain expansion $h^{(n,\omega)}_{\alpha\beta}(x^i)e^{-i \omega s}$ (with $\omega=\omega_m$).

We also note that the above description extends, with only minor changes, to the case of eccentric orbits in Schwarzschild. In that case there are two orbital phases, $\varphi_A=(\varphi_r,\varphi_\phi)$, and associated frequencies $\Omega_A=(\Omega_r,\Omega_\phi)$. The Fourier expansion~\eqref{hn Fourier} becomes $\sum_{m,k} h^{(n,m,k)}_{\alpha\beta}(x^i,{\cal J}_I)e^{-im\varphi_\phi-ik \varphi_r}$ [cf. the expansion~\eqref{generic waveform} in Kerr spacetime]. The adiabatically evolving parameters become ${\cal J}_I=(p_i,\delta M_A)$, for example, where $p_i=(p,e)$ are the semilatus rectum and eccentricity. The chain rule~\eqref{chain rule} becomes
\beq
\frac{\partial}{\partial y^\alpha} = e^i_\alpha \frac{\partial}{\partial x^i} + s_\alpha\left(\frac{d\varphi_A}{ds}\frac{\partial}{\partial\varphi_A} +\frac{d{\cal J}_I}{ds}\frac{\partial}{\partial {\cal J}_I}\right),
\eeq
meaning we make the following replacements: ${\cal V}_I\to (F^{(0)}_i,F^{(0)}_A)$ and the corresponding adjustment
\beq\label{param D eccentric}
\vec{\partial}_{\cal V}\to F_i^{(0)}\frac{\partial}{\partial p^i} + F_A^{(0)}\frac{\partial}{\partial\delta M_A};
\eeq
$\Omega\partial_{\phi_p}\to \Omega_A\partial_{\varphi_A}$ in Eqs.~\eqref{nabla0} and \eqref{s.del}; $F_{\Omega}^{(0)}\partial_{\phi_p}\to F^{(0)}_i\frac{\partial\Omega_A}{\partial p_i}\partial_{\varphi_A}$ in Eq.~\eqref{E1}; and $\omega_m\to \omega_{m,k}=m\Omega_\phi+k\Omega_r$ in Eq.~\eqref{del0 modes} [with $k$ the integer labeling $h^{(n,m,k)}_{\alpha\beta}$, not to be confused with the height function $k(r^*)$]. We refer to Ref.~\cite{Pound:2021qin} for more details.

%-------------------------------------------------------------------------------------------------------------------------------------------------------% 
\section{Lorenz-gauge field equations}\label{sec_twotimescale_expansion}
%-------------------------------------------------------------------------------------------------------------------------------------------------------%

In the calculations in Refs.~\cite{Pound-etal:19,Warburton:2021kwk,Wardell:2021fyy}, and in much of this paper, we work in the Lorenz gauge. Here we review the Lorenz-gauge field equations as formulated in Ref.~\cite{Miller:2020bft}.

\subsection{Four-dimensional form}

We impose the Lorenz gauge condition 
\beq\label{Lorenz gauge}
Z_\mu[\bar\hexact]=0, 
\eeq
where  $\hexact_{\mu\nu}:=\gexact_{\mu\nu}-g_{\mu\nu}$ is the total perturbation and $\bar\hexact_{\mu\nu}$ is its trace reverse with respect to $g_{\mu\nu}$. This reduces the linearized Einstein tensor to 
\beq
G^{(1)}_{\mu\nu}[\hexact] = -\frac{1}{2}E_{\mu\nu}[\bar \hexact].
\eeq
Following Ref.~\cite{Barack:2005nr}, in order to partially decouple the field equations, we use a modified operator
\beq\label{E damped}
\breve E_{\mu\nu} := E_{\mu\nu} - \frac{4M}{r^2}t_{(\mu}\breve Z_{\nu)},
\eeq
where $t_\alpha:=\partial_\alpha t$ and $\breve Z_\alpha = (Z_r,2Z_r,Z_\theta,Z_\phi)$ in Schwarzschild coordinates $(t,r,\theta,\phi)$; note $\breve Z_\alpha[\bar\hexact]$ vanishes if the gauge condition is satisfied. The complete Einstein equation is then
\beq
\breve E_{\mu\nu}[\bar\hexact] = -16\pi T_{\mu\nu} + 2\breve{G}^{(2)}_{\mu\nu}[\hexact,\hexact] + {\cal O}(|\hexact|^3).
\eeq
Here $\breve{G}^{(2)}_{\mu\nu}$ is given by $G^{(2)}_{\mu\nu}$ with $Z_\alpha$ set to zero. 

Performing a multiscale expansion leads to a slightly modified version of the hierarchy~\eqref{tt EFE1} and \eqref{tt EFE2},
\begin{align}
 \breve E^{(0)}_{\mu\nu}[\bar h^{(1)}] &= -16\pi T^{(1)}_{\mu\nu},\label{tt EFE1 Lorenz}\\
 \breve E^{(0)}_{\mu\nu}[\bar h^{(2)}] &= -16\pi T^{(2)}_{\mu\nu} + 2\breve G^{(2,0)}_{\mu\nu}[h^{(1)},h^{(1)}]\nonumber\\
 &\quad-\breve E^{(1)}_{\mu\nu}[\bar h^{(1)}].\label{tt EFE2 Lorenz}
\end{align}
The labels here have the same meaning as in the previous section: ``$(0)$'' on an operator indicates the operator with the replacement $\nabla_\alpha\to\nabla^{(0)}_\alpha$, and ``$(1)$'' indicates the term linear in ${\cal V}_I$. Explicitly, in terms of the operators in Eqs.~\eqref{E1} and~\eqref{Z1},
\beq
\breve E^{(1)}_{\mu\nu}[\bar h] = E^{(1)}_{\mu\nu}[\bar h] - \frac{4M}{r^2}t_{(\mu}\breve Z_{\nu)}^{(1)}[\bar h]
\eeq
with $\breve Z^{(1)}_\alpha = (Z^{(1)}_r,2Z^{(1)}_r,Z^{(1)}_\theta,Z^{(1)}_\phi)$ in Schwarzschild coordinates.

Similarly, after the multiscale expansion, the gauge condition~\eqref{Lorenz gauge} becomes
\begin{align}
Z^{(0)}_\mu[\bar h^{(1)}] &= 0,\label{tt gauge 1}\\
Z^{(0)}_\mu[\bar h^{(2)}] &= -Z^{(1)}_\mu[\bar h^{(1)}].\label{tt gauge 2}
\end{align}

%-------------------------------------------------------------------------------------------------------------------------------------------------------%
\subsection{Tensor-harmonic decomposition}
%-------------------------------------------------------------------------------------------------------------------------------------------------------% 

We next decompose the fields into tensor spherical harmonic modes, using the Barack-Lousto-Sago (BLS) basis of harmonics~\cite{Barack:2005nr,Barack:2007tm}: 
\begin{equation}
\bar h^{(n)}_{\alpha\beta}=\sum_{i\ell m} \frac{a_{i\ell}}{r}\bar h^{(n)}_{i\ell m}(r,{\cal J}_I)Y_{\alpha\beta}^{i\ell m}(r,\theta,\phi)e^{-im\phi_p},\label{modeDecomposition}
\end{equation}
where $i=1,\ldots,10$, $\ell\geq0$, $m=-\ell,\ldots,\ell$. The harmonics $Y^{i\ell m}_{\alpha\beta}$ provide an orthogonal basis for symmetric rank-2 tensors. They are given explicitly in Appendix~B of Ref.~\cite{Miller:2020bft}. $a_{i\ell}$  is a convenient numerical factor given by
\begin{equation}
a_{i\ell}=\begin{cases}
\frac{1}{\sqrt{2}} \quad &\text{ for }i=1,2,3,6,\\ 
\frac{1}{\sqrt{2\ell(\ell+1)}} \quad &\text{ for } i=4,5,8,9,\\
\frac{1}{\sqrt{2(\ell-1)\ell(\ell+1)(\ell+2)}} \quad &\text{ for }i=7,10.\label{a_il}
\end{cases}
\end{equation}
Following BLS, we have also pulled out a factor of $1/r$ in Eq.~\eqref{modeDecomposition} to simplify the field equations. 

We similarly decompose the source terms in Eqs.~\eqref{tt EFE1 Lorenz} and \eqref{tt EFE2 Lorenz}. Denoting the $n$th-order source term as $S^{(n)}_{\mu\nu}$, we write
\begin{align}
S^{(n)}_{\mu\nu} &= \sum_{i\ell m}S^{(n)}_{i\ell m}(r,{\cal J}_I)Y^{i\ell m}_{\mu\nu}e^{-im\phi_p}.\label{multiscale-S}
\end{align}

The mode number $m$ appearing in the Fourier decomposition~\eqref{hn Fourier} is the same as the azimuthal mode number in the spherical harmonics, such that each mode in Eqs.~\eqref{modeDecomposition} and ~\eqref{multiscale-S} has a simple dependence on $(\phi-\phi_p)$:
\beq
Y_{\alpha\beta}^{i\ell m}(r,\theta,\phi)e^{-im\phi_p} = Y_{\alpha\beta}^{i\ell m}(r,\theta,0)e^{im(\phi-\phi_p)}.
\eeq
This implies that when a mode of $\bar h^{(n)}_{\alpha\beta}$ is evaluated on the worldline (where $r=r_p$, $\phi=\phi_p$, and $\theta=\pi/2$), it reduces to a function of ${\cal J}_I$, with no dependence on $\phi_p$. The same is true of derivatives of $\bar h^{(n)}_{\alpha\beta}$, such as those that enter the self-force.

With these harmonic expansions, Eqs.~\eqref{tt EFE1 Lorenz} and \eqref{tt EFE2 Lorenz} each separate into a set of ten ODEs for the coefficients $\bar h^{(n)}_{i\ell m}$, which read \cite{Barack:2007tm} 
\begin{equation}\label{decomposed EFE}
E^{(0)}_{ij\ell m}\bar h^{(n)}_{j\ell m} = -\frac{r f}{4 a_{i\ell}}  S^{(n)}_{i\ell m}.
\end{equation}
Here the mode label $j$ is summed over. The decomposed wave operator is given by 
\beq\label{Eijlm}
E^{(0)}_{ij\ell m}\bar h_{j\ell m} := \Box^{(0)}_{\ell m}\bar h_{i\ell m} + {\cal M}_{ij}^{(0)}\bar h_{j\ell m},
\eeq
where 
\begin{multline}
\Box^{(0)}_{\ell m} := -\frac{1}{4}\Big[\partial_{r^*}^2 +i\omega_m  \left(2H \partial_{r^*}+H'\right)\\
		 +\left(1-H^2\right)\omega_m^2-4V_\ell(r)\Big].\label{tildeBox0}
\end{multline}
Here $H' :=\frac{dH}{dr^*}$, and 
\begin{equation} 
V_\ell(r)=\frac{f}{4}\left(\frac{2M}{r^3}+\frac{\ell(\ell+1)}{r^2} \right).\label{Vl}
\end{equation}
${\cal M}^{(0)}_{ij}$ with $i,j=1,\ldots,10$ are a set of matrices composed of first-order differential operators that couple between the various $\bar h_{j\ell m}$'s. Note that the coupling is only between different $j$'s; there is no coupling between modes of different $\ell$ and $m$. Also note that the only effect of our added $\breve Z_\nu$ term in Eq.~\eqref{E damped} is to alter these coupling matrices (reducing the coupling). The explicit form of the coupling matrices can be found in Appendix~\ref{sec_Mij0}. 

The source terms in the decomposed field equations~\eqref{decomposed EFE} are
\begin{align}
S^{(1)}_{i\ell m} &= -16\pi T^{(1)}_{i\ell m}, \label{Slm1}\\
S^{(2)}_{i\ell m} &= -16\pi T^{(2)}_{i\ell m} + 2G^{(2,0)}_{i\ell m}[h^{(1)},h^{(1)}] \nonumber\\
&\quad - E^{(1)}_{ij\ell m}\bar h^{(1)}_{j\ell m}. \label{Slm2}
\end{align}
The quadratic term, $G^{(2,0)}_{i\ell m}[h^{(1)},h^{(1)}] $, is discussed in detail in Ref.~\cite{coupling-paper}. Here we  highlight the term involving $E^{(1)}_{ij\ell m}$, which is the decomposition of $\breve E^{(1)}_{\mu\nu}$ in Eq.~\eqref{tt EFE2 Lorenz}. This term represents the system's slow evolution acting as a source for the second-order metric perturbation. Explicitly, 
\beq\label{E1ijlm}
E^{(1)}_{ij\ell m}\bar h_{j\ell m} := \Box^{(1)}_{\ell m}\bar h_{i\ell m} + {\cal M}_{ij}^{(1)}\bar h_{j\ell m},
\eeq
where
\begin{multline}
 \Box^{(1)}_{\ell m} = \frac{1}{4}\Big[\left(2 H\partial_{r^*}+ H'\right)\vec\partial_{\cal V}\\
			 -\left(1-H^2\right)\left(2i\omega_m\vec\partial_{\cal V}+i m F^{(0)}_{\Omega}\right)\Big].\label{tildeBox1}
\end{multline}
The coupling matrices ${\cal M}_{ij}^{(1)}$ are given in Eq.~\eqref{Mij1}.

Similarly, at the level of modes, the gauge conditions~\eqref{tt gauge 1} and \eqref{tt gauge 2} become
\begin{equation}
Z^{(0)}_{kj\ell}\bar h^{(1)}_{j\ell m} = 0\label{eq:gauges-w0}
\end{equation}
and
\begin{equation}
Z^{(0)}_{kj\ell}\bar h^{(2)}_{j\ell m} = - Z^{(1)}_{kj\ell}\bar h^{(1)}_{j\ell m},\label{eq:gauges-w1}
\end{equation}
where $k=1,\ldots,4$. We give the operators $Z^{(n)}_{kj\ell}$ in Eqs.~\eqref{eq:gauge1-w0} and \eqref{eq:gauge1-w1}.

For $H=0$ (i.e., $t$ slicing), $E^{(0)}_{ij\ell m}$ and $Z^{(0)}_{kj\ell}$ are the same operators that appear in the standard frequency-domain Lorenz-gauge linearized field equations for a metric perturbation $\bar h_{\mu\nu}=\frac{a_{i\ell}}{r}\bar h_{i\ell m}(r)Y^{i\ell m}_{\mu\nu}e^{-i\omega  t}$, as in Refs.~\cite{Akcay:2010dx,Wardell:2015ada}. 

The equations in this section apply equally well for eccentric orbits, with the replacements $\omega_m\to \omega_{m,k}=m\Omega_\phi+k\Omega_r$ and $imF_{\Omega}^{(0)}\to -F^{(0)}_i\frac{\partial\omega_{m,k}}{\partial p_i}$.

\subsection{Matrix form}

For each $\ell m$ mode, the field equations \eqref{decomposed EFE} represent 10 coupled ODEs. However, these ODEs partially decouple into a hierarchical structure; see Table~I of Ref.~\cite{Miller:2020bft}. Even-parity modes ($i=1,\ldots,7$) decouple from odd-parity modes ($i=8,9,10$). Moreover, the $i=1,3,5,6,7$ modes  decouple from the $i=2,4$ modes, and the $i=9,10$ modes decouple from the $i=8$ mode; this allows one to calculate the $i=2,4$ modes from the $i=1,3,5,6,7$ modes and the $i=8$ mode from the  the $i=9,10$ modes. One can also often use the gauge condition to algebraically obtain some modes from others. The number of relevant modes is further reduced by the facts that (i) even-parity (odd-parity) modes vanish for $\ell+m$ odd ($\ell+m$ even), and (ii) $m<0$ modes can be computed from $m>0$ modes using $\bar h_{i\ell m} = (-1)^m(\bar h_{i\ell,-m})^*$. 

It will be convenient to write these sets of equations in a canonical matrix form,
\begin{equation}
{\cal D}\psi:=\frac{d^2\psi}{dr^2}+B\frac{d\psi}{dr}+A\psi = J,\label{fieldEquations2}
\end{equation}
where $\psi(r)$ and $J(r)$ are column vectors with $d$ elements, and $A$ and $B$ are $r$-dependent $d\times d$ matrices. We further write this in first-order form as
\begin{equation}
\hat{\cal D}\hat\psi := \frac{d\hat\psi }{dr}+\hat A\hat \psi = \hat J,\label{fieldEquations3v1}
\end{equation}
where  $\hat\psi=\begin{pmatrix}\psi\\ \partial_r\psi\end{pmatrix}$, $\hat J(r)=\begin{pmatrix}\mathbf{0}_d\\ J(r) \end{pmatrix}$ are  $2d$-vectors and
\begin{equation}
\hat A=\begin{pmatrix}\mathbf{0}_{d\times d} & -\mathbf{1}_{d\times d}\\A &B\end{pmatrix}.\label{hat_A_here}
\end{equation}

For our Lorenz-gauge field equations, the column vector $\psi_{\ell m}$ is
\begin{equation}\label{psiColumnVector}
\psi = \begin{cases}
\left(\bar h_9\,\bar h_{10}\right)^T & \ell\geq2, \ell+m\text{ odd},\\
\left(\bar h_1\,\bar h_3\, \bar h_5\,\bar h_6\,\bar h_7\right)^T & \ell\geq2, \ell+m\text{ even}, m>0,\\
\left(\bar h_1\,\bar h_3\, \bar h_5\right)^T & \ell\geq2\text{ even}, m=0,\\
\left(\bar h_1\,\bar h_3\, \bar h_5\,\bar h_6\right)^T & \ell=1, m=1,\\
\bar h_9 & \ell=1, m=0,\\
\left(\bar h_1\,\bar h_3\right)^T & \ell=0,
\end{cases}
\end{equation}
where $T$ denotes the transpose and $\ell m$ labels are suppressed. We then define the $2d$-vector $\hat\psi_{\ell m}=\left(\psi_{\ell m},\partial_r\psi_{\ell m}\right)^T$. The matrices appearing in Eq.~\eqref{hat_A_here} are
\begin{subequations} \label{AB_gravity}
\begin{align}
A &= \frac{1}{f^2}\left[\omega_m^2\left(1-H^2\right) + i\omega_m H' - 4 V_\ell\right]\mathbf{1}_{d\times d}+{\cal M }_{h},\\
B &= \frac{1}{f}\left(\frac{2 M}{r^2} + 2i\omega_m H \right)\mathbf{1}_{d\times d} + {\cal M}_{\partial h},
\end{align} \end{subequations} 
where ${\cal M}_{h}$ and ${\cal M}_{\partial h} $ are $d\times d$ matrices given in Appendix~\ref{sec_Mij0}.

We translate the sources in the same way. If $\bar h_{i\ell m}$ satisfies an equation $E^{(0)}_{ij\ell m}\bar h_{j\ell m} = -\frac{rf}{4a_{i\ell}}S_{i\ell m}$, as in Eq.~\eqref{decomposed EFE}, then the sources $J$ in Eqs.~\eqref{fieldEquations2} and~\eqref{fieldEquations3v1} are
\begin{equation}\label{JColumnVector}
J = \frac{r}{a_{i\ell}f}\begin{cases}
\left(S_9\,S_{10}\right)^T & \ell\geq2, \ell+m\text{ odd}, m>0,\\
\left(S_1\,S_3\, S_5\,S_6\,S_7\right)^T & \ell\geq2, \ell+m\text{ even}, m>0,\\
\left(S_1\,S_3\,S_5\right)^T & \ell\geq2\text{ even}, m=0,\\
\left(S_1\,S_3\,S_5\,S_6\right)^T & \ell=1, m=1,\\
S_9 & \ell=1, m=0,\\
\left(S_1\,S_3\right)^T & \ell=0,
\end{cases}
\end{equation}
again with $\ell m$ labels suppressed.

For each $\ell m$, the modes that are missing from Eq.~\eqref{psiColumnVector} can be obtained from the listed modes using the gauge condition~\eqref{eq:gauges-w0} or \eqref{eq:gauges-w1}. These ``gauge modes'' are $\bar h_8$ (for $\ell+m$ odd), $\bar h_2$ and $\bar h_4$ (for $\ell>0$ and $\ell+m$ even), and $\bar h_2$ and $\bar h_6$ (for $\ell=0$).

In the case of the second-order field, it will be useful to further divide the field into two pieces,
\beq\label{psi2 split}
\hat\psi^{(2)} = \hat\psi^{(2,0)} + \hat\psi^{(1,1)},
\eeq
each satisfying its own field equation,
\begin{align}
\hat{\cal D}\hat\psi^{(2,0)} &= \hat J^{(2,0)},\label{fieldEquation20}\\
\hat{\cal D}\hat\psi^{(1,1)} &= \hat J^{(1,1)}.\label{fieldEquation11}
\end{align}
Here $\hat J^{(2,0)}$ is constructed from the subset of source terms in Eq.~\eqref{Slm2} that do not involve the forcing functions $F^{(0)}_I$,
\beq\label{S20}
S^{(2,0)}_{i\ell m} = -16\pi T^{(2,0)}_{i\ell m} + 2G^{(2,0)}_{i\ell m}[h^{(1)},h^{(1)}],
\eeq
and  $\hat J^{(1,1)}$ is constructed from the subset of source terms that are linear in the forcing functions,
\beq\label{S11}
S^{(1,1)}_{i\ell m} = -16\pi T^{(1,1)}_{i\ell m} - E^{(1)}_{ij\ell m}\bar h^{(1)}_{j\ell m}.
\eeq
In these expressions we have analogously split the stress-energy tensor into a piece ($T^{(2,0)}_{i\ell m}$) that is independent of $F^{(0)}_I$ and a piece ($T^{(1,1)}_{i\ell m}$) that is linear in it. In the present paper we will not require the explicit expressions for these two pieces; we only introduce the split to help organize discussions in later sections.

Despite the convenient split into $\hat\psi^{(2,0)}$ and $\hat\psi^{(1,1)}$, we stress that these fields are not actually independent: they are coupled through the gauge condition~\eqref{eq:gauges-w1}, which is only satisfied by the sum of the two fields.

\subsection{Boundary conditions, punctures, and slicing transformations}\label{sec_BCs}

Self-force calculations can encounter problematic divergent behavior in three regions: at the particle, near $\mathscr{H}^+$, and near $\mathscr{I}^+$.\footnote{Additionally, regular perturbation theory diverges on long time scales~\cite{Pound:2015wva}. This failure is overcome using a multiscale expansion, as we use here, or using a self-consistent expansion~\cite{Pound:10a} (if the latter is extended to account for the black hole's slow evolution, as described in Ref.~\cite{Miller:2020bft}).} The nature of the problem depends on the particular formulation of the small-$\e$ expansion, on the choice of gauge, and on the choice of time foliation. Punctures provide a practical way of enforcing physical boundary conditions in the presence of these divergences.  

To motivate the use of punctures, we first recall their use in controlling the divergence at the particle, which is the most familiar problem. Fundamentally, an expansion in the limit $\e\to0$ (at fixed external length scales) breaks down at distances $\sim \e$ from the small companion; there, the gravity of the small body dominates over the external gravity. Through a local analysis in that region, using the method of matched asymptotic expansions, one finds the correct local behavior of the physical solution~\cite{Barack:2018yvs}. The form of that solution, outside the body, is 
\beq\label{S-R split}
h^{(n)}_{\mu\nu}=h^{{\rm S}(n)}_{\mu\nu} +h^{{\rm R}(n)}_{\mu\nu}. 
\eeq
The self-field $h^{{\rm S}(n)}_{\mu\nu}$ captures local information about the body's multipole structure and diverges if analytically extended down to the body's representative worldline. The regular field $h^{{\rm R}(n)}_{\mu\nu}$ is a vacuum solution that depends on global boundary conditions and is smooth when analytically extended to the worldline. 

We then adopt an asymptotic matching condition, which is a type of boundary condition: near the representative worldline, the metric perturbations must recover the local form obtained from matched asymptotic expansions. The point-mass representation~\eqref{Detweiler T} and a puncture scheme are two differing ways to enforce this condition. Both methods use the analytical extension of Eq.~\eqref{S-R split} down to the worldline. The point-mass representation enforces the matching condition by defining source terms for $h^{(n)}_{\mu\nu}$ such that all solutions to the inhomogeneous field equation recover the correct local form~\eqref{S-R split}; we refer to Ref.~\cite{Upton:2021oxf} for further discussion. A puncture scheme instead imposes the matching condition by directly using the local form~\eqref{S-R split}. We  construct a local approximation to $h^{{\rm S}(n)}_{\mu\nu}$, called a puncture field $h^{{\cal P}(n)}_{\mu\nu}$, and then solve field equations for the residual field $h^{{\cal R}(n)}_{\mu\nu}:=h^{(n)}_{\mu\nu}-h^{{\cal P}(n)}_{\mu\nu}$.  In our generic matrix form, we write these field equations as
\begin{equation}
\hat{\cal D}\hat\psi^{\cal R} = \hat J - \hat{\cal D}\hat\psi^{\cal P} =: \hat J^{\rm eff}.\label{psiRfieldEquations}
\end{equation}
The puncture field is made to vanish outside some region around the particle, such that $\hat\psi^{\cal R}$ becomes the physical field outside that region. By solving Eq.~\eqref{psiRfieldEquations} with physical boundary conditions at $\mathscr{H}^+$ and $\mathscr{I}^+$ and adding the puncture, we then obtain a total field $\hat\psi^{\cal R}+\hat\psi^{\cal P}$ that necessarily satisfies the matching condition. We note that the punctures $h^{{\cal P}(n)}_{\mu\nu}$ and residual fields $h^{{\cal R}(n)}_{\mu\nu}$ possess the same multiscale form as $h^{(n)}_{\mu\nu}$ because $h^{{\cal P}(n)}_{\mu\nu}$ is an explicit function of the orbital trajectory.

This illustrates how a puncture scheme is simply a method of imposing a boundary condition. Suppose, more generally, we are given the boundary condition that the total physical solution to Eq.~\eqref{fieldEquations2} must take the form $\hat\psi^{\rm R}+\hat\psi^{{\rm S}}$ near some boundary, where $\hat\psi^{{\rm S}}$ is a specific particular solution (possibly singular at the boundary) and $\hat\psi^{\rm R}$ is a homogeneous solution that is regular at the boundary. If we construct a puncture $\psi^{\cal P}$ that approximates $\hat\psi^{{\rm S}}$ sufficiently well, and if we impose regular boundary conditions on the residual $\hat\psi^{\cal R}$, then solving Eq.~\eqref{psiRfieldEquations} and adding $\hat\psi^{\cal P}$ yields a solution to Eq.~\eqref{fieldEquations2} that satisfies the given boundary conditions. We apply this method at both outer boundaries, $\mathscr{H}^+$ and $\mathscr{I}^+$.

Sections~V.F--H in Ref.~\cite{Miller:2020bft} discuss the behavior of the second-order physical solution near $\mathscr{H}^+$ and $\mathscr{I}^+$. Here we briefly review and add some details to that discussion. We follow Ref.~\cite{Miller:2020bft} in using a label ``$[s]$'' to indicate the slicing in which a mode is defined. 

To frame the discussion, we first note how homogeneous solutions depend on slicing. For a homogeneous solution, the mode coefficients in generic $s$ slicing are related to those in $t$ slicing by
\beq
\psi_{[s]} = \psi_{[t]}e^{-i\omega k(r^*)},
\eeq
implying
\beq
\partial_r\psi_{[s]} = (\partial_r\psi_{[t]} -i\omega f^{-1}H\psi_{[t]})e^{-i\omega k(r^*)}.
\eeq
(We omit the subscript $m$ on $\omega$ for succinctness and because the discussion in this section applies equally well for eccentric orbits.) In matrix form,
\beq\label{hat psi transformation}
\hat\psi_{[s]} = \begin{pmatrix}\mathbf{1}_{d\times d}&\mathbf{0}_{d\times d}\\[0.4em]
			 \displaystyle - i\omega f^{-1} H\mathbf{1}_{d\times d}&\mathbf{1}_{d\times d}
\end{pmatrix}\hat\psi_{[t]}e^{-i \omega k(r^\ast)},
\eeq
where $d$ is the dimension of the vector $\psi$.

Homogeneous solutions regular at $\mathscr{H}^+$ behave like 
\beq\label{horizon-regular BCs - v}
\hat\psi_{[v]}\sim {f}^0
\eeq
for $r\to2M$ in $v$ slicing. Equation~\eqref{hat psi transformation} therefore implies that such solutions behave as 
\beq\label{horizon-regular BCs - t}
\hat\psi_{[t]}\sim \begin{pmatrix}f^0 \\ i\omega f^{-1}\end{pmatrix}e^{-i\omega r*} 
\eeq
in $t$ slicing, where it is understood that the two entries in the vector indicate the scaling of the top and bottom $d$ rows in $\hat\psi_{[t]}$, respectively. Homogeneous solutions regular at $\mathscr{I}^+$ behave like 
\beq\label{inf-regular BCs - u}
\hat\psi_{[u]}\sim r^0
\eeq
at large $r$ [corresponding to $h_{\mu\nu}\sim 1/r$ because of the rescaling by $r$ in Eq.~\eqref{modeDecomposition}]. Equation~\eqref{hat psi transformation} therefore implies 
\beq\label{inf-regular BCs - t}
\hat\psi_{[t]}\sim e^{+i\omega r*}
\eeq
in $t$ slicing.

At first order, outside the source region, the physical, retarded solutions are homogeneous. They therefore must satisfy the regularity conditions at $\mathscr{H}^+$ and $\mathscr{I}^+$ as described in the preceding paragraph. 

At second order, the boundary conditions are more complicated because of the behavior of the source terms. Away from the particle, the quadratic source term $G^{(2,0)}_{\mu\nu}$ is made up of products of homogeneous solutions. At large $r$, in generic $s$ slicing,
\beq
G^{(2,0)}_{\mu\nu} \sim \frac{e^{i\omega [r^*-k(r^*)]}}{r^2},
\eeq
implying a source $\hat J^{(2,0)}\sim \frac{e^{i\omega [r^*-k(r^*)]}}{r}$ in the equation~\eqref{fieldEquation20}. In $u$ slicing (for which $k=r^*$), the oscillations are eliminated, but the falloff is unaffected. All solutions are then singular at $\mathscr{I}^+$, behaving as 
\beq
\psi^{(2,0)}_{[u]}\sim \ln r
\eeq
for $\omega\neq0$ modes or as 
\beq
\psi^{(2,0)}_{[u]}\sim r\ln r
\eeq
for certain $\omega=0$ modes. This behaviour was discussed in detail in Ref.~\cite{Pound:2015wva} and will be returned to in a later paper.

At the opposite boundary, near the horizon,
\beq
G^{(2,0)}_{\mu\nu} \sim e^{-i\omega [r^*+k(r^*)]},
\eeq
meaning $\hat J^{(2,0)}\sim f^{-1}$ at the horizon in $v$ slicing (for which $k=-r^*$). The physical solution in the Lorenz gauge then turns out to be singular at the horizon despite the smoothness of the physical source $G^{(2,0)}_{\mu\nu}$. Again, this will be discussed in a future paper.

Next, we consider the source $G^{(1,1)}_{\mu\nu}[h^{(1)}]$ given in Eq.~\eqref{G11} and corresponding source $\hat J^{(1,1)}$ in Eq.~\eqref{fieldEquation11}. If we choose $t$ slicing, then $H=0$, implying 
\beq\label{J11t - FZ}
\hat J^{(1,1)}_{[t]}\sim \omega F^{(0)}_\Omega r^* e^{i\omega r^*}
\eeq
at large $r$, and so the field sourced by $\hat J^{(1,1)}_{[t]}$ behaves as $\hat\psi^{(1,1)}_{[t]}\sim r^2 e^{i\omega r^*}$. On the other hand, if we choose $u$ slicing, then $H=1$, implying
\beq\label{J11u - FZ}
\hat J^{(1,1)}_{[u]}\sim 1/r^2
\eeq
at large $r$, and so $\hat\psi_{[u]}\sim r^0$. Similarly, the source $\hat J^{(1,1)}_{[t]}$ is ill behaved at the horizon, scaling as
\beq\label{J11t - NH}
\hat J^{(1,1)}_{[t]} \sim r^* f^{-1}e^{-i\omega r^*},
\eeq
while the source $\hat J^{(1,1)}_{[v]}$ is smooth at the horizon, behaving as
\beq\label{J11v - NH}
\hat J^{(1,1)}_{[v]}\sim f^0.
\eeq
Fields sourced by $\hat J^{(1,1)}_{[t]}$ and fields sourced by $\hat J^{(1,1)}_{[v]}$ therefore have very different behavior near the horizon. We discuss that behavior in Sec.~\ref{sec_worldtube_G11}.

In cases where the physical solution is singular at a boundary, we introduce a puncture at that boundary as described above. Even in cases where a puncture is not strictly required, we can introduce one to increase the falloff rate of the effective source $\hat J^{\rm eff}$ toward the boundaries; this is beneficial because it improves the efficiency of integration over the source. In later sections we discuss the requirements on the puncture.

%-------------------------------------------------------------------------------------------------------------------------------------------------------% 
\section{Teukolsky equations}\label{sec_teukolsky_hyperboloidal_slicing}
%-------------------------------------------------------------------------------------------------------------------------------------------------------%
In the Teukolsky formalism, instead of directly dealing with metric
perturbations, one considers perturbations of a Weyl curvature scalar.  We shall focus on the curvature scalar $\psifour$, defined as
\beq\label{psi4 def}
	\psifour = C_{\alpha\beta\gamma\delta} n^{\alpha}
	\bar{m}^{\beta} n^{\gamma} \bar{m}^{\delta}.
\eeq
Here, the overbar denotes complex conjugation,
$C_{\alpha\beta\gamma\delta}$ is the Weyl curvature tensor,
and the vectors are elements of
a Newman-Penrose null tetrad $\{l^{\alpha}, n^{\alpha}, m^{\alpha},
\bar{m}^{\alpha}\}$~\cite{Newman:1961qr, Chandrasekhar:1975nkd}.

We will specifically focus on \emph{linear} perturbations of the Weyl scalar, meaning that for a given metric perturbation $h_{\alpha\beta}$, we consider $\psifour$ to be (with an abuse of notation) the piece of the Weyl curvature that is linear in $h_{\alpha\beta}$:
\beq
\psifour[h] = C^{(1,0)}_{\alpha\beta\gamma\delta}[h] n^{\alpha}
	\bar{m}^{\beta} n^{\gamma} \bar{m}^{\delta}. 
\eeq
Here the tetrad legs are defined in the background spacetime, and the linearized Weyl tensor $C^{(1,0)}_{\alpha\beta\gamma\delta}$ is defined in analogy with the linearized Einstein tensor $G^{(1,0)}_{\alpha\beta}[h]$ of previous sections. Our $\psifour$ corresponds to the quantity $\delta\psi_4$ in Ref.~\cite{coupling-paper}, but with the covariant derivative $\nabla$ replaced with its zeroth-order version $\nabla^{(0)}$ from Eq.~\eqref{nabla0}. Analogously, we define the linearized curvature scalar $\psi_0[h]=C^{(1,0)}_{\alpha\beta\gamma\delta}[h] l^{\alpha}m^{\beta} l^{\gamma} m^{\delta}$.

We note that our formulation in this section remains specialised to Schwarzschild spacetime (though the extension to Kerr, starting from Ref.~\cite{Spiers:2023cip}, is immediate).

\subsection{``Reduced'' Teukolsky equations}
\label{sec_reduced_teukolsky}

Given an equation $G^{(1,0)}_{\mu\nu}[h]=S_{\mu\nu}$ for a metric perturbation $h_{\mu\nu}$, we can obtain an associated spin-weight ${\sf s}=\pm2$ Teukolsky master equation~\cite{Teukolsky-Press:74},
\beq
	\,_{\sf s}{\cal O}\,_{\sf s}{\psi} = \,_{\sf s}{S},
	\label{eq:teukolsky_master_equation}
\eeq
where $\,_{\sf s}{\cal O}$ is the spin-weight ${\sf s}$ Teukolsky (wave) operator, ${}_{\sf s}{\psi}$ is constructed from $h_{\mu\nu}$, and ${}_{\sf s}{S}$ is constructed from $S_{\mu\nu}$.

The specific relationships between variables depends on the choice of tetrad.  We work with the Kinnersley tetrad~\cite{Kinnersley:1969zza}, in which $\,_{-2}\psi = \rho^{-4}\psifour$ and 
$\,_{-2}{\cal O} = 2r^{2}\rho^{-4}{\cal O}\rho^{4}$, where ${\cal O}$ is the second-order differential
operator
\begin{multline}
	{\cal O} := \left[\th^{\prime} - (2{\sf s} + 1) \rho^{\prime}\right](\th - \rho)
	- \edth^{\prime}\edth \\
	- \frac{1}{2}\left[(6{\sf s}+2) + 4{\sf s}^{2}\right]\psi_{2},\label{O_operator}
\end{multline}
with $\rho = -1/r$, $\rho^{\prime} = f/(2r)$, and $\psi_2=-M/r^3$. Similarly, the source for $\,_{-2}\psi$ is given by
$\,_{-2}T = 2 r^{2} \rho^{-4} {\cal S} [ S_{\alpha\beta} ]$, where\footnote{In \cite{Pound:2021qin}, this is denoted ${\cal S}_{4}$.}
\begin{align}
	{\cal S}[S_{\alpha\beta}] &= \frac{1}{2}\edth^{\prime}[(\th^{\prime} 
	- 2\bar{\rho}^{\prime})S_{(n\bar{m})} -
	\edth^{\prime}S_{nn}] \nn \\
	&\quad+ \frac{1}{2}(\th^{\prime} - 4\rho^{\prime} -
	\bar{\rho}^{\prime})[\edth^{\prime}S_{(n\bar{m})} - 
	(\th^{\prime} - \bar{\rho}^{\prime})S_{\bar{m}\bar{m}}].
	\label{eq:S4_operator} 
\end{align}
In these definitions, we adopt Geroch-Held-Penrose (GHP) notation, following the conventions of~\cite{Pound:2021qin} (simplified with $\tau = \tau^{\prime} = 0$ in a Schwarzschild background). The GHP derivatives $\th$, $\th'$, $\edth$, and $\edth'$, along with a brief review of the GHP formalism, can be found in Appendix~\ref{sec_ghp}.

The mode decomposition likewise depends on the choice of null tetrad.  Again working with the Kinnersley tetrad, we write our separation ansatz as 
\begin{align}
	\,_{\sf s}\psi &=
	r^{-(2{\sf s} + 1)}f^{-{\sf s}} \nonumber\\
	&\quad\times
	\sum^{\infty}_{\ell = 2} \sum^{\ell}_{m = -\ell} 
	\teukR{\sf s}{}{\lmw} (r,{\cal J}_I)  \Y{\sf s}{\ell m}(\theta,\phi)
	e^{-i m \phi_p},
	\label{eq:teukolsky_mode_ansatz}
\end{align}
where $\Y{\sf s}{\ell m}(\theta,\phi)$ are spin-weighted
spherical harmonics~\cite{Goldberg:1966uu}; these are straightforwardly related to the tensor harmonics $Y^{i\ell m}_{\mu\nu}$ that we use for the Lorenz-gauge metric perturbations~\cite{coupling-paper}. The radial Teukolsky function, $\teukR{\sf s}{}{\lmw} (r)$, satisfies the ordinary differential
equation
\begin{align}
	&\bigg( r^{2}f\sdiff{}{r} + 2\left[ M
	- {\sf s}(r - M) + i \omega r^{2} H \right]\diff{}{r} \nonumber\\
	&+ \frac{r^{2}}{f} \left[\omega^{2}(1 - {H}^{2}) + i\omega H^{\prime} - \teukpot{\sf s}{\lm}(r)\right] \bigg)
	\teukR{\sf s}{}{\lmw}(r)  \nonumber\\
	&= \Tgensource{\sf s}{\lmw}(r).
	\label{eq:radial_teukolsky_equation}
\end{align}
Here, $\omega=m\Omega$, the Teukolsky potential reads
\begin{multline}
	\teukpot{\sf s}{\lm}(r) := \frac{2i\,{\sf s}\,\omega}{r^{2}} \left[ fr (1 - H) - M(1 + H) \right] \\
	+ \frac{f}{r^{2}} \left[ \ell(\ell + 1) - {\sf s}({\sf s} + 1) + \frac{2M({\sf s} + 1)}{r} \right],
	\label{eq:teukolsky_potential}
\end{multline}
and the source mode coefficients are defined from
\begin{equation}
	\,_{\sf s}S = -
	\sum^{\infty}_{\ell = 2} \sum^{\ell}_{m = -\ell}
	\Tgensource{\sf s}{\lmw}(r,{\cal J}_I)
	\Y{\sf s}{\ell m}(\theta,\phi) 
	e^{-i m\phi_p}.
	\label{eq:S4_mode_source}
\end{equation}
The explicit form of the source for the case we shall explore in this work can be found in Appendix~\ref{sec_Tlmw}.

The factor $r^{-(2{\sf s} + 1)}f^{-{\sf s}}$ in the mode ansatz~\eqref{eq:teukolsky_mode_ansatz} is introduced for the 
purposes of numerical integration of the radial Teukolsky equation.  
This was first introduced in the context of hyperboloidal slicing for the Teukolsky equation in \cite{Zenginoglu:2011jz}, and has been further utilised in
the subsequent works \cite{Piovano:2021iwv, Nasipak:2021qfu, Nasipak:2022xjh}.
Without this rescaling, the potential would only fall off as $1/r$ toward infinity and would not vanish at the 
horizon.  
Therefore, the potential would be long-ranged, akin to the Coulomb potential. 
For any non-zero spin-weight, one could not accurately compute solutions of the homogeneous Teukolsky equation due to numerical round-off error 
either near the horizon or towards infinity.
Hence, in rescaling the master function in accordance with its asymptotic behaviour, we obtain a short-ranged 
potential in Eq.~(\ref{eq:teukolsky_potential}) that now falls off as $f$ near the horizon and $r^{-2}$ near null infinity.
Futhermore, our use of hyperboloidal slicing eliminates the oscillatory behaviour of the radial function toward infinity and the horizon, which increases the efficiency of the numerical solver.  

As we did for the Lorenz-gauge field equations, we express our radial Teukolsky equation in the form of Eq.~(\ref{fieldEquations2}). The column vector $\psi$ in Eq.~(\ref{fieldEquations2}) reduces to 1 element with $\psi(r) = \teukR{\sf s}{}{\lmw}(r)$, with $A$ and $B$ given by
\begin{align}
	A &= \frac{\omega^{2}(1 - {H}^{2}) - \teukpot{\sf s}{\lm}(r)}{f^{2}}, \nn \\
	B &= \frac{1}{f} \left( \frac{2M}{r^{2}} + 2 i\omega {H}
	+ \frac{2{\sf s}(r - M)}{r^{2}} \right).
	\label{eq:ab_teukolsky}
\end{align}
Similarly, the source reduces to $J(r) = r^{2{\sf s} - 1} f^{{\sf s} - 1}\Tgensource{\sf s}{\lmw}(r, {\cal J}_{I})$. 

All of the above formulas apply for each field
\begin{equation}
{}_{-2}\psi^{(n)} = \rho^{-4}\psi_4[h^{(n)}].
\end{equation}
In analogy with Eq.~\eqref{psi2 split}, we split the second-order field into two pieces,
\begin{align}
{}_{-2}\psi^{(2,0)} &= \rho^{-4}\psi_4[h^{(2,0)}],\\
{}_{-2}\psi^{(1,1)} &= \rho^{-4}\psi_4[h^{(1,1)}].
\end{align}
Again, all the formulas in this section apply to each of these pieces. In analogy with Eqs.~\eqref{fieldEquation20} and~\eqref{fieldEquation11}, the radial coefficients in the mode decompositions of ${}_{-2}\psi^{(2,0)}$ and ${}_{-2}\psi^{(1,1)}$ satisfy radial Teukolsky equations with sources constructed from
\begin{align}
S^{(2,0)}_{\mu\nu} &= 8\pi T^{(2,0)}_{\mu\nu} - G^{(2,0)}_{\mu\nu}[h^{(1)},h^{(1)}],\\
S^{(1,1)}_{\mu\nu} &= 8\pi T^{(1,1)}_{\mu\nu} - G^{(1,1)}_{\mu\nu}[h^{(1)}].
\end{align}

Note that our field ${}_{-2}\psi^{(2)}$ does not represent the second-order term in an expansion of the spacetime's full Weyl scalar. Such an expansion would include quadratic terms constructed from $h^{(1)}_{\mu\nu}$ and from perturbations of the tetrad legs, while our field involves only the piece that is linear in $h^{(1)}_{\mu\nu}$. We refer to Ref.~\cite{Spiers:2023cip} for a thorough discussion; there, we refer to the field equation for our linear ${}_{-2}\psi^{(2)}$ as the \emph{reduced} second-order Teukolsky equation. Similar comments apply to ${}_{-2}\psi^{(1,1)}$.

\subsection{Boundary conditions}

The core aspects of Sec.~\ref{sec_BCs} carry over to the Teukolsky case, including the use of punctures. However, solutions of the radial Teukolsky equation do behave in significantly different ways near the boundaries than Lorenz-gauge metric perturbations. Here we do not attempt a comprehensive summary of the behavior of sourced solutions, comparable to Sec.~\ref{sec_BCs}. Instead, we only highlight the behavior of a basis of homogeneous solutions, analogous to Eqs.~\eqref{horizon-regular BCs - v}--\eqref{inf-regular BCs - t}. 
This basis is made up of a pair of solutions that are respectively purely ingoing at $\mathscr{H}^+$ and purely outgoing at $\mathscr{I}^+$.
In the usual Teukolsky nomenclature these are referred to as `in' and `up' solutions~\cite{Pound:2021qin}.

The `in' solution is regular at the horizon and has the near-boundary behavior
\begin{equation}
	\hat{\psi}^{-}_{[t]} \sim
	\begin{cases}
		%r e^{-i\omega\rstar},\quad &r \to 2M,\\
		e^{-i\omega\rstar},\quad &r \to 2M,\\
		%f^{\sf s} e^{i\omega\rstar} + r^{2{\sf s}} f^{\sf s} e^{-i\omega\rstar},\quad &r \to \infty,
		e^{i\omega\rstar} + r^{2{\sf s}} e^{-i\omega\rstar},\quad &r \to \infty.
	\end{cases}
	\label{eq:teukolsky_boundary_conditions_in}
\end{equation}
Conversely, the `up' solution is regular at $\mathscr{I}^+$ and behaves as
\begin{equation}
	\hat{\psi}^{+}_{[t]} \sim
	\begin{cases}
		%r e^{-i\omega\rstar} + r^{2{\sf s}+1} f^{\sf s} e^{i\omega\rstar},\quad &r \to 2M,\\
		e^{-i\omega\rstar} + f^{\sf s} e^{i\omega\rstar},\quad &r \to 2M,\\
		%f^{\sf s} e^{i\omega\rstar},\quad &r \to \infty,
		e^{i\omega\rstar},\quad &r \to \infty.
	\end{cases}
	\label{eq:teukolsky_boundary_conditions_up}
\end{equation}
These limiting behaviors can be verified by applying the rescaling in Eq.~(\ref{eq:teukolsky_mode_ansatz}) to the form of the master functions in Table 1 of Ref.~\cite{1974ApJ...193..443T}. 

At the boundaries where these homogeneous solutions represent physical waves created by a compact source, they have the same behavior as the Lorenz-gauge metric perturbations. However, they differ in important ways at the opposite boundaries. For the spin weight we focus on (${\sf s}=-2$), Eq.~\eqref{eq:teukolsky_boundary_conditions_in} shows that in the `in' solution, the incoming portion of the solution at large $r$ decays rapidly, as $r^{-4}$; and Eq.~\eqref{eq:teukolsky_boundary_conditions_up} shows that in the `up' solution, the outgoing portion of the solution at the horizon blows up there. Homogeneous Lorenz-gauge perturbations, on the other hand, behave like ${\sf s}=0$ solutions in Eqs.~\eqref{eq:teukolsky_boundary_conditions_in} and \eqref{eq:teukolsky_boundary_conditions_up}: a solution that is a pure ingoing wave at the horizon is a mix of ingoing $(r^0 e^{-i\omega r^*}$) and outgoing $(r^0 e^{+i\omega r^*}$) waves at infinity; and a solution that is a pure outgoing wave at infinity is a mix of ingoing and (bounded) outgoing waves at the horizon. The more intricate behaviour of the homogeneous Teukolsky solutions has important knock-on effects for inhomogeneous solutions with noncompact sources, explained in Sec.~\ref{dh - vtu}.

%-------------------------------------------------------------------------------------------------------------------------------
%
\section{Puncture scheme with smooth slicing and windowed punctures}\label{sec_worldtube_scheme}
%
%------------------------------------------------------------------------------------------------------------------------------

Before considering our worldtube scheme with multiple distinct regions, we first consider a simpler but less computationally convenient method. We assume the time function $s$ is smooth, and if there are punctures, we use window functions to make them transition to zero at some distance from the particle or from the boundary where they are used. This is an extension of the window-function method that one of us applied to first-order Lorenz-gauge calculations in Ref.~\cite{Wardell:2015ada}, now allowing for alternative time functions and noncompact sources.

We keep our discussion generic in this section, making it equally valid for eccentric orbits as for quasicircular orbits.

%-------------------------------------------------------------------------------------------------------------------------------------------------------%
\subsection{Generic source}
%
%-------------------------------------------------------------------------------------------------------------------------------------------------------

We consider a generic set of coupled first-order radial ODEs written in the matrix form~\eqref{fieldEquations3v1}, reproduced here for convenience:
\begin{equation}
\hat{\cal D}\hat\psi := \frac{d\hat\psi }{dr}+\hat A\hat \psi = \hat J,\label{fieldEquations3}
\end{equation}
where  $\hat\psi=(\psi, \partial_r\psi)^T$ and $\hat J=(\mathbf{0}_d, J)^T$ are column vectors of length $2d$. These can be the Lorenz-gauge field equations, Teukolsky equations, or another set of equations. We let the domain of the solutions to (\ref{fieldEquations3}) be $r\in\left(2M,\infty\right)$, and we assume that $\hat J$ is integrable (or well defined as a distribution) in that domain and falls off sufficiently rapidly toward the boundaries (a condition that will be enforced below through the use of boundary punctures).

Now we seek a solution to Eq.~(\ref{fieldEquations3}) subject to physical boundary conditions at $r=2M$ and $r=\infty$. We can construct the solution using the method of variation of parameters~\cite{BoyceDiPrima}. Take the homogeneous version of Eq.~(\ref{fieldEquations3}) ($\hat J=0$), which has $2d$ independent solutions. We denote by $\hat\psi_{k-}$, with $k=1,\dots,d$, the $d$ independent homogeneous solutions that obey desired boundary conditions at $r=2M$, and by $\hat\psi_{k+}$ the $d$ independent solutions that obey desired boundary conditions at infinity. For concreteness, for $\omega\neq0$ modes, $\hat\psi_{k-}$ will represent ingoing waves regular at the future horizon, and $\hat\psi_{k+}$ will represent outgoing waves at future null infinity. For $\omega=0$ modes, $\hat\psi_{k-}$ will be homogeneous solutions regular at the horizon, and  $\hat\psi_{k+}$ will be asymptotically flat homogeneous solutions. Using these solutions we define a $2d\times 2d$ matrix of homogeneous solutions,
\begin{equation}
\Phi := \left(\hat\psi_{1-},\ldots,\hat\psi_{d-},\hat\psi_{1+},\ldots,\hat\psi_{d+}\right),\label{phi2}
\end{equation}
satisfying 
\beq\label{Phi ODE}
\frac{d}{dr}\Phi+\hat A\Phi=0. 
\eeq
Appendix~\ref{sec_basis} reviews the construction of the basis of homogeneous solutions.

In terms of $\Phi$  the general solution to Eq.~(\ref{fieldEquations3}) can be written as
\begin{equation}
\hat\psi = \Phi \left(\int_{2M}^{r} dr'\Phi^{-1}(r')\hat J(r')  + a \right),\label{generalSolution}
\end{equation}
with $a$ an arbitrary, constant $2d$-vector. Writing $\Phi$ in the form 
\beq
\Phi=\begin{pmatrix}\Phi_-,\Phi_+\end{pmatrix}
\eeq
with $\Phi_\pm=\left(\hat\psi_{1\pm},\ldots,\hat\psi_{d\pm}\right)$ a $2d\times d$ matrix, we can also write Eq.~(\ref{generalSolution}) as
\begin{align}
\hat\psi  &= \Phi_- \left(\int_{2M}^r dr'\Phi^{-1}_{\rm top}\hat J  + a_{\rm top}\right) \nonumber\\
&\quad + \Phi_+ \left(\int_{2M}^{r} dr' \Phi^{-1}_{\rm bot}\hat J  + a_{\rm bot}\right). \label{psi}
\end{align}
Here and below, `top' and `bot' refer to the top or bottom $d$ rows of a matrix with $2d$ rows, and $\Phi^{-1}_{\rm top/ bot}$ specifically denotes the top or bottom $d$ rows of the $2d\times 2d$ matrix~$\Phi^{-1}$.

%-------------------------------------------------------------------------------------------------------------------------------------------------------%
\subsection{Compact source}
%-------------------------------------------------------------------------------------------------------------------------------------------------------%

First consider the case in which the source has compact support. This is the typical situation at first order, in which the point-particle, frequency-domain source is confined to a single radius (for a circular orbit) or to the libration region (for an eccentric orbit~\cite{Akcay:2013wfa}), and the effective, punctured source is confined to a region around the particle. 

If the source is supported between some $r_{\rm min}$ and $r_{\rm max}$, then outside of that region, the physical, retarded field must reduce to a linear combination of the appropriate homogeneous solutions:
\beq
\hat\psi^{\rm ret} = \begin{cases}\Phi_+ c_+ & \text{for } r>r_{\rm max }, \\
					\Phi_- c_- & \text{for }  r<r_{\rm min},
			\end{cases}\label{BCs-compact}
\eeq
for some constant $d$-vectors  $c_\pm$.

Explicitly evaluating the general solution~\eqref{psi} outside the source region, we find
\begin{align}
\hat\psi(r<r_{\rm min}) &= \Phi_- a_{\rm top} + \Phi_+ a_{\rm bot},\\
\hat\psi(r>r_{\rm max}) &= \Phi_- \left(\int_{2M}^\infty dr'\Phi^{-1}_{\rm top}\hat J  + a_{\rm top}\right) \nonumber\\
									&\quad + \Phi_+ \left(\int_{2M}^\infty dr' \Phi^{-1}_{\rm bot}\hat J  + a_{\rm bot}\right). 
\end{align}
The boundary conditions~\eqref{BCs-compact} hence imply 
\begin{equation}
a = \left(-\displaystyle{ \int_{2M}^\infty dr' \Phi^{-1}_{\rm top}(r')\hat J(r') } \,,\, \mathbf{0}_d\right)^T.
\end{equation}
Therefore the retarded solution is
\begin{equation}
\hat\psi^{\rm ret}  = \Phi {\sf v}\label{psi_regular_with_r_0}, 
\end{equation}
where ${\sf v}$ is the $2d$-vector
\begin{equation}
{\sf v} = \left(- { \int_r^\infty dr'\Phi^{-1}_{\rm top} \hat J  }  \,,\, { \int_{2M}^r dr'   \Phi^{-1}_{\rm bot} \hat J  }\right)^T.\label{sfv}
\end{equation}
Note that ${\sf v}=(c_-,0)^T$ for $r<r_{\rm min}$ and ${\sf v}=(0,c_+)^T$ for $r>r_{\rm max}$; or, restated as an equation for $c^\pm$,
\begin{align}
c_+ &= \int_{2M}^{r_{\rm max}} dr'   \Phi^{-1}_{\rm bot} \hat J ,\label{c+}\\
c_- &= - \int_{r_{\rm min}}^\infty dr'\Phi^{-1}_{\rm top} \hat J.\label{c-}
\end{align}
For a given source $\hat J$, we will refer to \eqref{psi_regular_with_r_0} as the retarded integral of $\hat J$. If $\hat J$ has compact support, then this also represents the physical retarded solution; if $\hat J$ has noncompact support, then the retarded integral of it may or may not represent the physical retarded solution.

%-------------------------------------------------------------------------------------------------------------------------------------------------------%
\subsection{Noncompact, punctured source}
\label{boundary punctures}
%-------------------------------------------------------------------------------------------------------------------------------------------------------%

Now we consider the case where the source is noncompact, extending to the boundaries. This is the situation at second order. 

As described in Sec.~\ref{sec_BCs}, the boundary conditions on the physical field are
\beq
\hat\psi^{\rm ret} = \begin{cases} \hat\psi^{\rm S}_+ + \Phi_+ c_+ & \text{for }r\to\infty, \\
					\hat\psi^{\rm S}_- + \Phi_- c_- & \text{for }r\to2M,	
								\end{cases}\label{BCs-noncompact}
\eeq
for some constant $d$-vectors $c_\pm$. $\hat\psi^{\rm S}_\pm$ is a given particular solution to 
\beq
\hat{\cal D}\hat\psi^{\rm S}_\pm = \hat J,
\eeq
and it is typically singular at the boundary where it is used.

We can enforce the boundary conditions~\eqref{BCs-noncompact} using punctures. Let the punctures $\hat\psi^\P_\pm$ approximate $\hat\psi^{\rm S}_\pm$ near the boundaries and then transition smoothly to zero. Define $\hat\psi^\P = \hat\psi^\P_+ + \hat\psi^\P_-$ (plus a puncture at the particle if appropriate). The residual field $\hat\psi^{\res} = \hat\psi^{\rm ret} - \hat\psi^\P$ then satisfies
\beq
\hat{\cal D}\hat\psi^{\res} = \hat J - \hat{\cal D}\hat\psi^\P =: \hat J^{\rm eff}.
\eeq
To enforce~\eqref{BCs-noncompact}, we adopt the retarded solution to this equation, taking the retarded integral~\eqref{psi_regular_with_r_0} of $\hat J^{\rm eff}$:
\begin{align}
\hat\psi^{\res}  &= -\Phi_-\int_r^\infty dr'\Phi^{-1}_{\rm top} \hat J^{\rm eff} \nonumber\\
						&\quad + \Phi_+\int_{2M}^r dr' \Phi^{-1}_{\rm bot} \hat J^{\rm eff}.\label{retarded integral psi-res}
\end{align}
If  $\hat J^{\rm eff}$ falls off sufficiently quickly toward the boundaries, then this solution for $\hat\psi^{\res}$ approximates the homogeneous solutions $\Phi_\pm c_\pm$ near the boundaries, fixing the values of the coefficients $c_\pm$ and ensuring that the total field $\hat\psi^\P + \hat\psi^{\res}$ takes the form in~\eqref{BCs-noncompact}.

To see that we are correct in using the retarded integral~\eqref{retarded integral psi-res}, start by assuming we know the particular solutions $\hat\psi^{\rm S}_\pm$ exactly. In that case we can adopt punctures $\hat\psi^{\P {\rm exact}}_\pm = \hat\psi^{\rm S}_\pm{\cal W}_\pm$, where ${\cal W}_\pm$ is a window function that is identically equal to 1 in a neighborhood of the worldline and transitions to zero at some finite distance from the worldline. The effective source then has compact support, identically vanishing in a neighborhood of the boundaries. In this circumstance, we rename the $\hat\psi^\res$ in Eq.~\eqref{retarded integral psi-res} as $\hat\psi^{\res{\rm exact}}$. Near $r=2M$, where $\hat J^{\rm eff}=0$, $\hat\psi^{\res{\rm exact}}$ becomes the homogeneous solution
\beq
\hat\psi^{\res  {\rm exact}}  = - \Phi_-\int_{2M}^\infty dr'\Phi^{-1}_{\rm top} \hat J^{\rm eff},
\eeq
implying the unknown constant $c_-$ in Eq.~\eqref{BCs-noncompact} is $c_- = -\int_{2M}^\infty dr'\Phi^{-1}_{\rm top} \hat J^{\rm eff}$. Analogously, near $r\to\infty$ we have
\beq\label{psi-res_exact}
\hat\psi^{\res  {\rm exact}}  = \Phi_+\int_{2M}^\infty dr' \Phi^{-1}_{\rm bot} \hat J^{\rm eff},
\eeq
implying $c_+$ in Eq.~\eqref{BCs-noncompact} is $c_+ = \int_{2M}^\infty dr'\Phi^{-1}_{\rm bot} \hat J^{\rm eff}$.

Now consider the case we encounter in practice, in which we only know $\hat\psi^{\rm S}_\pm$ approximately, up to some finite order in a series expansion around $r\to\infty$ or around $r=2M$. Suppose we use such approximations as punctures $\hat\psi^{\P}_\pm$, such that
\beq
\hat\psi^{\P}_\pm = \hat\psi^{\P {\rm exact}}_\pm + \Delta\hat\psi^\P_\pm
\eeq
for some $\Delta\hat\psi^\P_\pm$. For the puncture scheme to be useful, the total field must be robust under this change in the punctures, at least so long as $\Delta\hat\psi^\P_\pm$ is sufficiently small in the limit to the boundary; the change in the punctures must be exactly counterbalanced by a commensurate change in the residual field, leaving the total field unaltered. 

Let us assess the restrictions this imposes on $\Delta\hat\psi^\P_\pm$, and whether we can safely use the retarded integral~\eqref{retarded integral psi-res} for $\hat\psi^\res$. The new residual field, $\hat\psi^{\res}=\hat\psi^{\rm ret} - \hat\psi^{\P}$, satisfies
\beq
\hat{\cal D}\hat\psi^{\res} = \hat J -\hat{\cal D}\hat\psi^{\P} = \hat J - \hat{\cal D}\hat\psi^{\P  {\rm exact}} - \hat{\cal D}\Delta\hat\psi^{\P}.
\eeq
Therefore the {\em change} in the residual field, $\Delta\hat\psi^{\res} = \hat\psi^{\res} - \hat\psi^{\res  {\rm exact}}$, satisfies
\beq
\hat{\cal D}\Delta\hat\psi^\res =  - \hat{\cal D}\Delta\hat\psi^{\P}.
\eeq
The retarded integral is
\begin{align}
\Delta\hat\psi^\res &=  \Phi_-\int_r^\infty dr'\Phi^{-1}_{\rm top}\hat{\cal D}\Delta\hat\psi^{\P} \nonumber\\
							&\quad - \Phi_+\int_{2M}^r dr' \Phi^{-1}_{\rm bot} \hat{\cal D}\Delta\hat\psi^{\P}.\label{Delta psi-res}
\end{align}
Using $\hat{\cal D}\Delta\hat\psi^{\P}=\frac{d}{dr}\Delta\hat\psi^{\P} + \hat A\Delta\hat\psi^{\P}$ and integrating by parts, we can rewrite the integrals as, for example, 
\begin{align}
\int_r^\infty dr'&\Phi^{-1}_{\rm top}\hat{\cal D}\Delta\hat\psi^{\P} \nonumber\\
						&= \int_r^\infty dr'\left(-\frac{d}{dr}\Phi^{-1}_{\rm top}+\Phi^{-1}_{\rm top}\hat A\right)\Delta\hat\psi^{\P}\nonumber\\
						&\quad +\left.\Phi^{-1}_{\rm top}\Delta\hat\psi^\P\right|^\infty_r.
\end{align}
It is straightforward to establish that
\beq\label{Phi inv ODE}
\frac{d}{dr}\Phi^{-1} - \Phi^{-1}\hat A = 0,
\eeq
starting from $\frac{d}{dr}(\Phi^{-1}\Phi)=0$ and using Eq.~\eqref{Phi ODE}. This simplifies the above result to
\begin{align}
\int_r^\infty dr'\Phi^{-1}_{\rm top}\hat{\cal D}\Delta\hat\psi^{\P} =  \left.\Phi^{-1}_{\rm top}\Delta\hat\psi^\P\right|^\infty_r.
\end{align}
Similarly evaluating the second integral in Eq.~\eqref{Delta psi-res}, we obtain 
\begin{align}
\Delta\hat\psi^\res &= - \Delta\hat\psi^{\P} + \Phi_- \lim_{r\to\infty}\left(\Phi^{-1}_{\rm top}\Delta\hat\psi^\P\right) \nonumber\\
								&\quad + \Phi_+ \lim_{r\to2M}\left(\Phi^{-1}_{\rm bot}\Delta\hat\psi^\P\right).\label{Delta psi-res 2}
\end{align}

We see that the change in the puncture is counterbalanced by the change in the residual field, and the retarded integral recovers the correct result, if\,\footnote{The same calculation also shows that the retarded integral yields the correct residual field if we change the puncture by a homogeneous solution near the boundaries, as in $\Delta\psi^\P_\pm = \Phi_\pm b_\pm {\cal W}_\pm$ for constant $d$-vectors $b_\pm$.  Equation~\eqref{Delta psi-res 2} in that case reads
\begin{align}
\Delta\hat\psi^\res &= - \Delta\hat\psi^{\P} + \Phi_- \lim_{r\to\infty}\left(\Phi^{-1}_{\rm top}\Phi_+\right)b_+ \nonumber\\
								&\quad + \Phi_+ \lim_{r\to2M}\left(\Phi^{-1}_{\rm bot}\Phi_-\right)b_-.
\end{align}
It follows from $\Phi^{-1}\Phi=\mathbf{1}_{2d\times 2d}$ that $\Phi^{-1}_{\rm top}\Phi_+=0=\Phi^{-1}_{\rm bot}\Phi_-$, and so $\Delta\hat\psi^\res = - \Delta\hat\psi^{\P}$. This simply trades the homogeneous solution between the puncture and the residual field, without altering the total, physical field. Equivalently, this trades a homogeneous solution between the two terms in the boundary conditions~\eqref{BCs-noncompact}.}
\beq
\lim_{r\to\infty}\left(\Phi^{-1}_{\rm top}\Delta\hat\psi^\P\right) = 0\label{infinity condition1}
\eeq
and
\beq
\lim_{r\to2M}\left(\Phi^{-1}_{\rm bot}\Delta\hat\psi^\P\right) = 0.\label{horizon condition1}
\eeq
We can also write this as 
\beq
\lim_{r\to\infty}\left[\Phi^{-1}_{\rm top}(\hat\psi^\P-\hat\psi^{\rm S})\right] = 0\label{infinity condition}
\eeq
and
\beq
\lim_{r\to2M}\left[\Phi^{-1}_{\rm bot}(\hat\psi^\P-\hat\psi^{\rm S})\right] = 0.\label{horizon condition}
\eeq
These conditions dictate the required order of a puncture. For example, if we work with a puncture that includes terms up to order $(r-2M)^n$ near the horizon, then Eq.~\eqref{horizon condition} tells us that $n$ must be large enough to ensure $\lim_{r\to2M}\left[\Phi^{-1}_{\rm bot}(r-2M)^{n+1}\right] = 0$. 

Equations~\eqref{infinity condition1} and \eqref{horizon condition1} also tell us the conditions under which we actually need a puncture. If we choose $\Delta\hat\psi^\P = -\hat\psi^{\P\rm exact}$, then $\hat\psi^{\P}=0$ and the residual field becomes simply the retarded integral of the physical source, $\hat\psi^{\R} = - \Phi_-\int_r^\infty dr'\Phi^{-1}_{\rm top} \hat J + \Phi_+\int_{2M}^r dr' \Phi^{-1}_{\rm bot} \hat J$. Equations~\eqref{infinity condition1} and \eqref{horizon condition1} become 
\beq
\lim_{r\to\infty}\left(\Phi^{-1}_{\rm top}\hat\psi^{\rm S}\right) = 0 \text{ and }  \lim_{r\to2M}\left(\Phi^{-1}_{\rm bot}\hat\psi^{\rm S}\right) = 0.\label{puncture conditions}
\eeq
If the conditions \eqref{puncture conditions} are satisfied, then
\begin{align}
\hat\psi^{\res\rm exact} &= -\hat\psi^{\P\rm exact} - \Phi_-\int_r^\infty dr'\Phi^{-1}_{\rm top} \hat J \nonumber\\
						&\quad + \Phi_+\int_{2M}^r dr' \Phi^{-1}_{\rm bot} \hat J.
\end{align}
This implies that no puncture is required: with or without a puncture, the total field $\hat\psi^\res +\hat\psi^{\P}$ is simply the retarded integral of the original source $\hat J$, meaning that this retarded integral automatically satisfies the correct boundary conditions. Conversely, if the conditions~\eqref{puncture conditions} are not satisfied, meaning
\beq
\lim_{r\to\infty}\left(\Phi^{-1}_{\rm top}\hat\psi^{\rm S}\right) \neq 0 \quad\text{or}\quad  \lim_{r\to2M}\left(\Phi^{-1}_{\rm bot}\hat\psi^{\rm S}\right) \neq 0,\label{puncture conditions 2}
\eeq
then a puncture \emph{is} required.

In summary, for a given particular solution $\hat\psi^{\rm S}$ in the boundary conditions~\eqref{BCs-noncompact}, the retarded integral~\eqref{retarded integral psi-res} yields a correct residual field so long as $\hat\psi^{\rm S}-\hat\psi^\P$ satisfies the conditions~\eqref{infinity condition} and \eqref{horizon condition}. We emphasise that those conditions are stronger than simply ensuring convergence of the retarded integral; two different punctures can both lead to convergent retarded integrals even if the difference between them violates~\eqref{infinity condition1} or \eqref{horizon condition1}, but in that case they will lead to two different total solutions $\hat\psi^\P+\hat\psi^\res$, satisfying different physical boundary conditions.

%-------------------------------------------------------------------------------------------------------------------------------------------------------% 
\section{Worldtube puncture scheme}\label{sec_general_formulation}
%
%--------------------------------------------------------------------------------------------------------------------------------------------------------

\begin{figure}[t]
\centering
\includegraphics[width=\columnwidth]{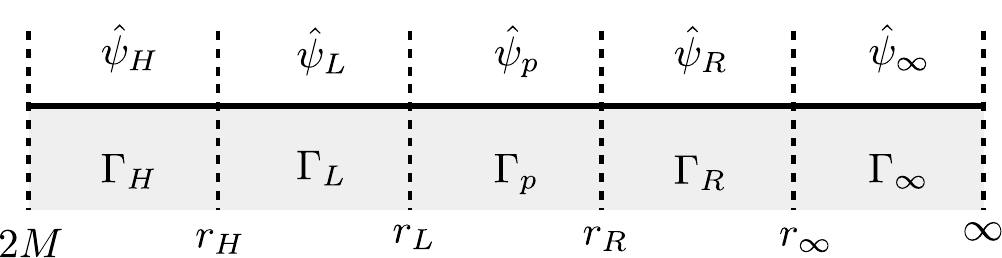}
\caption{Division of the numerical domain into regions $\Gamma_H$, $\Gamma_L$, $\Gamma_p$, $\Gamma_R$, and $\Gamma_\infty$. In each region $\Gamma_\placeholder$, $\placeholder\in\{H,L,p,R,\infty\}$, we use a corresponding field variable $\hat\psi_\placeholder$.}
\label{fig:regions}
\end{figure}

We now introduce our worldtube scheme. We split the domain into five regions: 
 a near-horizon region $\Gamma_H=(2M,r_H)$,
 a non-punctured region $\Gamma_L=(r_H,r_L)$ (where `L' stands for `left'),
 a worldtube\footnote{In three-dimensional space, each of the regions is a shell surrounding the large black hole, but we adopt traditional nomenclature by referring to the shell containing the particle as a worldtube.} around the particle, $\Gamma_p=(r_L,r_R)$  (where `R' stands for `right'),
 another non-punctured region $\Gamma_R=(r_R,r_\infty)$,
 and an asymptotic region $\Gamma_\infty=(r_\infty,\infty)$. These are illustrated in Fig.~\ref{fig:regions}. We assume there is a puncture at the particle, $\hat\psi^{\P}_{p}$, in $\Gamma_p$; a puncture at the horizon, $\hat\psi^{\P}_{ H}$, in $\Gamma_H$, and one at infinity, $\hat\psi^{\P}_{ \infty}$, in $\Gamma_\infty$. We also allow the operator $\hat{\cal D}$ to be different in the different regions, as it will be if we use different slicings in the different regions. We will ultimately obtain the solutions in all the regions by imposing junction conditions at the region boundaries. 
 
In many concrete calculations we omit one or both of the regions $\Gamma_L$ and $\Gamma_R$. However, for generality, we include all five regions in our description here.
 
Like in the preceding section, we keep our treatment generic, such that it applies both to eccentric and quasicircular orbits.
 
\subsection{General framework}  
  
In each region we define the field variable $\hat \psi_\placeholder$, with $\placeholder\in\{H,L,p,R,\infty\}$, as
\begin{subequations}\label{fields}
\begin{align}
\hat \psi_H &=\hat\psi^{\rm ret}_{H}-\hat\psi^{\P}_{ H},\\
\hat\psi_L &=\hat\psi^{{\rm ret}}_{L},\\
\hat \psi_p &=\hat\psi^{{\rm ret}}_{p}-\hat\psi^{\P}_{p},\\
\hat\psi_R &=\hat\psi^{{\rm ret}}_{R},\\
\hat \psi_\infty &=\hat\psi^{{\rm ret}}_{\infty}-\hat\psi^{\P}_{\infty},
\end{align}
\end{subequations}
where the domain of $\hat\psi_\placeholder$ is $\Gamma_\placeholder$. These fields satisfy the equations
\begin{equation}\label{ODE}
\hat{\cal D}_\placeholder\hat\psi_\placeholder := \partial_r\hat\psi_\placeholder+\hat A_\placeholder\hat\psi_\placeholder = \hat J^{{\rm eff}}_\placeholder,
\end{equation}
where $\hat A_\placeholder$ is in general different in each region,
and the sources are
$\hat J^{{\rm eff}}_{ H}=\hat J_H-\hat{\cal D}_H\hat\psi^{\P}_{ H}$,
$\hat J^{{\rm eff}}_{ L}=\hat J_L$,
$\hat J^{{\rm eff}}_{ p}=\hat J_p-\hat{\cal D}_p \hat\psi^{\P}_{ p}$,
$\hat J^{{\rm eff}}_{ R}=\hat J_R$,
and 
$\hat J^{{\rm eff}}_{\infty}=\hat J_\infty-\hat{\cal D}_\infty\hat\psi^{\P}_{ \infty}$, where the raw sources $\hat J_\placeholder$ are allowed to differ between regions.

The general solution in each region is
\begin{equation}\label{general_ODE_solution}
\hat\psi_\placeholder = \Phi_\placeholder\left(\int_{r_{\placeholder}}^r\Phi_\placeholder^{-1}\hat J^{{\rm eff}}_\placeholder dr + a_\placeholder\right),
\end{equation}
where $r_{\placeholder}\in\{2M,r_H,r_L,r_R,r_\infty\}$ is the left boundary of the domain $\Gamma_\placeholder$ of $\hat\psi_\placeholder$. $\Phi_\placeholder$ is the matrix of homogeneous solutions to each equation with an analogous form to (\ref{phi2}), satisfying $\partial_r \Phi_\placeholder+\hat A_\placeholder\Phi_\placeholder=0$. We assume that both these homogeneous solutions and the retarded solutions are related via transformations of the form
\begin{subequations}\label{general transformations}
\begin{align}
\Phi_L &=T_L\Phi_H,\\
\Phi_p &=T_p\Phi_L,\\
\Phi_R &=T_R\Phi_p,\\
\Phi_\infty &=T_\infty\Phi_R,
\end{align}
\end{subequations}
and analogously, $\hat\psi^{\rm ret}_{L} = T_L \hat\psi^{{\rm ret}}_{H}$, $\hat\psi^{{\rm ret}}_{p} =T_p\hat\psi^{{\rm ret}}_{L}$, etc. This will be the case for transformations between time slicings for the first-order field and the second-order field $\psi^{(2,0)}$ sourced by $G^{(2,0)}_{\mu\nu}$. We discuss other cases in Secs.~\ref{sec_worldtube_dr0_general_regular_solutn} and \ref{sec_worldtube_G11}.  Section~\ref{sec_worldtube_G11}, in particular, shows how junction conditions for second-order fields are derived from how the fields transform between slicings.

We now fix the constants in the general solution by imposing junction conditions and boundary conditions. From Eqs.~\eqref{fields} and \eqref{general transformations}, it follows that the junction conditions are
\begin{subequations}\label{matching_conditions}
\begin{align}
\hat\psi_L(r_H) &= T_L\left.\left(\hat\psi_H + \hat\psi^{\P}_{ H}\right)\right|_{r_H},\label{match-}\\
\hat\psi_p(r_L) &= \left.\left(T_p\hat\psi_L - \hat\psi^{\P}_{ p}\right)\right|_{r_L},\\
\hat\psi_R(r_R) &= T_R\left.\left(\hat\psi_p + \hat\psi^{\P}_{ p}\right)\right|_{r_R},\\
\hat\psi_\infty(r_\infty) &= \left.\left(T_\infty\hat\psi_R - \hat\psi^{\P}_{ \infty}\right)\right|_{r_\infty}.
\end{align}
\end{subequations}

We assume boundary conditions of the form~\eqref{BCs-noncompact},
\beq
\hat\psi^{\rm ret} = \begin{cases} \hat\psi^{\rm S}_\infty + \Phi_{\infty+} c_+ & \text{for }r\to\infty, \\
													\hat\psi^{\rm S}_H + \Phi_{H-} c_- & \text{for }r\to2M,	
								\end{cases}
\eeq
for some constant $d$-vectors $c_\pm$, where  
\beq
\hat{\cal D}\hat\psi^{\rm S}_\placeholder = \hat J_\placeholder.
\eeq
We also assume that the punctures $\hat\psi_H$ and $\hat\psi_\infty$ satisfy the analogues of~\eqref{infinity condition} and~\eqref{horizon condition}:
\beq
\lim_{r\to\infty}\left[\Phi^{-1}_{\infty\rm top}(\hat\psi^\P_\infty-\hat\psi^{\rm S}_\infty)\right] = 0\label{infinity condition repeat}
\eeq
and
\beq
\lim_{r\to2M}\left[\Phi^{-1}_{H\rm bot}(\hat\psi^\P_H-\hat\psi^{\rm S}_H)\right] = 0.\label{horizon condition repeat}
\eeq
We can then impose retarded boundary conditions on $\hat\psi_H$ at $r=2M$ and on $\hat\psi_\infty$ at $r\to\infty$, which fixes the constants in the outermost regions to be
\begin{align}
a_H &= \left(a^H_1,\ldots,a^H_{k},\mathbf 0_d\right)^T,\label{regularity_1} \\
a_\infty &=\left( -\int_{r_\infty}^\infty\Phi_\infty^{-1}\hat J^{{\rm eff}}_{\infty} dr ,a^\infty_{1},\ldots,a^\infty_{k}\right)^T.\label{regularity_2}
\end{align}

By combining Eqs.~(\ref{general transformations})~and~(\ref{matching_conditions}), we derive the jump conditions
\begin{subequations}\label{jump_conditions}
\allowdisplaybreaks
\begin{align}
a_L - a_H &= \int_{2M}^{r_H}dr\, \Phi_H^{-1}\hat J^{{\rm eff}}_{H}   + C^H,\label{rH_jump}\\
a_p - a_L &= \int_{r_H}^{r_L} dr\,\Phi_L^{-1}\hat J_L  +C^L,\label{r-_jump}\\
a_R - a_p &= \int_{r_L}^{r_R} dr\,\Phi_p^{-1}\hat J^{{\rm eff}}_{p}  +C^R,\label{r+_jump}\\
a_\infty - a_R &= \int_{r_R}^{r_\infty}dr\, \Phi_R^{-1}\hat J_{R}  +C^\infty,\label{rinf_jump}
\end{align}
\end{subequations}
where 
\begin{subequations}\label{Cs}
\begin{align}
C^H &= \Phi_H^{-1}\hat\psi^{\P}_{ H}\Bigr|_{r_H},\\
C^L &= -\Phi_L^{-1}T^{-1}_p\hat\psi^{\P}_{ p}\Bigr|_{r_L},\\
C^R &= \Phi_p^{-1}\hat\psi^{\P}_{ p}\Bigr|_{r_R},\\
C^\infty &= -\Phi_R^{-1}T^{-1}_\infty\hat\psi^{\P}_{\infty}\Bigr|_{r_\infty}.
\end{align}
\end{subequations}

The boundary and jump conditions provide enough equations to determine the $a_\placeholder$'s. We find that
\begingroup
\allowdisplaybreaks
\begin{subequations}\label{the_as}
\begin{align}
a_H &= \begin{pmatrix} -\displaystyle \int_{2M}^\infty \!\!\! dr\,\Phi_{\rm top}^{-1}\hat{ J}^{{\rm eff}}\\
				\mathbf 0_d\end{pmatrix}\nonumber\\*
		&\quad +\begin{pmatrix} - C^H_{\rm top} - C^L_{\rm top} - C^R_{\,\rm top} - C^\infty_{\rm top}\\
					\mathbf 0_d\end{pmatrix}\!,\\[0.5em]
a_L &= \begin{pmatrix} -\displaystyle \int_{r_H}^\infty \!\!\! dr\,\Phi_{\rm top}^{-1}\hat{ J}^{{\rm eff}}\\[1em]
									\displaystyle \int_{2M}^{r_H}\!\!\! dr\,\Phi_{\rm bot}^{-1}\hat{ J}^{{\rm eff}}\end{pmatrix}\nonumber\\*
		&\quad +\begin{pmatrix} - C^L_{\rm top} - C^R_{\,\rm top}-C^\infty_{\rm top}\\
					  C^H_{\rm bot}\end{pmatrix}\!,\\[0.5em]
a_p &= \begin{pmatrix} -\displaystyle \int_{r_L}^\infty \!\!\! dr\,\Phi_{\rm top}^{-1}\hat{ J}^{{\rm eff}}\\[1em]
									\displaystyle \int_{2M}^{r_L}\!\!\! dr\,\Phi_{\rm bot}^{-1}\hat{ J}^{{\rm eff}}\end{pmatrix}\nonumber\\*
		&\quad +\begin{pmatrix} - C^R_{\,\rm top} - C^\infty_{\rm top}\\
					 C^H_{\rm bot} + C^L_{\rm bot}\end{pmatrix}\!,\\[0.5em]
a_R &= \begin{pmatrix} -\displaystyle \int_{r_R}^\infty \!\!\! dr\,\Phi_{\rm top}^{-1}\hat{ J}^{{\rm eff}}\\[1em]
										\displaystyle \int_{2M}^{r_R}\!\!\! dr\,\Phi_{\rm bot}^{-1}\hat{ J}^{{\rm eff}} \end{pmatrix}\nonumber\\*
		&\quad +\begin{pmatrix} -C^\infty_{\rm top}\\
					 C^H_{\rm bot} + C^L_{\rm bot} + C^R_{\,\rm bot}\end{pmatrix}\!,\\[0.5em]
a_\infty &= \begin{pmatrix} -\displaystyle \int_{r_\infty}^\infty \!\!\! dr\,\Phi_{\rm top}^{-1}\hat{ J}^{{\rm eff}}\\[1em]
											\displaystyle \int_{2M}^{r_\infty}\!\!\!dr\,\Phi_{\rm bot}^{-1}\hat{ J}^{{\rm eff}} \end{pmatrix} \nonumber\\*
		&\quad + \begin{pmatrix}\mathbf 0_d\\
						C^H_{\rm bot} + C^L_{\rm bot} + C^R_{\,\rm bot}+C^\infty_{\rm bot}\end{pmatrix}\!,
\end{align}
\end{subequations}
\endgroup
where we have defined $\Phi(r) := \Phi_\placeholder(r)$ for $r\in\Gamma_\placeholder$ and $\hat J^{\rm eff}(r):=\hat J^{\rm eff}_\placeholder(r)$ for $r\in\Gamma_\placeholder$.

With Eq.~\eqref{general_ODE_solution}, Eq.~\eqref{the_as} gives the global solution:
\begin{align}
\hat\psi &= \Phi \Bigg[{\sf v} - \begin{pmatrix}C^H_{\rm top}+C^L_{\rm top}+C^R_{\rm top}+C^\infty_{\rm top}\\\mathbf{0}_d\end{pmatrix}  \nonumber\\
			&\quad + C^H\theta(r-r_H)+C^L\theta(r-r_L)+C^R\theta(r-r_R)\nonumber\\
			&\quad +C^\infty\theta(r-r_\infty)\Bigg],\label{psi_worldtube_solution}
\end{align}
where ${\sf v}$ is given by Eq.~\eqref{sfv} with the replacement $\hat{J}\to\hat{J}^{\rm eff}$, and where, following the convention just above, we have defined $\Phi(r) := \Phi_\placeholder(r)$ and $\hat\psi(r) := \hat\psi_\placeholder(r)$ for $r\in\Gamma_\placeholder$.

Equation~\eqref{psi_worldtube_solution} with Eqs.~\eqref{sfv} and \eqref{Cs} give the solution in each region $\Gamma_\placeholder$ in a form close to that of Eq.~\eqref{psi_regular_with_r_0}, but with the junction conditions across region boundaries accounted for by the additive constants $C^\placeholder$. In the next two sections, we describe two specific examples of this general framework.

%-------------------------------------------------------------------------------------------------------------------------------------------------------%
\subsection{Example 1: $t$ slicing}\label{sec_punctures_at_the_horizon_and_infinity}
%-------------------------------------------------------------------------------------------------------------------------------------------------------%

We first consider calculations on constant-$t$ slices (i.e., $H=0$), with punctures allowed at the horizon, at the particle, and at infinity.

In this case, the operators $\hat{\cal D}_\placeholder$ are the same for all regions, equal to $\hat{\cal D}_{[t]} = \mathbf{1}_{2d\times 2d}\frac{d}{dr}+\hat A_{[t]}$, where $\hat A_{[t]}$, defined in Eq.~\eqref{hat_A_here}, is given by~\eqref{AB_gravity} (Lorenz) or \eqref{eq:ab_teukolsky} (Teukolsky) with $H=0$. The matrices $\Phi_\placeholder$ are also all the same, equal to $\Phi_{[t]}$, the matrix of homogeneous solutions satisfying $\hat{\cal D}_{[t]}\hat\psi^{[t]}_{k\pm}=0$ subject to the boundary conditions~\eqref{BCs_radiative} and \eqref{BCs_stationary} with $k(r^*)=0$. 

The solution is hence given by Eq.~\eqref{psi_worldtube_solution} with $\Phi=\Phi_{[t]}$ and $\hat{\cal D}$ (which appears in $\hat J^{\rm eff}$) given by $\hat{\cal D}_{[t]}$. The constants $C^s$ are given by Eq.~\eqref{Cs} with  $T_p = T_L = T_R = T_\infty = \mathbf{1}_{2d\times 2d}$.

%-------------------------------------------------------------------------------------------------------------------------------------------------------%
\subsection{Example 2: sharp $v$-$t$-$u$ slicing}\label{sec_discontinuous_slicing}
%-------------------------------------------------------------------------------------------------------------------------------------------------------%

Next we consider a sharp hyperboloidal slicing of the type described in Sec.~\ref{sec_multiscale_expansion}, with the sharp transitions occurring at boundaries between regions. 

As an example, we consider using $s=v$ in $\Gamma_H$; $s=t$ in $\Gamma_L$, $\Gamma_R$, and $\Gamma_p$; and $s=u$ in $\Gamma_\infty$. The operators $\hat{\cal D}_\placeholder$ in this case are $\hat{\cal D}_H  = \hat{\cal D}_{[v]}$, $\hat{\cal D}_L=\hat{\cal D}_R=\hat{\cal D}_p = \hat{\cal D}_{[t]}$, and  $\hat{\cal D}_\infty = \hat{\cal D}_{[u]}$. Here  $\hat{\cal D}_{[s]} = \mathbf{1}_{2d\times 2d}\frac{d}{dr}+\hat A_{[s]}$ ($s=t$, $v$, or $u$) with $\hat A_{[s]}$ given by Eqs.~\eqref{hat_A_here} and \eqref{AB_gravity} with $H=-1$ ($s=v$), $H=0$ ($s=t$), or $H=+1$ ($s=u$). Similarly, the matrices of homogeneous solutions in this case are $\Phi_H  = \Phi_{[v]}$, $\Phi_L=\Phi_R=\Phi_p = \Phi_{[t]}$, and $\Phi_\infty = \Phi_{[u]}$, and are constructed from the homogeneous solutions satisfying $\hat{\cal D}_{[s]}\hat\psi^{[s]}_{k\pm}=0$ subject to the boundary conditions~\eqref{BCs_radiative}, \eqref{BCs_stationary}, or \eqref{BCs_teukolsky} with the corresponding choice of height function $k(r^*)$.

Referring to the discussion around Eq.~\eqref{hat psi transformation}, we find that the homogeneous solutions in the different regions are related as
\begin{subequations}\label{vtu transformations}
\begin{align}
\Phi_{[v]} &= P_-\Phi_{[t]}, &\Phi_{[u]} &= P_+\Phi_{[t]},\label{Phiuv_to_Phit}\\
\Phi_{[t]} &= P_+\Phi_{[v]}, &\Phi_{[t]} &= P_-\Phi_{[u]} ,\label{Phit_to_Phiuv}
\end{align}
\end{subequations}
where
\begin{equation}
P_\pm = e^{\mp i \omega r^\ast}\begin{pmatrix}\mathbf{1}_{d\times d}&\mathbf{0}_{d\times d}\\[0.4em]
																\mp \displaystyle  i\omega f^{-1} \mathbf{1}_{d\times d}&\mathbf{1}_{d\times d}
															\end{pmatrix}.
\end{equation} 
Note that $P_\pm^{-1}=P_\mp$. The transformation matrices in Eq.~\eqref{general transformations} are therefore $T_L=T_\infty= P_+$ and $T_p = T_R = \mathbf{1}_{2d\times 2d}$.

The solution in each region is given by Eq.~\eqref{psi_worldtube_solution} with $\Phi$ and $\hat{\cal D}$ as described above. Equations~\eqref{Cs} become
\begin{subequations}\label{Cs-vtu}
\begin{align}
C^H &= \Phi^{-1}_{[v]}\hat\psi^{\P}_{ H}\Bigr|_{r_H},\\
C^L &= -\Phi^{-1}_{[t]}\hat\psi^{\P}_{ p}\Bigr|_{r_L},\\
C^R &= \Phi^{-1}_{[t]}\hat\psi^{\P}_{ p}\Bigr|_{r_R},\\
C^\infty &= -\Phi^{-1}_{[u]}\hat\psi^{\P}_{\infty}\Bigr|_{r_\infty},
\end{align}
\end{subequations}
where the inverse matrices are evaluated at the relevant boundary between regions.

%-------------------------------------------------------------------------------------------------------------------------------------------------------% 
%
\section{Derivative of the field with respect to an orbital parameter}\label{sec_worldtube_dr0_general_regular_solutn}
%
%-------------------------------------------------------------------------------------------------------------------------------------------------------

As discussed in Sec.~\ref{sec_multiscale_expansion}, one of the required ingredients in the multiscale expansion is the parametric derivative $\vec\partial_{\cal V}h^{(1)}_{\mu\nu}$, where $\vec\partial_{\cal V}$ is defined in Eq.~\eqref{param D quasicircular} for quasicircular orbits and Eq.~\eqref{param D eccentric} for eccentric orbits. This has two types of essential input: derivatives with respect to orbital parameters $p^i$, and derivatives with respect to black hole parameters $\delta M_A$. Here we will only consider the first type. The second type is trivial because the contribution to $h^{(1)}_{\mu\nu}$ from $\delta M$ and $\delta J$ are simple analytical functions~\cite{Miller:2020bft}, while the dependence of $h^{(1)}_{\mu\nu}$ on orbital parameters is (in general) only known numerically.

We keep our discussion in this section generic by writing a derivative of $\psi$ with respect to an orbital parameter as $\delta\psi$. However, our treatment is slightly less generic than in the previous two sections: we assume that $\psi$ has a compact source bounded between some minimum and maximum radius, as is the case at first order for bound orbits. $\psi$ is then given by a retarded solution~\eqref{psi_regular_with_r_0} that reduces to the form~\eqref{BCs-compact} outside the source region.

As shown in Ref.~\cite{Durkan:2022fvm}, the most efficient way to calculate $\delta \psi$ is to formulate a field equation for it. As we explain in this section, that field equation can be solved using the puncture scheme developed in the previous two sections. However, new junction conditions must be introduced at the boundaries between regions, and punctures must often be introduced at the outer boundaries.

%-------------------------------------------------------------------------------------------------------------------------------------------------------%
\subsection{Smooth slicing and windowed punctures}
%
%-------------------------------------------------------------------------------------------------------------------------------------------------------

To introduce the structure of the problem, we return to the case of smooth slicing and windowed punctures. Using the same notation as in previous sections, we assume that $\hat A,\hat J$ are functions of both $r$ and $p^i$, $\hat A=\hat A(r,p^i)$ and $\hat J=\hat J(r,p^i)$. As a result, the matrix of homogeneous solutions $\Phi$ will also depend on $p^i$, $\Phi=\Phi(r,p^i)$. 

$\hat\psi$ satisfies Eq.~\eqref{fieldEquations3}. By differentiating that equation with respect to an orbital parameter, we obtain a field equation for $\delta \hat \psi$,
\begin{equation}
\hat{\cal D}\hat \varphi = \hat K, \label{dr0_field_equation}
\end{equation} 
where we have defined the field variable
\beq
\hat\varphi := \delta\hat\psi
\eeq
and the source
\begin{equation} 
\hat K   = - \delta \hat A\, \hat\psi + \delta \hat J.
\label{dSourcePointParticle} 
\end{equation} 
We note that $\hat A$ only depends on $p^i$ through a dependence on $\omega(p^i)$, meaning
\beq
\delta\hat A = 0 \quad \text{for } \omega=0.
\eeq 

Equation~(\ref{dr0_field_equation}) for $\hat\varphi$ has the same form as Eq.~\eqref{fieldEquations3}, just with a different source. However, the source is now always noncompact, due to the term $\delta \hat A\, \hat\psi $. Hence, in general the retarded integral may not yield the correct solution (or indeed, even converge); we will in fact find that is the case if we use $t$ as our time function. To allow for that possibility, we introduce punctures at the boundaries, $\delta \hat \psi^{\P}_{ H}$ and $\delta \hat \psi^{\P}_{\infty}$. For now we take these punctures to include windows, making them go to zero at some distance from the boundaries, and we define the total puncture $\delta \hat \psi^{\P}=\delta \hat \psi^{\P}_{ H}+\delta \hat \psi^{\P}_{\infty}$. We then have
\begin{equation}
\hat{\cal D}\hat \varphi^\res = \hat{K} -\hat{\cal D}\hat \varphi^\P =:  \hat{K}^{\rm eff}. \label{dr0_field_equation-eff}
\end{equation} 

We can now solve for $\hat \varphi^\res$ using the same methods we used to solve (\ref{fieldEquations3}). The retarded integral of Eq.~(\ref{dr0_field_equation-eff}) can be read off Eq.~(\ref{psi_regular_with_r_0}) by substituting
$\hat{K}^{\rm eff} $ for $\hat J$, yielding
\begin{equation}
\hat \varphi^\res =\Phi\,{\sf v}_\varphi,\label{d_r_0_psi_regular2}
\end{equation}
where ${\sf v}_\varphi$ is given by Eq.~\eqref{sfv} with the replacement $\hat J\to\hat{K}^{\rm eff}$: 
\begin{equation}
{\sf v}_\varphi = \left(- { \int_r^\infty dr'\Phi^{-1}_{\rm top} \hat{K}^{\rm eff}  }  \,,\, { \int_{2M}^r dr'   \Phi^{-1}_{\rm bot} \hat{K}^{\rm eff}  }\right)^T.
\end{equation}
It will be useful to write this as
\beq\label{sfv_phi}
{\sf v}_\varphi = {\sf v}_1+{\sf v}_2
\eeq
with ${\sf v}_1$ and ${\sf v}_2$ given by
\begin{align}
{\sf v}_1 &= \begin{pmatrix}\displaystyle\int_r^\infty \!\!dr' \Phi^{-1}_{\rm top} \left(\delta \hat A\hat\psi + \hat{\cal D}\hat \varphi^\P\right)\\[1em]
					-\displaystyle\int_{2M}^r \!\!dr'  \Phi^{-1}_{\rm bot}\left(\delta \hat A\hat\psi + \hat{\cal D}\hat \varphi^\P\right)\end{pmatrix}\!,\label{def_v1}\\
{\sf v}_2 &= \begin{pmatrix}-\displaystyle\int_r^\infty \!\!dr' \Phi^{-1}_{\rm top}  \delta \hat J\\[1em]
					\displaystyle\int_{2M}^r \!\!dr'  \Phi^{-1}_{\rm bot} \delta \hat J\end{pmatrix}\!.\label{def_v2}
\end{align}

%----------------------------------------------------------------------------------------------------------------------------%
\subsection{Worldtube method}
%----------------------------------------------------------------------------------------------------------------------------%

We can reformulate the calculation of $\hat \varphi$ in precisely the same way we did the calculation of $\hat\psi$ in Sec.~\ref{sec_general_formulation}. In place of Eq.~\eqref{dr0_field_equation-eff}, we have equations in each region $\Gamma_\placeholder$:
\begin{equation}\label{dr0_ODE}
\hat{\cal D}_\placeholder\hat \varphi_\placeholder = \hat{\cal  J}^{{\rm eff}}_\placeholder.
\end{equation}
The field variable $\hat\varphi_a$ in $\Gamma_\placeholder$ can be either the physical field $\delta\hat\psi^{\rm ret}$ or a residual field $\delta\hat\psi^{\rm ret}-\hat\varphi^\P$.

The general solution to this equation in each region is
\begin{equation}\label{general_solutions_to_effective_source_equations_again}
\hat \varphi_\placeholder = \Phi_\placeholder\left(\int_{r_{\placeholder}}^r\Phi_\placeholder^{-1}\hat{K}^{{\rm eff}}_\placeholder dr + b_\placeholder\right).
\end{equation}
We can find the constants $b_\placeholder$ from junction conditions and boundary conditions, in the same manner we found the $a_\placeholder$'s in Sec.~\ref{sec_general_formulation}. 

We readily derive the junction conditions for $\hat \varphi$ by taking a parametric derivative of the conditions $\hat\psi^{\rm ret}_L(r_H) = T_L\hat\psi^{\rm ret}_H(r_H)$, $\hat\psi^{\rm ret}_p(r_L) = T_p\hat\psi^{\rm ret}_L(r_L)$, $\hat\psi^{\rm ret}_R(r_R) = T_R\hat\psi^{\rm ret}_p(r_R)$, and $\hat\psi^{\rm ret}_\infty(r_\infty) = T_\infty\hat\psi^{\rm ret}_R(r_\infty)$. The results are
\begin{subequations}\label{dr0_matching_conditions}
\begin{align}
\hat \varphi_L(r_H) &= \left[T_L\left(\hat \varphi_H +\hat \varphi^{\P}_{ H}\right) +\delta T_L\hat\psi^{\rm ret}_H\right]_{r_H},\\
\hat \varphi_p(r_L) &= \left[T_p\hat \varphi_L +\delta T_p\hat\psi_L - \hat \varphi^{\P}_{ p}\right]_{r_L},\\
\hat \varphi_R(r_R) &= \left[T_R\left(\hat \varphi_p +\hat \varphi^{\P}_{p}\right)+\delta T_R\hat\psi^{\rm ret}_p\right]_{r_R},\\
\hat \varphi_\infty(r_\infty) &= \left[T_\infty\hat \varphi_R
					+\delta T_\infty\hat\psi_R - \hat \varphi^{\P}_{\infty}\right]_{r_\infty}.
\end{align}\end{subequations}

Using these conditions to derive the analogues of~\eqref{jump_conditions}, and imposing retarded boundary conditions, we obtain enough equations to fix the $b_\placeholder$'s. The result is that $b_\placeholder$ is identical to $a_\placeholder$, as given in Eq.~\eqref{the_as}, with the replacements $\hat J^{\rm eff}\to\hat{K}^{\rm eff}$ and $C^\placeholder\to{D}^\placeholder$, where
\begin{subequations}
\begin{align}
{D}^H &= \left[\Phi_H^{-1}\hat \varphi^{\P}_{ H} +\Phi_H^{-1}T_L^{-1}\delta T_{L}\hat\psi^{\rm ret}_H\right]_{r_H},\\
{D}^L &= \left[-\Phi_L^{-1}T^{-1}_p\hat \varphi^{\P}_{p} +\Phi_L^{-1}T_p^{-1}\delta T_{p}\hat\psi_L\right]_{r_L},\\
{D}^R &= \left[\Phi_p^{-1}\hat \varphi^{\P}_{p}+\Phi_p^{-1}T_R^{-1}\delta T_{R}\hat\psi^{\rm ret}_p\right]_{r_R},\\
{D}^\infty &= \left[-\Phi_R^{-1}T^{-1}_\infty\hat \varphi^{\P}_{\infty} +\Phi_R^{-1}T_\infty^{-1}\delta T_{\infty}\hat\psi_R\right]_{r_\infty}.
\end{align}
\end{subequations}

Substituting the $b_\placeholder$'s into Eq.~\eqref{general_solutions_to_effective_source_equations_again}, we obtain
\begin{align}
\hat \varphi &= \Phi \Bigg[{\sf v}_\varphi - \begin{pmatrix}{D}^H_{\rm top}+{D}^L_{\rm top}+{D}^R_{\rm top}+{D}^\infty_{\rm top}\\\mathbf{0}_d\end{pmatrix}  \nonumber\\
			&\quad + {D}^H\theta(r-r_H)+{D}^L\theta(r-r_L)+{D}^R\theta(r-r_R)\nonumber\\
			&\quad + {D}^\infty\theta(r-r_\infty)\Bigg],\label{dpsi_worldtube_solution}
\end{align}
with ${\sf v}_\varphi$ given by Eq.~\eqref{sfv_phi}. As in Eq.~\eqref{psi_worldtube_solution}, we have defined $\hat \varphi(r) = \hat \varphi_\placeholder(r)$, $\Phi(r)=\Phi_\placeholder(r)$, $\hat{K}^{\rm eff}(r)=\hat{K}^{\rm eff}_\placeholder(r)$ for $r\in\Gamma_\placeholder$.

Equation \eqref{dpsi_worldtube_solution} yields $\hat \varphi$ in each region for generic slicings and punctures. We next consider the more specific cases of $t$ slicing and sharp $v$-$t$-$u$ slicing.

%----------------------------------------------------------------------------------------------------------------------------
\subsection{Example 1: $t$ slicing}\label{sec_worldtube_method_t_slicing}
%----------------------------------------------------------------------------------------------------------------------------

First we specialize to $t$ slicing. As we shall see, punctures are required at the boundaries in this case.

$\hat{\cal D}_\placeholder$ and $\Phi_\placeholder$ are the same for all regions, which means the transformation matrices are all $T_\placeholder=1$. We leave punctures at the horizon and infinity, but we use a point source at the particle instead of a puncture. As a result, we can combine the regions $\Gamma_L=(r_H,r_L)$, $\Gamma_p=(r_L,r_R)$, and $\Gamma_R=(r_R,r_\infty)$ into an enlarged $\Gamma_p=(r_H,r_\infty)$. With this setup, Eq.~\eqref{dpsi_worldtube_solution} reduces to
\begin{align}
\hat \varphi &= \Phi_{[t]} \Bigg[{\sf v}_\varphi - \begin{pmatrix} {D}^H_{\rm top}+ {D}^\infty_{\rm top}\\\mathbf{0}_d\end{pmatrix} +  {D}^H\theta(r-r_H)  \nonumber\\
			&\quad  + {D}^\infty\theta(r-r_\infty)\Bigg],\label{dpsi_t_slicing}
\end{align} 
where ${D}^\infty=-\left.\Phi^{-1}_{[t]}\hat \varphi^{{\cal P}}_{\infty}\right|_{r_\infty}$ and ${D}^H=\left.\Phi^{-1}_{[t]}\hat \varphi^{{\cal P}}_{H}\right|_{r_H}$.  Again we stress that here we define $\hat \varphi(r) = \hat \varphi_\placeholder(r)$ for $r\in\Gamma_\placeholder$, meaning $\hat \varphi$ is to be interpreted as $\hat \varphi_\infty = \hat \varphi^\res$ in $\Gamma_\infty$, for example.
 
To assess the need for punctures, we first analyze the integrands $\Phi^{-1}_{[t]\rm top}\delta\hat A_{[t]}\hat\psi^{\rm ret}_{[t]}$ and $\Phi^{-1}_{[t]\rm bot}\delta\hat A\hat\psi^{\rm ret}_{[t]}$ in Eq.~\eqref{sfv_phi}. Our analysis appeals to the concrete form of $\hat A$ in Eqs.~\eqref{hat_A_here}, \eqref{AB_gravity}, and \eqref{eq:ab_teukolsky}. 

Recalling that $\delta\hat A_{[t]}=0$ for $\omega=0$ modes, we examine $\omega\neq0$ modes at large $r$. We have $\delta\hat A_{[t]}\sim r^0$ and $\hat\psi^{\rm ret}_{[t]}\sim e^{+ i\omega r^*}$. $\Phi^{-1}_{[t]}$ is made up of quantities that all behave as $\sim e^{i k\omega r^*}$ for some $k$ at large $r$. So in principle, $\Phi^{-1}_{[t]\rm top}\delta\hat A_{[t]}\hat\psi^{\rm ret}_{[t]}$ asymptotes to a sum of terms $\sim e^{i p\omega r^*}$ with different $p$'s. For $p=0$, $\int^\infty_{r} e^{i p\omega r'^*}dr' \sim\lim_{R\to\infty}R$; for $p\neq0$, $\int^\infty_{r} e^{i p\omega r'^*}dr' \sim\lim_{R\to\infty}e^{i p\omega R}$. In either case, the limit does not exist, indicating that the integral in the top row of Eq.~\eqref{sfv_phi} does not converge without a puncture. 

For $\omega\neq0$ modes near $r=2M$, we have $\delta\hat A_{[t]}\sim f^0$ and $\hat\psi^{\rm ret}_{[t]}\sim e^{- i\omega r^*}$. $\Phi^{-1}_{[t]}$ is made up of quantities that behave as $\sim e^{i q\omega r^*}$ or $\sim f^{-1}e^{i q\omega r^*}$, for some integer $q$ at $r\to 2M$. So in principle, $\Phi^{-1}_{[t]\rm bot}\delta\hat A_{[t]}\hat\psi^{\rm ret}_{[t]}$ could possess a power-law divergence at the horizon, indicating that the integral in the bottom row of Eq.~\eqref{sfv_phi} would diverge without a puncture.

These analytical scalings suggest the need for punctures at both the horizon and at infinity. We have  confirmed this requirement numerically. We can also confirm it by considering that $\hat \varphi$ is a parametric derivative of a retarded field. Since $\hat\psi^{\rm ret}_{[t]}\sim e^{- i\omega r^*}$ for $r\to 2M$ and $\hat\psi^{\rm ret}_{[t]}\sim e^{+ i\omega r^*}$ for $r\to\infty$, this implies
\begin{align}
\hat \varphi^{\rm ret} &\sim (\delta\omega) \ln\left(\frac{r}{2M}-1\right) e^{- i\omega r^*} \quad \text{for } r\to 2M,\\
\hat \varphi^{\rm ret} &\sim (\delta\omega) r^* e^{+ i\omega r^*} \quad \text{for } r\to \infty.
\end{align}
These behaviors clearly violate the analogue of Eqs.~\eqref{puncture conditions}, verifying the need for punctures.

We can obtain the punctures in a practical way from the large-$r$ and near-horizon expansions~\eqref{BCs_radiative} or \eqref{BCs_teukolsky}. To construct the puncture at infinity, we define 
\begin{equation}
\hat\chi_{\infty}=\Phi^{\rm out} {\sf v},\label{psi_P_infinity}
\end{equation}
with ${\sf v}$ given by Eq.~\eqref{sfv} and 
\begin{equation}
\Phi^{\rm out} = \begin{pmatrix}\begin{array}{c|c} \mathbf{0}_{2d\times d}\,&\, \hat\Psi^{1+}\cdots\hat\Psi^{d+}\end{array}\end{pmatrix},\label{Phi_out_again}
\end{equation}
where $\hat\Psi^{k+}=\left(\Psi^{k+},\partial_r\Psi^{k+}\right)^T$, $k=1,\dots,d$, and $\Psi^{k+}_{\ell m}=a^{\ell m}_{k,0}(p^i)e^{i \omega r^*}$ is the leading term in the large-$r$ expansion~\eqref{eq:BCout}. The parametric derivative is then
\begin{align}
\delta\chi_{\infty} &= \delta\Phi^{\rm out}  {\sf v} + \Phi^{\rm out} \delta{\sf v}.
\label{dr0_psi_P_infinity}
\end{align}
For the Lorenz-gauge case, we only need to remove the leading large-$r$ behavior. It therefore suffices to take $\hat\varphi^{{\cal P}}_{\infty}$ to be the leading term in Eq.~(\ref{dr0_psi_P_infinity}), 
\begin{align}\label{dpsiP infty}
\hat\varphi^{\P}_{\infty} = i(\delta\omega) r^*\Phi^{\rm out}\cdot\begin{pmatrix}{\bf 0}_d\\ c_+\end{pmatrix},%\\
\end{align}
where $c_+$ is given by Eq.~\eqref{c+}. For the Teukolsky case, the general construction is the same, but three more orders must be included in the puncture to obtain a convergent retarded integral of the effective source.

The puncture at the horizon is derived analogously. We define
\begin{equation}
\hat\chi_{H} = \Phi^{\rm in} {\sf v},\label{psi_PH}
\end{equation}
with 
\begin{align}\label{Phi_in}
\Phi^{\rm in} &= \begin{pmatrix}\begin{array}{c|c}\hat\Psi^{1-}\cdots\hat\Psi^{d-}   \,&\, \mathbf{0}_{2d\times d}\end{array}\end{pmatrix},
\end{align}
where $\hat\Psi^{k-}_{\ell m}=\left(\Psi^{k-}_{\ell m},\partial_r\Psi^{k-}_{\ell m}\right)^T$, $k=1,\dots,d$, and  $\Psi^{k-}_{\ell m}=b^{\ell m}_{k,0}(p^{i})e^{-i \omega r^*}$, the leading term in the near-horizon expansion~(\ref{eq:BCin}). The parametric derivative is 
\begin{align}
\delta\hat\chi_{H} &= \delta\Phi^{\rm in}  {\sf v} + \Phi^{\rm in} \delta{\sf v}.
\label{dr0_psi_PH}
\end{align}
Again it suffices to include just the leading term,
\beq\label{dpsiP H}
\hat\varphi^{\P}_{H} = -i(\delta\omega) r^*\Phi^{\rm in} \cdot\begin{pmatrix}c_-\\ {\bf 0}_d\end{pmatrix},
\eeq 
where $c_-$ is given by Eq.~\eqref{c-}.

In summary, with $t$ slicing, the parametric derivative of the retarded field, $\delta\hat\psi^{\rm ret}$, is given by 
\beq
\delta\hat\psi^{\rm ret} = \hat\varphi_H + \hat\varphi^{\P}_H
\eeq
for $2M<r<r_H$, by 
\beq
\delta\hat\psi^{\rm ret} = \hat\varphi_\infty + \hat\varphi^{\P}_\infty
\eeq
for $r>r_\infty$, and by 
\beq
\delta\hat\psi^{\rm ret} = \hat\varphi_p
\eeq
for $r_H<r<r_\infty$, where $\hat\varphi_a$ is given by Eq.~\eqref{dpsi_t_slicing}, %with Eqs.~\eqref{DdpsiPH}--\eqref{dAt}, \eqref{dr0_psi_P_infinity}, and \eqref{dr0_psi_PH}, 
with the punctures $\hat\varphi^\P_\infty$ and $\hat\varphi^{\P}_H$ given by Eqs.~\eqref{dpsiP infty} and \eqref{dpsiP H}.

%----------------------------------------------------------------------------------------------------------------------------%
\subsection{Example 2: $v$-$t$-$u$ slicing}\label{dh - vtu} 
%----------------------------------------------------------------------------------------------------------------------------%
We next consider sharp $v$-$t$-$u$ slicing. This is the slicing used in our numerical calculations of $\varphi$, and our description in this section focuses in on the particular choices we make in our numerical implementation. Unlike in previous sections, here we also divide the discussion between Lorenz-gauge and Teukolsky calculations as they differ in important ways.

Like in the case of $t$ slicing, we do not use a puncture at the particle.

\subsubsection{Lorenz Gauge}
\label{dh_vtu_lorenz}

In our Lorenz-gauge calculations, we merge $\Gamma_L$ and $\Gamma_R$ into $\Gamma_p$. The matrices of homogeneous solutions in the three regions $\Gamma_H$, $\Gamma_p$, $\Gamma_\infty$ are $\Phi_H=\Phi_{[v]}$, $\Phi_p=\Phi_{[t]}$, and $\Phi_\infty=\Phi_{[u]}$, as described in Sec.~\ref{sec_discontinuous_slicing}. 

We show below that a puncture is not required with this setup. Equation~\eqref{dpsi_worldtube_solution} therefore reduces to
\begin{align}
\hat\varphi &= \Phi \Bigg[{\sf v}_\varphi - \begin{pmatrix}{D}^H_{\rm top}+{D}^\infty_{\rm top}\\\mathbf{0}_d\end{pmatrix}  + {D}^H\theta(r-r_H)\nonumber\\
			&\quad  + {D}^\infty\theta(r-r_\infty)\Bigg],\label{dpsi_vtu_slicing}
\end{align}
where ${D}^H=\Phi^{-1}_{[v]}P_-\delta P_+\hat\psi^{\rm ret}_{[v]}$ and ${D}^\infty=\Phi^{-1}_{[t]}P_-\delta P_+\hat\psi^{\rm ret}_{[t]}$. Explicitly,
\begin{align}
{D}^H &= -i\delta \omega\Phi^{-1}_{[v]} \begin{pmatrix} r^*{\bf 1}_{d\times d} & {\bf 0}_{d\times d} \\ f^{-1}{\bf 1}_{d\times d} & r^*{\bf 1}_{d\times d}\end{pmatrix}\hat \psi^{\rm ret}_{[v]}\bigg|_{r=r_H},\\
{D}^\infty &= -i\delta \omega\Phi^{-1}_{[t]} \begin{pmatrix} r^*{\bf 1}_{d\times d} & {\bf 0}_{d\times d} \\ f^{-1}{\bf 1}_{d\times d} & r^*{\bf 1}_{d\times d}\end{pmatrix}\hat \psi^{\rm ret}_{[t]}\bigg|_{r=r_\infty}.
\end{align}

To justify the conclusion that no punctures are required, we first consider the integrands that appear in Eq.~\eqref{sfv_phi}. From the large-$r$ behavior~\eqref{eq:BCout}, we have $\hat\psi^{\rm ret}_{[u]} \sim (r^0,\ldots, r^0,1/r^2,\ldots, 1/r^2)^T$; the absence of the phase factor $e^{i\omega r^*}$ in $u$ slicing means that the $r$ derivative of the leading term vanishes, leading to the $\sim 1/r^2$ behavior in $\partial_r\psi^{\rm ret}_{[u]}$. We also have 
\beq
\delta\hat A_{[u]}\sim \begin{pmatrix}\mathbf{0}_{d\times d} & \mathbf{0}_{d\times d}\\ r^{-2}\mathbf{1}_{d\times d} & r^0 \mathbf{1}_{d\times d} \end{pmatrix}.
\label{deltaA_u_lorenz}
\eeq
Hence, the source in $\Gamma_\infty$, $\hat{K}_\infty = -\delta\hat A_{[u]}\hat\psi^{\rm ret}_{[u]}$, behaves as $\sim r^{-2}$. To assess the falloff of the integrand $\Phi^{-1}_{[u]\rm top}\delta\hat A_{[u]}\hat \psi^{\rm ret}_{[u]}$, we also require the falloff of $\Phi^{-1}_{[u]\rm top}$. From the large-$r$ behavior~\eqref{eq:BCout}, we have the block form 
\beq
\Phi_{[u]} \sim \begin{pmatrix} (1 + e^{-2i\omega r^*})\mathbf{1}_{d\times d}  & r^0 \mathbf{1}_{d\times d} \\ e^{-2i\omega r^*}\mathbf{1}_{d\times d}  & r^{-2}\mathbf{1}_{d\times d}\end{pmatrix}, 
\eeq
from which we can derive $\Phi^{-1}_{[u]\rm top}\sim r^0$ (possibly with oscillatory terms). Therefore the integrand in the upper half of Eq.~\eqref{sfv_phi} falls off as $1/r^2$ (again, possibly with oscillatory terms), and the integrals converge without a need for a puncture. 

The core of this sketch is that in hyperboloidal slicing, the outgoing modes at infinity do not contain an oscillatory factor. A similar sketch applies for the integral in the lower half of Eq.~\eqref{sfv_phi}, using the fact that the ingoing modes at the horizon likewise contain no oscillatory factor. 

We have also numerically verified that the stronger (but necessary) conditions~\eqref{puncture conditions} are met. The key reason is again the lack of oscillatory factors. The correct boundary conditions are provided by the parametric derivative of the retarded field $\hat\psi^{\rm ret}_{[vtu]}$. Because there are no oscillatory factors in $\hat\psi^{\rm ret}_{[vtu]}$, we have that $\delta\hat\psi^{\rm ret}_{[u]}$ has the same falloff as $\hat\psi^{\rm ret}_{[u]}$ as $r\to\infty$, and $\delta\hat\psi^{\rm ret}_{[v]}$ has the same behavior as $\hat\psi^{\rm ret}_{[v]}$ as $r\to 2M$; this contrasts with the behavior in $t$ slicing, illustrated in Eqs.~\eqref{dr0_psi_P_infinity} and \eqref{dr0_psi_PH}, where the parametric derivative introduces irregularities at the boundaries. 

In summary, with $v$-$t$-$u$ slicing, $\delta\hat\psi^{\rm ret}$ is given by Eq.~\eqref{dpsi_vtu_slicing} with Eq.~\eqref{sfv_phi} and vanishing punctures in ${\sf v}_\varphi$.

\subsubsection{Teukolsky}
\label{dh_vtu_teukolsky}

We now consider the equivalent calculation in our Teukolsky formalism of Sec.~\ref{sec_reduced_teukolsky}.  
Similar to the Lorenz-gauge calculations, we consolidate $\Gamma_L$ and $\Gamma_R$ into a single region, $\Gamma_{p}$.
But, unlike in the Lorenz gauge, it is necessary to include a puncture, $\hat{\varphi}_{\infty}$ in the asymptotic regime of $\Gamma_{\infty}$.  
The solution in Eq.~(\ref{dpsi_worldtube_solution}) reduces to 
\begin{align}
\hat \varphi &= \Phi \Bigg[{\sf v}_\varphi - \begin{pmatrix} {D}^H_{\rm top}+ {D}^\infty_{\rm top}\\\mathbf{0}_d\end{pmatrix} +
	{D^H\theta(r-r_H)}   \nonumber\\
	&\quad  + {D}^\infty\theta(r-r_\infty)\Bigg],
	\label{dpsi_teuk_vtu_slicing}
\end{align} 
where 
\begin{align}
	{D}^H &= \Phi^{-1}_{[v]}P_-\delta P_+\hat\psi^{\rm ret}_{[v]} \nonumber\\
	{D}^\infty &= \left[-\Phi_{[t]}^{-1}P_{-}\hat \varphi^{\P}_{\infty} + \Phi_{[t]}^{-1}P_{-}\delta 
	P_{+}\hat\psi^{\rm ret}_{[t]} \right]_{r_\infty},
	\label{D_teuk_vtu_slicing}
\end{align}
with $\Phi_H=\Phi_{[v]}$, $\Phi_p=\Phi_{[t]}$, and $\Phi_\infty=\Phi_{[u]}$ in
the regions $\Gamma_{H}$, $\Gamma_{p}$, $\Gamma_{\infty}$ respectively.  
In this section we wish to demonstrate the need for an appropriate puncture within the asymptotic region $\Gamma_{\infty}$ despite the
introduction of hyperboloidal slicing.

Once more, let us consider the integrals over the extended source term that appear in Eq.~(\ref{sfv_phi}).
The boundary conditions of the homogeneous solutions to the Teukolsky equation results imply, under rescaling, $\hat{\psi}^{\rm ret}_{[u]} \sim (r^{0}, 1/r^{2})^{T}$, like in the Lorenz gauge.

In the Teukolsky form of our worldtube method, $\delta \hat{A}_{[u]}$ is given, at leading order, by
\beq
	\delta\hat A_{[u]}\sim \begin{pmatrix}\mathbf{0} & \mathbf{0}\\
	r^{-2} & r^0 \end{pmatrix}.
\label{deltaA_u_teukolsky}
\eeq
Therefore the source behaves as $\hat{K}_{\infty} \sim r^{-2}$ in the asymptotic region $\Gamma_{\infty}$, again just as in the Lorenz gauge. However, the falloff of the entire integrand is where the similarities with the Lorenz gauge end.  If we consider the final factor in the integrand, $\Phi^{-1}_{[u]\rm top}$, we find that Eqs.~(\ref{eq:teukolsky_boundary_conditions_in}) and (\ref{eq:teukolsky_boundary_conditions_up}) imply
\beq\label{Phiu Teukolsky}
\Phi_{[u]} \sim \begin{pmatrix} r^0 + r^{2{\sf s}}e^{-2i\omega r^*}  & r^0\\ 
			r^{-2} + r^{2{\sf s}}e^{-2i\omega r^*} & r^{-2}\end{pmatrix}, 
\eeq
which leads to $\Phi^{-1}_{[u]\rm top} \sim r^{-2{\sf s}}(r^{-2}, r^{0})^{T}$ (neglecting oscillatory factors).\footnote{For a generic matrix of the form~\eqref{Phiu Teukolsky}, the large-$r$ growth of $\Phi^{-1}_{[u]\rm top}$ is slower than our displayed scaling. Our scaling relies on the fact that the determinant of the leading-order large-$r$ term in $\Phi_{[u]}$ vanishes. We can write that matrix as $\begin{pmatrix} a r^0  & b r^0\\ c r^{-2} & d r^{-2}\end{pmatrix}$. Each column here comes from the large-$r$ expansion of an outgoing wave solution $[{}_{\sf s} R^+_{\lmw},\partial_r({}_{\sf s} R^+_{\lmw})]^T$, multiplied by a constant $a$ or $b$, with ${}_{\sf s} R^+_{\lmw}=r^0+a_1 r^{-1}+{\cal O}(r^{-2})$ for some constant $a_1$. We therefore have $c=-a a_1$ and $d=-b a_1$, which makes the determinant vanish.} The integrands in the upper half of the solution~\eqref{sfv_phi}
 therefore diverge as $\Phi^{-1}_{[u]\rm top}\delta\hat A_{[u]}\hat \psi^{\rm ret}_{[u]} \sim r^{-2({\sf s} + 1)}$.  For the spin-weight 
${\sf s} = - 2$ that we are considering here, $\Phi^{-1}_{[u]\rm top}\delta\hat A_{[u]}\hat \psi^{\rm ret}_{[u]} \sim r^{2}$.  Hence the integrals
diverge, and to obtain a physical solution one requires a suitable puncture in $\Gamma_{\infty}$.

In our Teukolsky calculation, we obtain punctures in the same manner as in Sec.~\ref{sec_worldtube_method_t_slicing}, by appealing to
the large-$r$ expansions in Appendix~\ref{sec_teukolsky_basis}.  To construct a suitable puncture in $\Gamma_{\infty}$, we again use the definition in
Eq.~(\ref{psi_P_infinity}), with $\Phi^{\rm out} = (\mathbf{0}_{2\times 1} | \hat{\Psi}^{+})$.  Here $\hat{\Psi}^{+}$ is given by terms derived from the asymptotic 
expansion in Eq.~(\ref{BCs_teukolsky}),
\begin{equation}
	\Psi^{+}_{\ell m} = f^{-2} \sum^{j_{\rm max}}_{j = 0} \frac{a^{\ell m}_{j}(p^{i})}{(\omega r)^{j}}.  
	\label{Psi_plus_expansion}
\end{equation}
Here $j_{\rm max} \geq 2$; since the divergence is $\sim r^{2}$ in the integrand, one must consider an expansion at least up to ${\cal O}(r^{-2})$.
This is to ensure the effective source ${\hat K}^{\rm eff}_{\infty}$ falls off sufficiently quickly for the integral to converge.  
The puncture, $\hat{\varphi}^{\cal P}_{\infty}$, is therefore defined by taking the first term of the parametric derivative, 
$\delta \chi_{\infty}$, in Eq.~(\ref{dr0_psi_P_infinity}) with $\delta\Phi^{\rm out}$ derived from the asymptotic expansion in Eq.~(\ref{Psi_plus_expansion}).
This yields
\begin{equation}
	\delta\Phi^{\rm out} = \frac{\delta \omega}{f^2 r} \sum^{j_{\rm max}}_{j = 0} j \frac{a^{\ell m}_{j}(p^{i})}{(\omega r)^{j-1}} 
	\begin{pmatrix}
		0 & {r^2 } \\ 0 & -j r - 4M f^{-1}
	\end{pmatrix},
%	\label{Psi_plus_expansion}
\end{equation}
such that the final puncture is given by
\begin{equation}
	\hat\varphi^{\P}_{\infty} = \delta \Phi^{\rm out}\cdot\begin{pmatrix}0\\ c_+\end{pmatrix},
\end{equation}
with $c_+$ given by Eq.~\eqref{c+}.

\section{Field equations with parametric-derivative sources}\label{sec_worldtube_G11}

As a final case, we consider a field sourced by a parametric derivative of a lower-order field. This is the type of source in the field equation~\eqref{fieldEquation11}, which we rewrite here as
\begin{equation}
\hat{\cal D}_a\hat \psi^{(1,1)}_a = \hat {J}^{(1,1)}_{{\rm eff},a}. \label{J11_field_equation}
\end{equation} 
We restrict our analysis to quasicircular orbits for simplicity, but the extension to eccentric orbits is immediate. For simplicity, we also assume the falloff properties of the Lorenz-gauge $\Phi^{-1}$, but the discussion is straightforwardly extended to allow for the Teukolsky falloff behavior.

We organize our analysis somewhat differently here than in the preceding three sections. Rather than first considering a generic formulation and then examining the scheme in $t$ slicing and in $v$-$t$-$u$ slicing, here we begin with the fact that no punctures at the boundaries are required in $v$-$t$-$u$ slicing (for fields exhibiting the Lorenz-gauge falloff); this follows from the scaling of the sources in $u$ and $v$ slicing, given in Eqs.~\eqref{J11u - FZ} and \eqref{J11v - NH}, and the arguments in Sec.~\ref{dh - vtu}. We then analyse the transformation between slicings in order to derive punctures in $t$ slicing and junction conditions in $v$-$t$-$u$ slicing. Finally, we summarize the solution in $v$-$t$-$u$ slicing.

\subsection{Junction conditions and punctures}

We first consider the transformation from $u$ to $t$ slicing. We consider a field $\psi_{[u]}[{\cal J}_I(u)]e^{-im\phi_p(u)}$ in $u$ slicing, suppressing the dependence on $r$ and $\e$. Expanding functions of $u$ around their values at $t$, we obtain 
\begin{align}
\phi_p(u) &= \phi_p(t) - r^*\Omega(t) +\frac{\e}{2}(r^*)^2 F_\Omega^{(0)}(t) +O(\e^2), \\
{\cal J}_I(u) &= {\cal J}_I(t) -\e r^* F_I^{(0)}(t)+O(\e^2),
\end{align}
and therefore 
\begin{align}
&\psi_{[u]}[{\cal J}_I(u)]e^{-im\phi_p(u)} \nonumber\\
&= \bigg\{\psi_{[u]}[{\cal J}_I(t)]  -\e\Big[r^*\vec{\partial}_{\cal V}\psi_{[u]}\nonumber\\
&\quad\ +\frac{i}{2}(r^*)^2 mF_\Omega^{(0)}\psi_{[u]}\Big]+O(\e^2)\bigg\}e^{im[\Omega r^*-\phi_p(t)]},\label{psi[u] expansion}
\end{align}
where all functions on the right are evaluated at time~$t$. Equating the right-hand side of Eq.~\eqref{psi[u] expansion} to $\psi_{[t]}[{\cal J}_I(t)]e^{-im\phi_p(t)}$, and writing the expansions
\beq
\psi_{[t]} = \e \psi^{(1)}_{[t]}+\e^2 \psi^{(2)}_{[t]}+O(\e^3)
\eeq
and
\beq
\psi_{[u]} = \e\psi^{(1)}_{[u]}+\e^2\psi^{(2)}_{[u]}+O(\e^3), 
\eeq
we find $\psi^{(1)}_{[u]}=\psi^{(1)}_{[t]}e^{-im\Omega r^*}$ and
\beq\label{psi2u vs psi2t}
\psi^{(2)}_{[u]} = \left(\psi^{(2)}_{[t]} + \Delta\psi^{(2)}_{[t]} \right)e^{-im\Omega r^*},
\eeq
where 
\beq
\Delta\psi^{(2)}_{[t]} = r^*\vec{\partial}_{\cal V}\psi^{(1)}_{[t]} - \frac{i}{2} (r^*)^2mF_\Omega^{(0)}\psi^{(1)}_{[t]}.
\eeq
Here all functions are evaluated at the same values of their arguments.

We relate 
\beq
\hat\psi^{(2)}_{[u]}=\left(\psi^{(2)}_{[u]},\partial_r\psi^{(2)}_{[u]}\right)^T
\eeq
to 
\beq
\hat\psi^{(2)}_{[t]}=\left(\psi^{(2)}_{[t]},\partial_r\psi^{(2)}_{[t]}\right)^T
\eeq
by taking a radial derivative of Eq.~\eqref{psi2u vs psi2t}. This yields
\beq
\hat\psi^{(2)}_{[u]} = P_+\left(\hat\psi^{(2)}_{[t]} +\Delta\hat\psi^{(2)}_{[t]}\right), 
\eeq
where
\begin{align}
\Delta\hat\psi^{(2)}_{[t]} &= r^*\vec{\partial}_{\cal V}\hat\psi^{(1)}_{[t]}- \frac{i}{2} (r^*)^2mF_\Omega^{(0)}\hat\psi^{(1)}_{[t]}\nonumber\\
&\quad + f^{-1}\left({\bf 0}_{d},\vec{\partial}_{\cal V}\psi^{(1)}_{[t]}-imF_\Omega^{(0)} r^*\psi^{(1)}_{[t]}\right)^T.
\end{align}
Noting that every term in the transformation involves a forcing function, we can also write
\begin{align}\label{t-u junction condition}
\hat\psi^{(1,1)}_{[u]} = P_+\left(\hat\psi^{(1,1)}_{[t]} +\Delta\hat\psi^{(2)}_{[t]}\right), 
\end{align}
while the rest of $\hat\psi^{(2)}$ transforms in the trivial way: 
\beq
\hat\psi^{(2,0)}_{[u]}= P_+ \hat\psi^{(2,0)}_{[t]}. 
\eeq

Equation~\eqref{t-u junction condition} is the junction condition at a boundary between $t$ and $u$ slicings. The same equation holds at a boundary between $v$ and $t$ slicings, with the relabeling $t\to v$, $u\to t$:
\beq\label{v-t junction condition}
\hat\psi^{(1,1)}_{[t]} = P_+\left(\hat\psi^{(1,1)}_{[v]} +\Delta\hat\psi^{(2)}_{[v]}\right), 
\eeq
where
\begin{align}
\Delta\hat\psi^{(2)}_{[v]} &= r^*\vec{\partial}_{\cal V}\hat\psi^{(1)}_{[v]}- \frac{i}{2} (r^*)^2mF_\Omega^{(0)}\hat\psi^{(1)}_{[v]}\nonumber\\
&\quad + f^{-1}\left({\bf 0}_{d},\vec{\partial}_{\cal V}\psi^{(1)}_{[v]}-imF_\Omega^{(0)} r^*\psi^{(1)}_{[v]}\right)^T.
\end{align}

In addition to providing a junction condition, Eq.~\eqref{v-t junction condition} can be used to construct a puncture at the horizon in $t$~slicing. The singularity at the horizon comes from the second term, which then serves as a puncture, 
\beq
\hat\psi^{\P(1,1)}_{H[t]} = P_+\Delta\hat\psi^{(2)}_{[v]}.
\eeq
A puncture at infinity can be constructed in the same way. Following the same steps that led to Eq.~\eqref{t-u junction condition}, we find
\beq\label{u-t junction condition}
\hat\psi^{(1,1)}_{[t]} = P_-\left(\hat\psi^{(1,1)}_{[u]} +\Delta\hat\psi^{(2)}_{[u]}\right), 
\eeq
where
\begin{align}
\Delta\hat\psi^{(2)}_{[u]} &= - r^*\vec{\partial}_{\cal V}\hat\psi^{(1)}_{[u]} - \frac{i}{2} (r^*)^2mF_\Omega^{(0)}\hat\psi^{(1)}_{[u]}\nonumber\\
&\quad - f^{-1}\left({\bf 0}_{d},\vec{\partial}_{\cal V}\psi^{(1)}_{[u]}+imF_\Omega^{(0)} r^*\psi^{(1)}_{[u]}\right)^T.
\end{align}
A valid puncture at infinity is therefore
\beq
\hat\psi^{\P(1,1)}_{\infty[t]} = P_-\Delta\hat\psi^{(2)}_{[u]}.
\eeq

\subsection{Example: $v$-$t$-$u$ slicing}

As a concrete example, we now specialize to the following setup: 
\begin{enumerate}[label=\textnormal{(\roman*)}]
\item in $\Gamma_H$, we use $v$ slicing and a puncture $\hat\psi^{\P}_{H}$
\item in $\Gamma_L$, we use $v$ slicing and no puncture
\item in $\Gamma_p$, we use $t$ slicing and a puncture $\hat\psi^{\P}_{p}$
\item we omit $\Gamma_R$
\item in $\Gamma_\infty$, we use $u$ slicing and a puncture $\hat\psi^{\P}_{\infty}$.
\end{enumerate}
This is the arrangement used in Refs.~\cite{Pound-etal:19,Warburton:2021kwk,Wardell:2021fyy}.

Following the same steps as in previous sections, starting from a solution of the form~\eqref{general_ODE_solution}, we arrive at the first three subequations in Eq.~\eqref{jump_conditions} with
\begin{subequations}\label{Cs-slowtime1}
\begin{align}
C^H_{(1,1)} &= \Phi_{[v]}^{-1}\hat\psi^{\P(1,1)}_{ H}\bigr|_{r_H},\\
C^L_{(1,1)} &= \Phi_{[t]}^{-1}\left[P_+\Delta\hat\psi^{(2)}_{[v]} - \hat\psi^{\P(1,1)}_{p}\right]\Bigr|_{r_L},\\
%C^+ &= \Phi_p^{-1}\hat\psi^{\P}_{ p}(r_+),\\
C^\infty_{(1,1)} &= \Phi_{[u]}^{-1}\Big\{P_+\!\left[\Delta\hat\psi^{(2)}_{[t]}+\hat\psi^{\P(1,1)}_{p}\right]-\hat\psi^{\P(1,1)}_{\infty}\Big\}\Bigr|_{r_\infty}
\end{align}
\end{subequations}
and with the replacements $r_R=r_\infty$, $C^R\to C^\infty_{(1,1)}$, and $a_R\to a_\infty$. The solution can then be put in the form~\eqref{psi_worldtube_solution} with $r_R=r_\infty$ and $C^R$ set to zero. 

%We can write Eqs.~\eqref{Cs-slowtime1} in the form of \eqref{Cs-vtu} with two additional terms:
%\begin{subequations}\label{Cs-slowtime2}
%\begin{align}
%C^H_{(1,1)} &= \Phi^{-1}_{[v]}\hat\psi^{\P(1,1)}_{H}\bigr|_{r_H},\\
%C^L_{(1,1)} &= \left[-\Phi^{-1}_{[t]}\hat\psi^{\P(1,1)}_{ p}+ \Phi^{-1}_{[v]}\Delta \hat\psi^{(2)}_{[v]}\right]_{r_L},\\
%C^R_{(1,1)} &= \Phi^{-1}_{[t]}\left[\hat\psi^{\P(1,1)}_{ p} + \Delta\hat\psi^{(2)}_{[t]}\right]\Bigr|_{r_R},\\
%C^\infty_{(1,1)} &= -\Phi^{-1}_{[u]}\hat\psi^{\P(1,1)}_{\infty}\bigr|_{r_\infty}.
%\end{align}
%\end{subequations}
Here, for generality, we allow for punctures at infinity and the horizon. Though they are not needed in $v$-$t$-$u$ slicing, they can be used to accelerate convergence of integrals.

%-------------------------------------------------------------------------------------------------------------------------------------------------------%
\section{Demonstration 1: Lorenz-gauge calculations for quasicircular orbits}\label{sec_demonstration_Lorenz}

%-------------------------------------------------------------------------------------------------------------------------------------------------------%

As a demonstration of our method, we consider the Lorenz-gauge field equations for a point mass on a quasicircular orbit. In that context, we calculate the first-order metric perturbation $\bar h^{(1)}_{i\ell m}$ and the parametric derivative $\partial_{r_0}\bar h^{(1)}_{i\ell m}$, where $r_0=M(M\Omega)^{-2/3}$ is the leading-order orbital radius.

\subsection{Calculation of $\bar h^{(1)}_{i\ell m}$}

In $t$ slicing, the field equations for $\bar h^{(1)}_{i\ell m}$ are identical to Ref.~\cite{Akcay:2010dx}'s Lorenz-gauge frequency-domain field equations for a particle on a circular geodesic of radius $r_0$. We write them for generic slicing in matrix form as
\beq
\hat{\cal D}_a\hat\psi^{(1)}_a = \hat J^{(1)}_a,
\label{matrix_field_eqn}
\eeq
where $\hat{\cal D}$ is defined by Eq.~\eqref{fieldEquations3v1} with Eqs.~\eqref{hat_A_here} and \eqref{AB_gravity}.

The point source takes the form
\begin{equation}
\hat J^{(1)}=\hat J^{\rm pp}\left(r_0\right)\delta\left(r-r_0\right),\label{J_point_particle_1}
\end{equation}
where $\hat J^{\rm pp}=(\mathbf{0}_d, J^{pp})^T$ and
\begin{equation}\label{Jpp}
\!J^{pp} = \beta\begin{cases}
\left(t_1\,t_3\, t_5\right)^T&\ell>0, m=0,\ell\text{ even},\\
t_8 &\ell>0,m=0,\ell\text{ odd},\\
\left(t_1\,t_3\, t_5\,t_6\right)^T&\ell=1,m=1,\\
\left(t_9\,t_{10}\right)^T&\ell, m>0,\ell+m\text{ odd},\\
\left(t_1\,t_3\, t_5\,t_6\,t_7\right)^T&\ell,m>0,\ell+m\text{ even},
\end{cases}
\end{equation}
with $\beta=64\pi M/f^2_0$ and~\cite{Barack:2007tm, Akcay:2010dx}
\begin{equation}\label{t0ilm}
t^0_{i\ell m} = -\frac{1}{4}{\cal E}_0 \alpha_{i\ell m}\begin{cases}Y^*_{\ell m }(\pi/2,0 )& i=1,\ldots, 7 \\ 
																						\partial_{\theta} Y^*_{\ell m}(\pi/2, 0) & i=8,9,10. 
																					\end{cases} 
\end{equation}
Here ${\cal E}_0 = f_0/\sqrt{1-3M/r_0}$, with $f_0= 1-2M/r_0$, is the specific energy of a point mass on a circular geodesic of radius $r_0$, and the $\alpha_{i\ell m}$'s are given by (suppressing $\ell m$ labels)
\begin{subequations}\label{alphas}
\begin{align}
\alpha_1 &=  f^2_0/r_0, &\alpha_{2,5,9} &=0, \\
\alpha_3 &= f_0/r_0, &\alpha_4 &= 2 i f_0  m \Omega, \\
\alpha_6 &= r_0 \Omega^2,  &\alpha_7 &= r_0 \Omega^2 [\ell(\ell+1)-2m^2], \\
\alpha_8 &= 2 f_0 \Omega,  &\alpha_{10} &=2 i m r_0\Omega^2.
\end{align}
\end{subequations}
Note that the $ i = 2,5,9 $ equations are sourceless.

With this source, if we use a continuous slicing, we can immediately write the solution in the form~\eqref{psi_regular_with_r_0}. Equation~(\ref{sfv}) reduces to
\begin{equation}
{\sf v}^{(1)}={\sf v}^-\theta\left(r_0-r\right)+{\sf v}^+\theta\left(r-r_0\right),\label{v^pp}
\end{equation}
where 
\begin{equation}\label{v+v-}
{\sf v}^- = \begin{pmatrix}-\Phi^{-1}_{0\,\rm top}\, \hat J^{\rm pp}\\[1em]\mathbf 0_d\end{pmatrix},\quad
{\sf v}^+ = \begin{pmatrix}\mathbf 0_d\\[1em]\Phi^{-1}_{0\,\rm bot}\,\hat J^{\rm pp} \end{pmatrix}.
\end{equation}
Here we use $\Phi_0:=\Phi(r_0)$ for brevity. The solution~(\ref{psi_regular_with_r_0}) thus becomes 
\begin{equation}
\hat\psi^{(1)}=\hat\psi^{(1)}_{-}\theta\left(r_0-r\right)+\hat\psi^{(1)}_{+}\theta\left(r-r_0\right), \label{retarded_solution_point_particle}
\end{equation}
where $\hat\psi^{(1)}_{\pm}=\Phi\,{\sf v}^\pm$. This is the standard point-particle solution in, e.g., Ref.~\cite{Akcay:2010dx}.

In sharp $v$-$t$-$u$ slicing, Eq.~\eqref{retarded_solution_point_particle} remains valid, with a simple change: in $\hat\psi^{(1)}_{\pm}=\Phi\,{\sf v}^\pm$, the matrix $\Phi$ becomes $\Phi_{[v]}$, $\Phi_{[t]}$, or $\Phi_{[u]}$, depending on the region where $\hat\psi^{(1)}_{\pm}$ is evaluated.

\begin{figure*}[t]
	\centering
	\includegraphics[width=\textwidth]{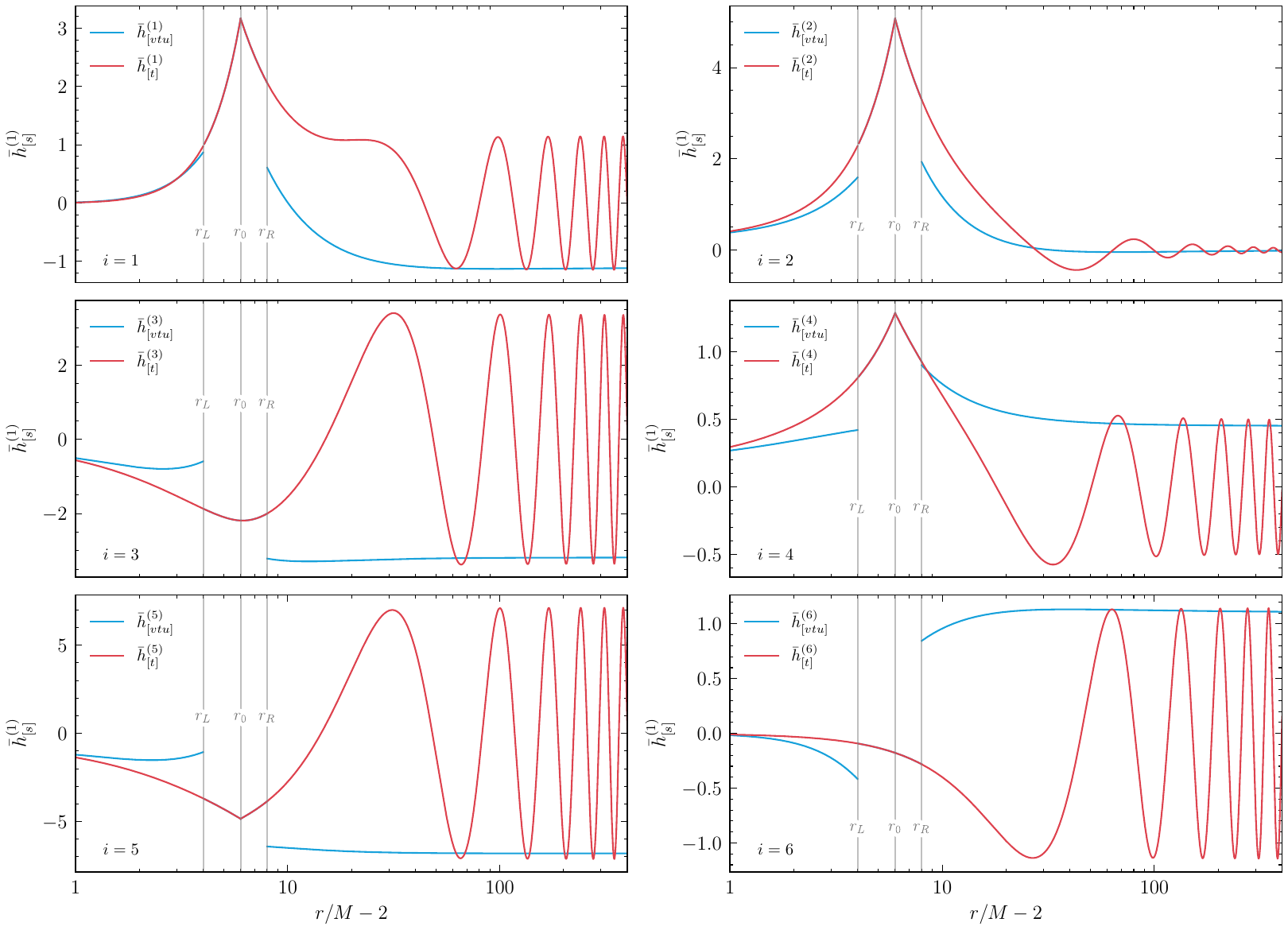}
	\caption{${\rm Re}\bigl(\bar h^{(1)}_{[vtu]}\bigr)$ (blue line)  and ${\rm Re}\bigl(\bar h^{(1)}_{[t]}\bigr)$ (red line) for all nonvanishing $i$ modes with $\ell=2, m=2$, $r_0=8M$. Note that we have not included the $i = 7$ BLS mode in this figure for brevity as this is qualitatively the same as the $i = 6$ mode.}
	\label{fig_h1_vtu}
\end{figure*}

%-------------------------------------------------------------------------------------------------------------------------------------------------------%
%\subsection{Computation of $\tilde h^{1}_{i\ell m}$}
%-------------------------------------------------------------------------------------------------------------------------------------------------------%

We evaluate this solution using the following method:
\begin{enumerate}
\item Fix a zeroth-order orbital radius $r_0$.
\item For each $\ell m$ mode, construct the matrix $\Phi$ of homogeneous solutions as reviewed in Appendix~\ref{sec_lorenz_gauge_basis}. $\Phi$ is output and stored on a grid determined by an adaptive solver.
\item Calculate the retarded field for the column vector~\eqref{psiColumnVector} using Eq.~(\ref{retarded_solution_point_particle}). In the calculation of ${\sf v}^{(1)}$, we invert the $\Phi$ matrix using the lower-upper (LU) decomposition method. Integrations and matrix inversions are performed on the same grid as in step 2.
\item For the gauge modes, we calculate the retarded field from the gauge conditions~(\ref{eq:gauge1-w0}). For $\ell=m=2$, these gauge modes are $\bar h^{(1)}_2$ and $\bar h^{(1)}_4$. We found significant numerical errors in the region close to the inner boundary $r_{\rm in}$ due to a loss of precision when subtracting one large number from another in the gauge conditions~(\ref{eq:gauge1-w0}). We used long double variables when computing these modes to resolve this issue.
\end{enumerate}

We compared our results for several modes against the same computation performed in {\it Mathematica} to validate our code. We also compared our results for $\bar h^{1}_{[t]}$ with data from the code in Ref.~\cite{Akcay:2013wfa} and found relative differences of $\lesssim 10^{-12}$, except at points near the horizon, where we found we achieved more accurate results through our use of the greater-than-machine-precision routine described in Appendix~\ref{sec_basis}.

In Fig.~\ref{fig_h1_vtu} we  compare the $\ell=2,m=2$, even-parity mode of $\bar h^{(1)}_{i\ell m}$ with $t$ slicing and with $v$-$t$-$u$ slicing. The jumps in ${\rm Re}(\bar h^{(1)}_{[vtu]})$ occur where the slicing changes from $v$ to $t$ or from $t$ to $u$. Note that $\bar h^{(1)}_{i\ell m}$ on different slices can only differ by a complex phase, such that the modulus $|\bar h^{(1)}_{[vtu]}|$ is continuous across slices. Because $v$-$t$-$u$ slicing follows wavefronts, $\bar h^{(1)}_{[vtu]}$ contains no oscillations, while $\bar h^{(1)}_{[t]}$ contains constant-amplitude oscillations at large $r$.  However, the oscillations near the horizon would only become visible at points much closer to the horizon.

To compare our results for $\bar h^{(1)}_{[t]}$ and $\bar h^{(1)}_{[vtu]}$, we transform the $\bar h^{(1)}_{[t]}$ data onto $v$-$t$-$u$ slices using the relationship
\begin{equation}
\bar h^{(1)}_{[vtu]} = e^{-i\omega  k(r^*)}\bar h^{(1)}_{[t]}.\label{rotation}
\end{equation}
Here $k(r^*)$ is given by Eq.~\eqref{sharp_slicing}, which we restate here for convenience: $k=-r^*$ for $v$ slicing, $k=0$ for $t$ slicing, and $k=+r^*$ for $u$ slicing. After performing this transformation, we find that the results in the different slicings agree to within a relative difference of $\lesssim 10^{-12}$.

%--------------------------------------------------------------------------------------------------------------------------------------------------------%
\subsection{Calculation of $\partial_{r_0}\bar h^{(1)}_{i\ell m}$}\label{sec_calculation_dr0h1}
%-------------------------------------------------------------------------------------------------------------------------------------------------------%

\subsubsection{Overview}

We next consider the field $\delta\hat\psi^{(1)}$, where we now let 
\beq
\delta := \partial_{r_0}.
\eeq
For our quasicircular orbits, the field equation satisfied by $\hat\varphi^{\rm ret} = \delta\hat\psi^{(1)}$ is 
\beq\label{dvphi Lorenz}
\hat{\cal D}\hat\varphi^{\rm ret} = \hat K^{(1)},
\eeq
where the source is
\beq\label{K Lorenz}
\hat K^{(1)} = -\delta\hat A\hat\psi^{(1)} + \delta\hat J^{(1)}.
\eeq

The first term in $\hat K^{(1)}$ is an extended source,
\begin{align}\label{dA psi1}
\delta\hat A\,\hat\psi^{(1)} &= \delta\hat A\, \hat\psi^{(1)}_-\theta(r_0-r) \nonumber\\												&\quad + \delta \hat A\,\hat\psi^{(1)}_{+}\theta(r-r_0),
\end{align}
where $\hat\psi^{(1)}_{\pm}=\Phi\,{\sf v}^\pm$ with ${\sf v}^\pm$ as given in Eq.~\eqref{v+v-}. $\hat A$ is given in terms of $A$ and $B$ in Eq.~\eqref{hat_A_here}, where $A$ and $B$ are given by \eqref{AB_gravity}. Taking a parametric derivative, we obtain
\begin{equation}
\delta \hat A = \begin{pmatrix}\mathbf{0}_{d\times d} & \mathbf{0}_{d\times d}\\ \delta A & \delta B\end{pmatrix},\label{dr0hat_A_here}
\end{equation}
where
\begin{subequations}\label{dr0_AB_gravity}
\begin{align}
\delta  A &= f^{-2} \left[2\left(1-H^2\right)\omega_m \delta \omega_m -2\omega_m^2 H \delta  H\right.\nonumber\\
						&\quad \left. +i  \delta \omega_m H'+i \omega_m \delta H'\right]\mathbf{1}_{d\times d} \nonumber\\
						&\quad +\delta {\cal M }_{h}, \\
\delta  B &= 2 i f^{-1}\left( \delta\omega_m H + \omega_m \delta H\right)\mathbf{1}_{d\times d}.
\end{align}
\end{subequations}
Here $\delta\omega_m=m\delta\Omega = -\frac{3}{2}m\sqrt{M/r_0^5}$, and we recall the notation $H'=dH/dr^*$. $\delta {\cal M}_{h}$ is given explicitly in Eqs.~\eqref{dr0_Mh_w_even} and \eqref{dr0_Mh_w_dipole}. We have allowed the height function to depend on $r_0$, in the case that the slicing evolves along with the orbit. 

Given  that $\hat J^{(1)}=\hat J^{\rm pp}\delta(r-r_0)$, the second source term in Eq.~\eqref{K Lorenz} is restricted to $r=r_0$:
\begin{align}
\delta\hat J^{(1)} &= \delta\hat J^{\rm pp} \delta(r-r_0) - \hat J^{\rm pp} \delta'(r-r_0). \label{J_dpsi_pp}
\end{align}

We solve the field equation~\eqref{dvphi Lorenz} for $\partial_{r_0}\bar h^{(1)}_{i\ell m}$ on $t$ and $v$-$t$-$u$ slices. In all cases, we evaluate integrals over extended sources using a Gauss-Kronrod quadrature routine.

\subsubsection{$\partial_{r_0}\bar h^{(1)}_{i\ell m}$ on $t$ slices}

We obtain the solution for $\partial _{r_0}\bar h^{(1)}_{[t]}$ using Eq.~(\ref{dpsi_t_slicing}), with punctures at the horizon and at infinity constructed according to Eqs.~(\ref{dpsiP infty}) and (\ref{dpsiP H}).

The main input to the solution is ${\sf v}_\varphi={\sf v}_1+{\sf v}_2$, where ${\sf v}_1$ and ${\sf v}_2$ are the integrals of source terms defined in Eqs.~\eqref{def_v1} and \eqref{def_v2}.

Given Eq.~\eqref{dA psi1}, it follows that
\begin{align}
{\sf v}_1  &= \begin{pmatrix} I_1+I_2 \\
	-I_3 \end{pmatrix} \theta\left(r_0-r\right)\nonumber\\
	&\quad	+\begin{pmatrix}   I_4 \\ 
			-I_5 - I_6 \end{pmatrix} \theta\left(r-r_0\right),\label{def_v1_pp_1_split}
\end{align}
with
\begin{subequations}\label{I1-6}
\begin{align}
I_1 &=  \int ^\infty_{r_0} dr'  \Phi^{-1}_{\rm top}\left(\delta\hat A\,\hat\psi^{\rm ret }_{+}+\hat{\cal D}\hat \varphi^{\P}_{\infty}\right)\!,\\
I_2 &= \int^{r_0}_{r} dr'  \Phi^{-1}_{\rm top}\left(\delta\hat A\,\hat\psi^{\rm ret }_{-}+\hat{\cal D}\hat \varphi^{\P}_{ H}\right)\!, \\
I_3 &= \int^{r}_{ 2M} dr' \Phi^{-1}_{\rm bot} \left(\delta\hat A\,\hat\psi^{\rm ret }_{-}+\hat{\cal D}\hat \varphi^{\P}_{ H}\right)\!, \\
I_4 &= \int^\infty_{r} dr'  \Phi^{-1}_{\rm top} \left(\delta\hat A\,\hat\psi^{\rm ret }_{+}+\hat{\cal D}\hat \varphi^{\P}_{\infty}\right)\!, \\
I_5 &= \int^{r_0}_{2M} dr' \Phi^{-1}_{\rm bot} \left(\delta\hat A\,\hat\psi^{\rm ret }_{-}+\hat{\cal D}\hat \varphi^{\P}_{ H}\right)\!, \\ 
I_6 &= \int^{r}_{r_0} dr' \Phi^{-1}_{\rm bot} \left(\delta\hat A\,\hat\psi^{\rm ret }_{+}+\hat{\cal D}\hat \varphi^{\P}_{ \infty}\right)\!.
\end{align}
\end{subequations}
In $t$ slicing, $\delta\hat A$ reduces to
\beq
\delta\hat A_{[t]} = 2\omega\delta\omega\begin{pmatrix} \mathbf{0}_{d\times d} & \mathbf{0}_{d\times d}\\
														 \mathbf{1}_{d\times d} & \mathbf{0}_{d\times d}\end{pmatrix}.\label{dAt}
\eeq

Given Eq.~\eqref{J_dpsi_pp}, the contribution ${\sf v}_2$ simplifies more significantly. After some manipulations involving integration by parts, we find
\begin{equation}
{\sf v}_2={\sf v}_2^-\theta\left(r_0-r\right)+{\sf v}_2^+\theta\left(r-r_0\right)-\Phi^{-1}_0\hat J^{\rm pp}\delta(r-r_0),\label{def_v2_pp_0}
\end{equation}
with 
\begin{subequations}\label{v2s}
\begin{align}
{\sf v}_2^- &= \begin{pmatrix}-\Phi^{-1}_{0,\rm top}\left[\delta\hat J^{\rm pp} + \hat A_0\hat J^{\rm pp}\right]\\
												\mathbf 0_d\end{pmatrix},\\
{\sf v}_2^+ &= \begin{pmatrix}\mathbf 0_d ,\\
											\Phi^{-1}_{0,\rm bot}\left[\delta\hat J^{\rm pp} + \hat A_0\hat J^{\rm pp}\right]\end{pmatrix}.
\end{align}
\end{subequations}
Here the factor $\hat A_0:=\hat A(r_0)$ has appeared after applying the identity~\eqref{Phi inv ODE}.

\begin{figure*}[t]
\centering
\includegraphics[width=\textwidth]{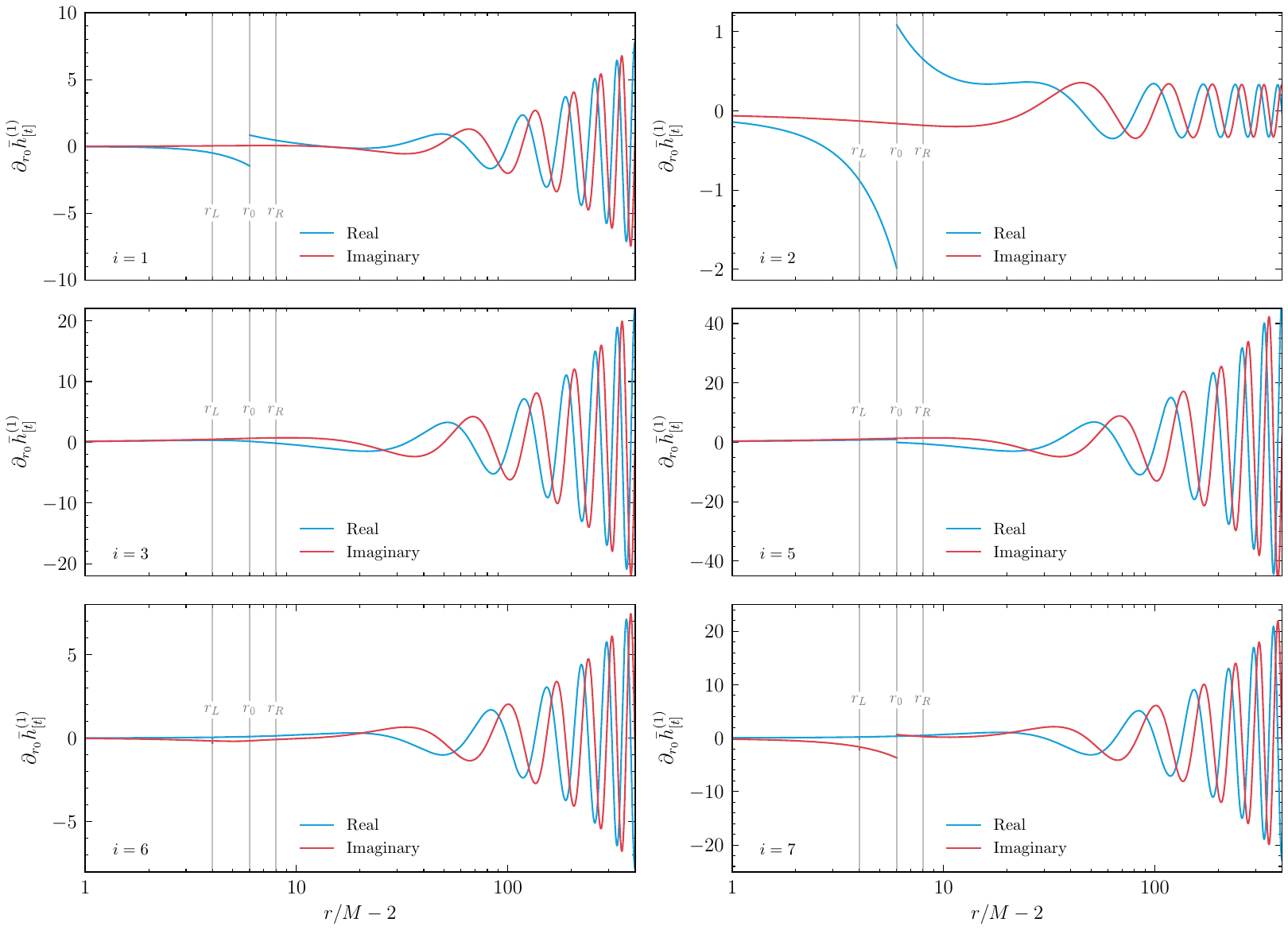}
\caption{Real and imaginary parts of $\partial_{r_0}\bar h^{(i)}_{[t]}$ for all nonvanishing $i$ modes with $\ell=2$, $m=2$, $r_0=8M$.  Note that we have not included the $i = 4$ BLS mode in this figure for brevity as this is qualitatively the same as the $i = 1$ mode.}
\label{fig_dr0_h1_t}
\end{figure*}

\begin{figure*}[t]
\centering
\includegraphics[width=\textwidth]{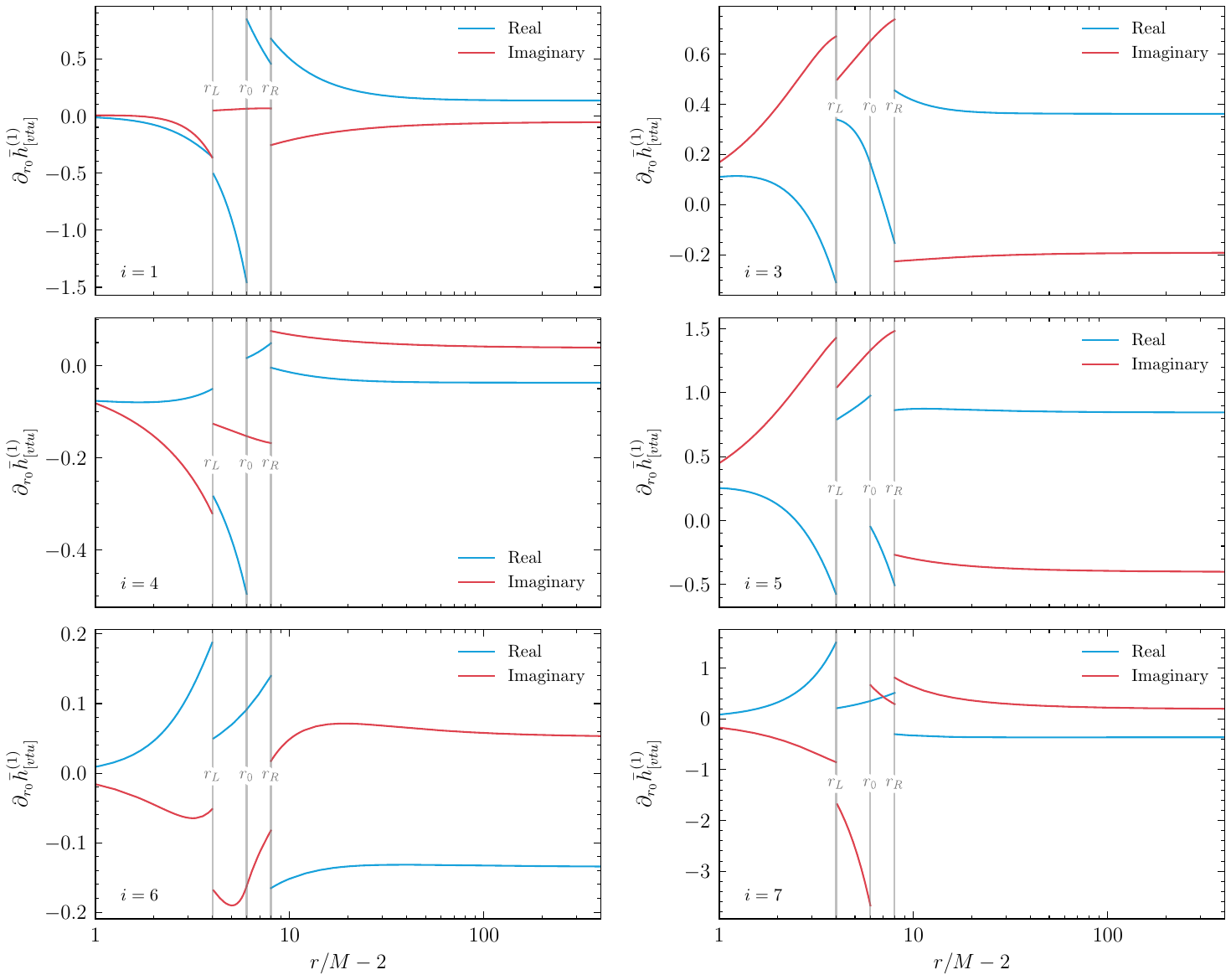}
\caption{Real and imaginary parts of $\partial_{r_0}\bar h^{(1)}_{[vtu]}$ for all nonvanishing $i$ modes with $\ell=2$, $m=2$, $r_0=8M$.  Note that we have not included the $i = 2$ BLS mode in this figure for brevity as this is qualitatively the same as the $i = 1$ mode.}
\label{fig_dr0_h1_vtu}
\end{figure*}

\begin{figure}[tb]
\centering
\includegraphics[width=0.48\textwidth]{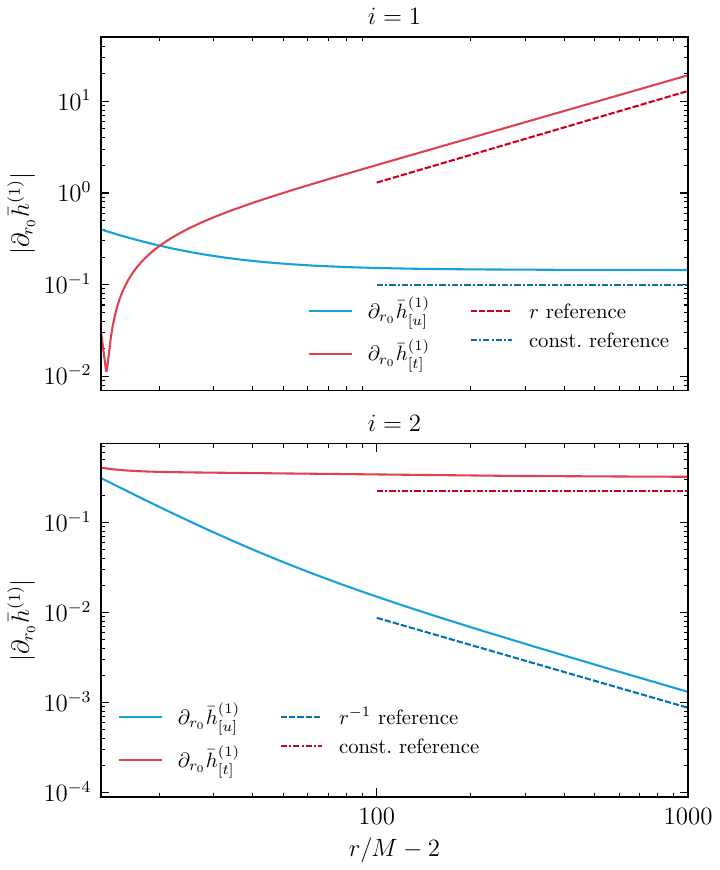}
\caption{Comparison between $|\partial_{r_0}\bar h^1_{[u]}|$ (blue line)  and $|\partial_{r_0}\bar h^1_{[t]}|$ (red line) for $i=1$ (top panel) and $i=2$ (bottom panel) with $\ell=2, m=2$, $r_0=8M$.}
\label{fig_dr0h1_t_and_u}
\end{figure}

\subsubsection{$\partial_{r_0}\bar h^{(1)}_{i\ell m}$ on $v$-$t$-$u$ slices}

We obtain $\partial _{r_0}\bar h^{(1)}_{[vtu]}$ using Eq.~(\ref{dpsi_vtu_slicing}). With this slicing, no punctures are required.

In ${\sf v}_\varphi={\sf v}_1+{\sf v}_2$, ${\sf v}_1$ is again given by Eq.~\eqref{def_v1_pp_1_split}, but now with $\hat\varphi^\P_{a}=0$.  $\delta\hat A$ is now given by
\begin{align}
\delta\hat A_{[v]} &= \begin{pmatrix}\mathbf{0}_{d\times d} & \mathbf{0}_{d\times d}\\ \delta{\cal M }_{h} & -2 i m f^{-1}\delta\Omega \mathbf{1}_{d\times d}\end{pmatrix} \text{ for } r\in\Gamma_H, \label{dAv}\\
\delta\hat A_{[u]} &= \begin{pmatrix}\mathbf{0}_{d\times d} & \mathbf{0}_{d\times d}\\ \delta{\cal M }_{h} & 2 i m f^{-1}\delta\Omega \mathbf{1}_{d\times d}\end{pmatrix} \text{ for } r\in\Gamma_\infty,\label{dAu}
\end{align}
and by Eq.~\eqref{dAt} for $r\in\Gamma_p$. $\delta{\cal M}_{h}$ is given explicitly in Eqs.~\eqref{dr0_Mh_w_even} and \eqref{dr0_Mh_w_dipole}, with $H=-1$ in $\delta\hat A_{[v]}$ and $H=+1$ in $\delta\hat A_{[u]}$.

The contribution ${\sf v}_2$ is given by Eq.~\eqref{def_v2_pp_0}, unchanged from $t$ slicing.

\subsubsection{Results and comparison between slicings}

Figures~\ref{fig_dr0_h1_t} and~\ref{fig_dr0_h1_vtu}  show our results for the $\ell=2,m=2$ mode of $\partial_{r_0} \bar h^{(1)}_{[t]}$ and $\partial_{r_0}\bar h^{(1)}_{[vtu]}$. These fields are generically discontinuous at $r=r_0$ due to the $\delta'(r-r_0)$ source. $\partial_{r_0}\bar h^{(1)}_{[vtu]}$ additionally contains discontinuities at the boundaries between slicings, as $\bar h^{(1)}_{[vtu]}$ did. At large $r$, $\partial_{r_0} \bar h^{(1)}_{[t]}$ oscillates with growing amplitude, while $\partial_{r_0} \bar h^{(1)}_{[vtu]}$ goes to a constant. This is reinforced in Fig.~\ref{fig_dr0h1_t_and_u}, which shows the absolute value $\Bigl|\partial_{r_0} \bar h^{(1)}_{[t]}\Bigr|$ growing at large $r$ while $|\partial_{r_0} \bar h^{(1)}_{[vtu]}|$ decays to a constant. Similar differences in behavior would appear near the horizon if the plots were to zoom in on that region.

To compare our results in the two slicings, we transform from $t$ to $v$-$t$-$u$ slicing using 
\begin{align}
\delta \bar h^{(1)}_{[vtu]} &= \delta\!\left(e^{-i\omega_m  k(r^*)}\bar h^{(1)}_{[t]}\right)\nonumber\\
					&= e^{-i\omega_m  k(r^*)}\!\left(\delta\bar h^{(1)}_{[t]} -i \delta\omega_m  k(r^*) \tilde h^{(1)}_{[t]}\right)\!.\label{dr0_rotation}
\end{align}
We find a relative difference $\lesssim 10^{-11}$ after performing this transformation, confirming the consistency of our results for different slicings.

%----------------------------------------------------------------------------------------------------------------------------%
\section{Demonstration 2: Teukolsky calculations for quasicircular orbits}\label{sec_demonstration_Teukolsky}
%----------------------------------------------------------------------------------------------------------------------------%
In this section we apply our scheme to the calculation of the first-order ${\sf s}=-2$ Teukolsky master function and its derivative with respect to an orbital parameter. This problem is slightly different in structure than the Lorenz-gauge problem explored in Sec.~\ref{sec_demonstration_Lorenz}, but the generic method is still applicable. The equations we solve still have the forms of Eqs.~\eqref{matrix_field_eqn} and~(\ref{dvphi Lorenz}), simply with different differential operators and source terms.

%The avid reader will have seen in Sec.~\ref{sec_worldtube_dr0_general_regular_solutn} that the practical approach to compute such quantities is to calculate them directly through deriving a field equation.
%In fact, in terms of the matrix formulation that has been shown here, the form of the equation we are solving is identical to Eq.~(\ref{dvphi Lorenz}) with a different differential operator and source term.

\subsection{Calculation of $\teukR{-2}{}{\lmw}$}

As in the Lorenz-gauge case, we first review the calculation of the first-order retarded solution. 

For a particle on a quasicircular orbit of (leading-order) radius $r_0$, the Teukolsky master function for the first-order perturbed Weyl scalar, $\psifour^{(1)}=\psifour[h^{(1)}]$, is given by
Eq.~(\ref{matrix_field_eqn}) with $\hat{\psi}^{(1)} = (\teukR{-2}{}{\lmw}, \partial_{r}(\!\teukR{-2}{}{\lmw}))^{T}$ and with $\hat{\cal D}$ now defined by the matrices $A$ and $B$ given in Eq.~(\ref{eq:ab_teukolsky}).
The point-particle source for the Teukolsky master function has further
distributional content than the Lorenz-gauge source in Eq.~(\ref{J_point_particle_1}), such that
\begin{multline}
	\hat{J}^{(1)} = \hat{J}^{(A)}_{\rm pp}(r_0) \delta (r - r_0)
	+ \hat{J}^{(B)}_{\rm pp}(r_0) \delta^{\prime} (r - r_0) \\
	+ \hat{J}^{(C)}_{\rm pp}(r_0) \delta^{\prime\prime} (r - r_0),\label{J1 Teukolsky}
\end{multline}
where $\hat{J}^{(i)}_{\rm pp} = (0, {J}^{(i)}_{\rm pp})^{T}$ with $i \in \{A, B, C\}$. ${J}^{(i)}_{\rm pp}$ are the source terms
given in Appendix \Ref{sec_Tlmw}.

The retarded point-particle solution for the Teukolsky master function is given by
Eq.~(\ref{psi_regular_with_r_0}).
But the additional distributional content in the source leads to $\sf v$ having a similar schematic form to Eq.~(\ref{def_v2_pp_0}) for $\partial_{r_0}\bar h^{(1)}_{i\ell m}$, as opposed to Eq.~\eqref{v^pp} for $\bar h^{(1)}_{i\ell m}$. We write this as ${\sf v}={\sf v}_{\theta}+{\sf v}_{\delta}$, where
\begin{align}
	{\sf v}_\theta &= {\sf v}_\theta^-\theta\left(r_0-r\right)+{\sf v}_\theta^+\theta\left(r-r_0\right),\label{def_v1_pp_0_Teuk}\\
	{\sf v}_\delta &= \Phi^{-1}_0\left[ \mathbf{1}\delta^{\prime}(r - r_0) - \mathbf{R}\,\delta(r - r_0)\right] \hat J^{(C)}_{\rm pp}.
	\label{def_v2_pp_0_Teuk}
\end{align}
Here $\mathbf{R} = \begin{pmatrix} 0 & 1 \\ 1 & 0 \end{pmatrix}$ is a reflection matrix such that $ \mathbf{R}\,\hat J^{(C)}_{\rm pp} = (J^{(C)}_{\rm pp}, 0)^{T}$.
The quantities ${\sf v}_\theta^-$ and ${\sf v}_\theta^+$ are found through integration by parts, which yields
\begin{align}
	{\sf v}_\theta^- &= \begin{pmatrix}-\Phi^{-1}_{0,\rm top} \left[\hat J^{(A)}_{\rm pp} - \hat A_{0}\hat J^{(B)}_{\rm pp} + (\hat A^{2}_{0} + \partial_{r}\hat A_{0})\hat J^{(C)}_{\rm pp}\right]\\
													0\end{pmatrix},\\
	{\sf v}_\theta^+ &= \begin{pmatrix}0\\
												\Phi^{-1}_{0,\rm bot}\left[\hat J^{(A)}_{\rm pp} - \hat A_{0}\hat J^{(B)}_{\rm pp} + (\hat A^{2}_{0} + \partial_{r}\hat A_{0})\hat J^{(C)}_{\rm pp}\right]\end{pmatrix},
\end{align}
where we have used the relation in Eq.~(\ref{Phi inv ODE}) and its derivative: $\frac{d^{2} \Phi^{-1}}{dr^{2}} = \Phi^{-1} \left( \hat A^{2} + \frac{d \hat A}{dr} \right)$.

The solution is evaluated in the following manner:
\begin{enumerate}
	\item Fix a zeroth-order orbital radius $r_{0}$.
	\item For each $\ell m$ mode, we construct a matrix of homogeneous solutions, $\Phi$, as explained in Appendix~\ref{sec_teukolsky_basis}.
	\item Calculate the retarded field for the column vector $\hat{\psi}^{(1)}$ using Eq.~\eqref{psi_regular_with_r_0} with 
	${\sf v} = {\sf v}_{\theta} + {\sf v}_{\delta}$ given by Eqs.~(\ref{def_v1_pp_0_Teuk})-(\ref{def_v2_pp_0_Teuk}).  
	The inversion of the matrix $\Phi$ for the calculation of ${\sf v}$ is facilitated by \emph{Mathematica's} \texttt{Inverse} routine.
	\item Although written in a different manner, this solution is entirely equivalent to the solutions in the 
	literature~\cite{Cutler:1993vq,Cutler:1994pb,Hughes:2001jr,Hughes:2018qxz,Hughes:2021exa} and can be readily computed using the 
	\texttt{Teukolsky} package in the Black Hole Perturbation Toolkit (BHPToolkit) \cite{BHPToolkit}.  
	This provides a robust check of our numerical results for the retarded field and we find relative differences of $\lesssim 10^{-12}$ between
	our calculation and the BHPToolkit.
\end{enumerate}

\subsection{Calculation of $\partial_{r_0}(\teukR{-2}{}{\lmw})$}

\subsubsection{Overview}

After reviewing the retarded point-particle solution, we now move on to the calculation of its derivative with respect to an orbital parameter. Like in the Lorenz-gauge case, the field equation we consider is Eq.~(\ref{dvphi Lorenz}), but now with $\varphi^{\rm ret} = \delta \hat \psi^{(1)} = (\delta \teukR{-2}{}{\lmw}, \partial_{r}(\delta\! \teukR{-2}{}{\lmw}))^{T}$.
The source term, $K^{(1)}$, has the same form as Eq.~(\ref{K Lorenz}), but the first, extended term has added distributional content,
\begin{align}\label{dA_psi1}
	\delta\hat A\,\hat\psi^{(1)} &= \delta\hat A\, \hat\psi^{(1)}_-\theta(r_0-r)											
	+ \delta \hat A\,\hat\psi^{(1)}_{+}\theta(r-r_0) \nonumber\\	
	&\quad - \delta \hat A \hat J^{(C)}_{\rm pp} \delta(r - r_0).
\end{align}
The extended support from the source term again originates from the retarded point particle (Teukolsky) solution, $\hat\psi^{(1)}_{\pm}=\Phi\,{\sf v}^\pm$.  Explicitly, $\delta A$ and $\delta B$ for the Teukolsky problem have the following form:
\begin{subequations}\label{dr0_AB_Teukolsky}
	\begin{align}
	\delta  A &= 2 f^{-2} \big[\left(1-H^2\right)\omega_m \delta \omega_m \nn \\ 
	&\qquad\qquad -2\omega_m^2 H \delta  H - \delta\!\teukpot{-2}{\lm}(r)\big],\\
	\delta  B &= 2 i f^{-1} \left( \delta\omega_m H + \omega_m \delta H\right),
	\end{align}
\end{subequations}
where
\begin{multline}
	\delta\! \teukpot{-2}{\lm}(r) = 4 i r^{-2} \big\{ \delta \omega_m [ r (1 - H)f  -  M(1 + H) ]\\
	- r f \delta H - M \delta H \big\}.
\end{multline}
The secondary source term arising from the derivative of the point-particle Teukolsky source is written concisely as
\begin{align}
	\delta \hat {J}^{(1)} &= 
	\bigg[ \delta \hat J^{(A)}_{\rm pp} \delta(r - r_0)
	+ \delta \hat J^{(B)}_{\rm pp} \delta^{\prime}(r - r_0) \nn \\
	&+ \delta \hat J^{(C)}_{\rm pp} \delta^{\prime\prime}(r - r_0)
	- \hat J^{(A)}_{\rm pp} \delta^{\prime}(r - r_0) \nn \\
	&- \hat J^{(B)}_{\rm pp} \delta^{\prime\prime}(r - r_0)
	- \hat J^{(C)}_{\rm pp} \delta^{\prime\prime\prime}(r - r_0)
	\bigg].
	\label{J_dpsi_pp_teuk}
\end{align}

The solution is given by Eq.~\eqref{dpsi_teuk_vtu_slicing}. Due to the more complicated source, the quantity ${\sf v}_{\varphi}$, defined in Eq.~\eqref{sfv_phi}, has a more complicated form than in the Lorenz gauge:
\begin{align}
	{\sf v}_{\varphi} &= {\sf v}_{3} + {\sf v}_{4} + {\sf v}_{5} + {\sf v}_{6} \nonumber\\
	&\quad + \Phi^{-1}_0\left[\mathbf{R}\, \delta(r - r_0) + \mathbf{1}\delta^{\prime}(r - r_0) \right] 
	\left( \hat J^{(B)}_{\rm pp} - \delta\hat J^{(C)}_{\rm pp} \right) \nonumber\\
	&\quad + \Phi^{-1}_0\left[\mathbf{R}\, \delta^{\prime}(r - r_0) + \mathbf{1}\delta^{\prime\prime}(r - r_0) \right] \hat J^{(C)}_{\rm pp}.
	\label{def_vphi_pp_0_teuk}
\end{align}
The terms ${\sf v}_{3}$ and ${\sf v}_{6}$ originate from the source  $\delta \hat A\hat\psi^{(1)}$ [i.e., they comprise ${\sf v}_{1}$ in Eq.~\eqref{sfv_phi}], while all other terms originate from the source $\delta \hat J^{(1)}$ [i.e., they comprise ${\sf v}_{2}$ in Eq.~\eqref{sfv_phi}]. 

${\sf v}_{3}$ specifically corresponds to the integral over the Heaviside terms in Eq.~\eqref{dA_psi1}, meaning it has the form given in Eq.~(\ref{def_v1_pp_1_split}) and accounts for the integration
over the extended piece of the source. ${\sf v}_{6}$ is then the integral over the delta term in Eq.~(\ref{dA_psi1}):
\begin{equation}
	{\sf v}_{6} = {\sf v}_{6}^-\theta\left(r_0-r\right)+{\sf v}_{6}^+\theta\left(r-r_0\right).\label{def_v6_pp_0}
\end{equation}
Here ${\sf v}_{6}^\pm$ are given concisely by
\begin{equation}\label{v6+v6-}
	{\sf v}_{6}^- = \begin{pmatrix}-\Phi^{-1}_{0,\,\rm top}\, \delta \hat A_{0} \hat J^{(C)}_{\rm pp}\\[1em] 0\end{pmatrix}\!,\quad
	{\sf v}_{6}^+ = \begin{pmatrix} 0\\[1em]\Phi^{-1}_{0,\,\rm bot}\, \delta \hat A_{0} \hat J^{(C)}_{\rm pp} \end{pmatrix}\!.
\end{equation}

The rest of the terms follow immediately from the general formula~\eqref{def_v2} for ${\sf v}_{2}$ with $\delta\hat J$ given by Eq.~(\ref{J_dpsi_pp_teuk}). ${\sf v}_{4}$ represents the integral of the first set of delta functions in Eq.~(\ref{J_dpsi_pp_teuk}), from which one finds
\begin{equation}
	{\sf v}_{4} = {\sf v}_{4}^-\theta\left(r_0-r\right)+{\sf v}_{4}^+\theta\left(r-r_0\right),\label{def_v4_pp_0}
\end{equation}
where
\begin{align}
	{\sf v}_4^- &=\! \begin{pmatrix}\!-\Phi^{-1}_{0,\rm top}\!\left[\delta \hat J^{(A)}_{\rm pp} - \hat A_{0}\delta\hat J^{(B)}_{\rm pp} + (\hat A^{2}_{0} + \partial_{r}\hat A_{0})\delta \hat J^{(C)}_{\rm pp}\right]\\
													 0\end{pmatrix}\!,\\
	{\sf v}_4^+ &=\! \begin{pmatrix} 0,\\
		\Phi^{-1}_{0,\rm bot}\!\left[\delta \hat J^{(A)}_{\rm pp} - \hat A_{0}\delta\hat J^{(B)}_{\rm pp} + (\hat A^{2}_{0} + \partial_{r}\hat A_{0})\delta \hat J^{(C)}_{\rm pp}\right]\end{pmatrix}\!.
\end{align}
The next term, ${\sf v}_{5}$, arises from the second grouping of delta functions in Eq.~(\ref{J_dpsi_pp_teuk}) and hence follows the same split
as Eq.~(\ref{def_v4_pp_0}),
\begin{equation}
	{\sf v}_{5} = {\sf v}_{5}^-\theta\left(r_0-r\right)+{\sf v}_{5}^+\theta\left(r-r_0\right).\label{def_v5_pp_0}
\end{equation}
However, the higher-order derivatives of the delta functions than seen previously requires one higher derivative of Eq.~(\ref{Phi inv ODE}), leading to
\beq
\frac{d^{3} \Phi^{-1}}{dr^{3}} = \Phi^{-1} \left( \hat A \frac{d \hat A}{d r} + \frac{d \hat A}{d r} \hat A + \hat A^{3} + \hat A \frac{d \hat A}{dr} + \frac{d^{2} \hat A}{dr^{2}} \right). 
\eeq
 Using this relation one finds
\begin{align}
	{\sf v}_5^- &=\! \begin{pmatrix}\Phi^{-1}_{0,\rm top} \Big[\hat A_{0}\hat J^{(A)}_{\rm pp} - (\hat A^{2}_{0} + \partial_{r}\hat A_{0})\hat J^{(B)}_{\rm pp} 
		+ \bigl(\hat A_{0} \partial_{r} \hat A_{0} \\ + \partial_{r} \hat A_{0}\, \hat A_{0}
	    + \hat A^{3}_{0} + \hat A_{0}\partial_{r} \hat A_{0} + \partial_{r}^{2} \hat A_{0}\bigr)\hat J^{(C)}_{\rm pp} \Big]\\
													 0\end{pmatrix}\!,\\
	{\sf v}_5^+ &=\! \begin{pmatrix} 0\\
		\!-\Phi^{-1}_{0,\rm bot}\Big[\hat A_{0}\hat J^{(A)}_{\rm pp} - (\hat A^{2}_{0} + \partial_{r}\hat A_{0})\hat J^{(B)}_{\rm pp} 
		+ \bigl(\hat A_{0} \partial_{r} \hat A_{0} \\ + \partial_{r} \hat A_{0}\, \hat A_{0}
	    + \hat A^{3}_{0} + \hat A_{0}\partial_{r} \hat A_{0} + \partial_{r}^{2} \hat A_{0}\bigr)\hat J^{(C)}_{\rm pp} \Big]\end{pmatrix}\!.
\end{align}

\subsubsection{$\partial _{r_0}R_{[vtu]}$ on $v$-$t$-$u$ slices}
Our calculation of $\partial _{r_0}R_{[vtu]}$ is done through Eq.~(\ref{dpsi_teuk_vtu_slicing}), with ${\sf v}_\varphi$ given by Eq.~(\ref{def_vphi_pp_0_teuk}), but within the integrals of the source terms 
in ${\sf v}_{3}$, we include no puncture in the regions where $r \in \Gamma_{p}$ as we are only considering the retarded solution.  Furthermore, we choose not to include a puncture toward the horizon,
where $r \in \Gamma_{H}$, as we find Eq.~\eqref{puncture conditions} is satisfied for a fiducial horizon puncture.
Hence in the integrals $I_{i}$ that enter ${\sf v}_3$, with $i \in\{1, \dots, 6\}$, we set $\hat\varphi^\P_{H} = \hat\varphi^\P_{p} = 0$.

\begin{figure}[tb]
	\centering
	\includegraphics[width=0.48\textwidth]{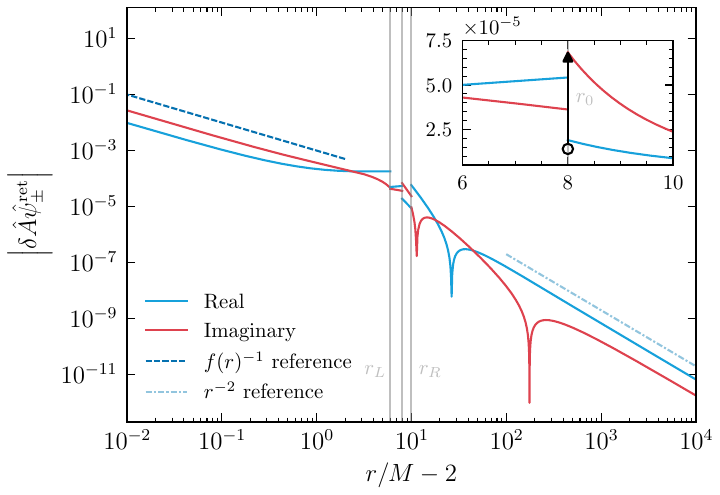}
	\caption{Real and imaginary parts of the extended source terms appearing 
	in the integrands within Eq.~(\ref{def_v1_pp_1_split}) for ${\sf v}_{3}$.
	In the asymptotic regions, the source decays
	as $\sim r^{-2}$ toward spatial infinity  but  grows as $\sim f^{-1}$
	toward the horizon.  Here $\ell = 2$, $m = 2$ and the secondary is at
	$r_0 = 10M$.   The vertical arrow in the inset of the plot signifies the
	presence of additional distributional pieces that must be taken into account
	when calculating the full solution.}
	\label{fig_dr0R1_source}
\end{figure}

In the Teukolsky framework (with $\sf s = -2$), $\delta\hat A$ is now given in the various slicings by
\begin{align}
\delta\hat A_{[v]} &= 2 m f^{-1}\delta\Omega \begin{pmatrix}0 & 0\\ 4r^{-1} & 1\end{pmatrix}, r\in\Gamma_H, \label{dAv_Teukolsky}\\
\delta\hat A_{[t]} &= 2 m f^{-2} \delta\omega_m \begin{pmatrix}0 & 0\\ 2r ^{-2}( r f - M ) + i \omega_m  & 0\end{pmatrix}, r\in\Gamma_p, \label{dAt_Teukolsky}\\
\delta\hat A_{[u]} &= -2 m f^{-1}\delta\Omega \begin{pmatrix}0 & 0\\ 4r^{-2} & 1\end{pmatrix}, r\in\Gamma_\infty.\label{dAu_Teukolsky}
\end{align}
$\delta\hat A$, shown in Fig.~\ref{fig_dr0R1_source}, forms a central component of 
the overall source term.  We see that this piece diverges as $\sim f^{-1}$ toward 
the horizon but converges toward infinity.  However, to determine whether the retarded integrals converge, we must analyse the entire integrands.

\begin{figure}[tb]
	\centering
	\includegraphics[width=0.48\textwidth]{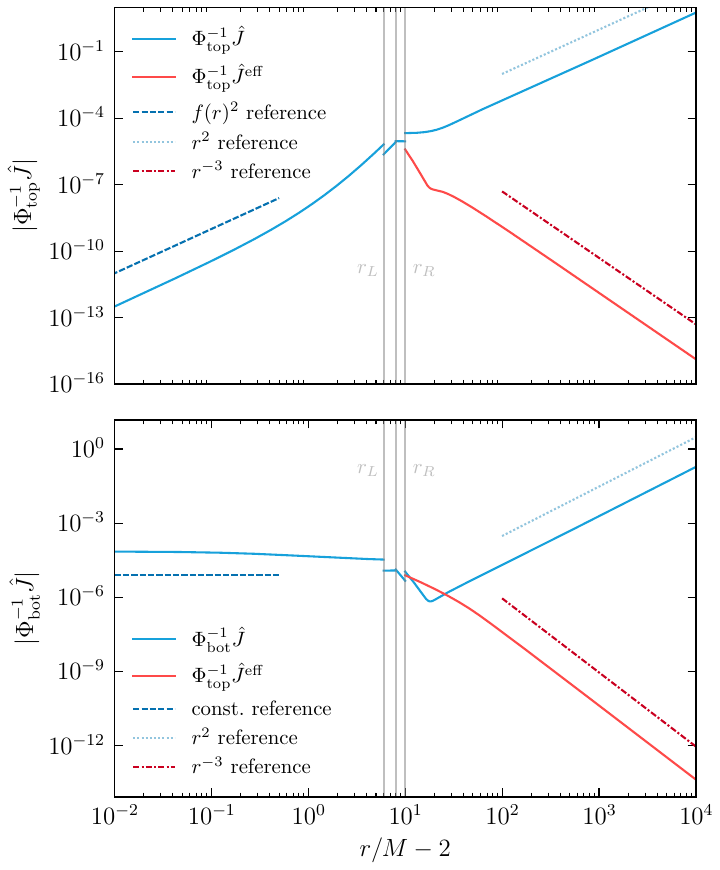}
	\caption{Comparison of the non-punctured and punctured integrands for the extended source 
	terms within the integrals appearing in Eq.~(\ref{def_v1_pp_1_split}) for ${\sf v}_{3}$. 
	\emph{Top panel:} The integrands of the weighting coefficient for
	the homogeneous solutions $\Phi_{-}$ throughout the entire
	numerical domain. In the absence of a puncture, the integrand falls off as $\sim f^{2}$
	toward the horizon but grows as $\sim r^{2}$ toward null infinity. 
	Therefore in the region $\Gamma_{\infty}$ we apply a suitable puncture to
	make the integral converge.  The puncture used in the figure ensures
	the integrand now falls off as $\sim r^{-3}$ toward null infinity.
	\emph{Bottom panel:} The integrands of the weighting coefficient for
	the homogeneous solutions $\Phi_{+}$ throughout the entire
	numerical domain.  Without a puncture (blue curve), the integrand tends to a constant
	toward the horizon but again grows as $\sim r^{2}$ toward null infinity.
	With the puncture (red curve) applied in $\Gamma_{\infty}$, the integrand again falls off as $\sim r^{-3}$.}
	\label{fig_dr0R1_integrands}
\end{figure}

Figure~\ref{fig_dr0R1_integrands} plots the integrands in Eq.~(\ref{I1-6}), both with and without punctures. As shown by the reference curves in the two
plots, the integrands $\Phi^{-1}_{\rm top}\delta\hat A \hat \psi^{\rm ret}_-$ and $\Phi^{-1}_{\rm bot}\delta\hat A \hat \psi^{\rm ret}_-$ converge toward the horizon as $\sim f^{2}$ and $\sim f^0$, respectively.  
The analogous integrands, however, diverge as
$\sim r^{2}$ toward null infinity, verifying the need for
a puncture in the region $\Gamma_{\infty}$.

Construction of appropriate punctures in this context was discussed in
Sec.~\ref{dh_vtu_teukolsky}.
We show the application of these punctures in improving the falloff of the integrands
in the region $\Gamma_{\infty}$ in Fig.~\ref{fig_dr0R1_integrands}.  Here we used the 
asymptotic expansion in Eq.~(\ref{Psi_plus_expansion}) with $j_{\rm max} = 4$.
The plot shows how the inclusion of the puncture now forces the integrands to fall
off as $\sim r^{-3}$ and therefore leave us with a finite integral.
Punctures could be constructed in a similar manner in the other asymptotic
region, $\Gamma_{H}$.  This would speed up the convergence of the integration over this region, but it is not required.

\begin{figure}[tb]
	\centering
	\includegraphics[width=0.48\textwidth]{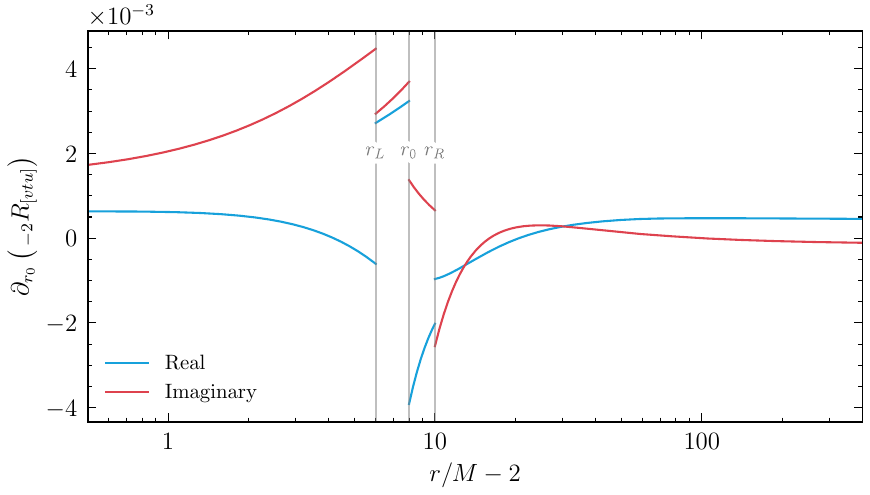}
	\caption{Real and imaginary parts of $\partial_{r_0} \left(\,_{-2}R_{[vtu]}\right)$  
	with $\ell=2$, $m=2$, $r_0=10M$.}
	\label{fig_dr0R1_vtu}
\end{figure}

In Fig.~\ref{fig_dr0R1_vtu}, we present results for the $\ell = 2$, $m = 2$ mode of 
$\partial_{r_0} \left(\,_{-2}R_{[vtu]}\right)$.  As is evident from the form of the solution in
Eq.~(\ref{def_vphi_pp_0_teuk}), the solution is discontinuous at the particle's location, owing to the
Dirac delta primes that appear in the source.  Furthermore, there are also jumps at the
boundaries of the regions $r_{L}$ and $r_R$ due to the change in slicing there.

\subsubsection{$\partial _{r_0}R_{[t]}$ on $t$ slices}
\begin{figure}[h]
	\centering
	\includegraphics[width=0.48\textwidth]{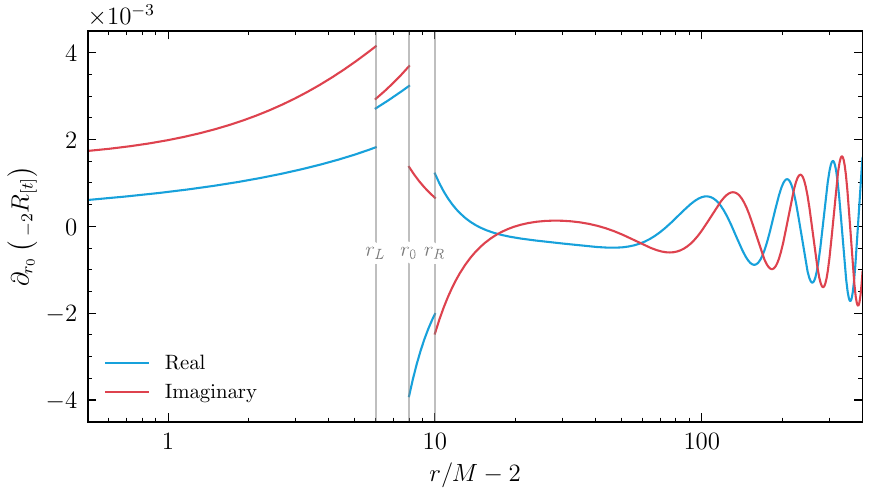}
	\caption{Real and imaginary parts of $\partial_{r_0} \left(\,_{-2}R_{[t]}\right)$  
	with $\ell=2$, $m=2$, $r_0=10M$.}
	\label{fig_dr0R1_t}
\end{figure}

For comparison purposes we also present in Fig.~\ref{fig_dr0R1_t} the same results after transforming to $t$ slicing throughout the 
numerical domain.
As we observed in Fig.~\ref{fig_h1_vtu}, we see that $\partial_{r_0} \left(\,_{-2}R_{[t]}\right)$ contains 
constant-amplitude oscillations at large $r$ that are not present in the solution for $\partial_{r_0} \left(\,_{-2}R_{[vtu]}\right)$.
Also, in $t$ slicing, there will be oscillations toward the horizon, but the aspect ratio and choice of radial coordinate preclude them from being seen in Fig.~\ref{fig_h1_vtu}.

%-------------------------------------------------------------------------------------------------------------------------------------------------------%
\section{Conclusion and outlook}\label{sec_conclusion}
%
%-------------------------------------------------------------------------------------------------------------------------------------------------------

In this paper we have formulated a worldtube puncture scheme for self-force calculations in the Fourier domain. We have specifically focused on the types of field equations that arise in a multiscale expansion of the Einstein equation, but the method applies equally well in an ordinary frequency-domain calculation. 

We have also demonstrated our scheme's utility and flexibility by numerically implementing it both for the Lorenz-gauge field equations and the Teukolsky equation. Moreover, we note that although it is described here for the first time, our method has already been successfully employed more broadly; it underlay all second-order calculations to date~\cite{Pound-etal:19,Warburton:2021kwk,Wardell:2021fyy}. 

However, in the time since our method was first formulated and implemented, at least two alternatives have been presented that offer some clear advantages~\cite{PanossoMacedo:2022fdi,Durkan:2022fvm}. It is therefore worth making a careful assessment of the relative merits of these various approaches. It is also important to point out some aspects of our method that can be usefully carried over to those other schemes (and vice versa).

We first observe that our scheme is substantially more general than earlier worldtube puncture schemes. It is intrinsically a multi-domain method, and it exploits that flexibility by (i) accommodating punctures in multiple domains and (ii) allowing different choices of time slicing (and therefore different field equations) in different domains. This type of generality could be beneficial in any approach.

In terms of practical implementation, the key difference between our scheme and the alternatives is that we apply the method of variation of parameters for sources with spatially unbounded support. This approach obtains inhomogeneous solutions by convolving homogeneous solutions against the source, a procedure with substantial drawbacks when the source has unbounded support. One drawback is that the homogeneous solutions need to be known at all radii.
This is problematic because the `up' solutions (i.e., the homogeneous solutions that are regular at $\mathscr{I}^+$) need to be calculated near the horizon, and similarly the `in' solutions (i.e., the homogeneous solutions that are regular at $\mathscr{H}^+$) need to be calculated at large radius.
It can be difficult to accurately compute these homogeneous solutions far away from where their numerical boundary conditions are specified. A second drawback is that the method sacrifices some of the advantages of hyperboloidal slicing. On these slices, the retarded inhomogeneous solution varies slowly across the domain, with no oscillations at large $r$ or near the horizon; this means, in principle, no oscillations need to be numerically resolved. The `up' solutions share this property at large $r$, and the `in' solutions share it near the horizon. However, each of these homogeneous solutions oscillates in the opposite domain: `up', near the horizon; and `in', at large $r$. This means that to evaluate the variation-of-parameters integrals, we must resolve the oscillations even though we are guaranteed that they do not appear in the ultimate, retarded solution.

These two drawbacks can be tempered by the use of higher-order punctures in the horizon and infinity regions to force the source to fall off more rapidly and thus reduce the contributions from the undesirable homogeneous solutions in each region. We do take that approach in Refs.~\cite{Pound-etal:19,Warburton:2021kwk,Wardell:2021fyy}. But the alternative approaches in Refs.~\cite{PanossoMacedo:2022fdi,Durkan:2022fvm} have more elegantly circumvented the issues that arise in variation of parameters.

The first alternative approach uses the method of partial annihilators \cite{Hopper:2012ty}.
This method can be applied when there exists an operator which when applied to the source takes it from unbounded support to pointlike (i.e., measure-zero) support.
Acting with this operator on the whole field equation results in a higher-order differential equation with a distributional source.
This new equation can then be solved with variation of parameters, and each homogeneous solution is only required in the region where it is well behaved. %for quasicircular orbits this reduces to matching conditions on the particle's worldline which involve the homogeneous solutions and their derivatives evaluated only on the worldline.
Such a partial annihilator operator exists for the calculation of parametric derivatives, and this approach has been employed for the $r_0$ derivatives of the Regge-Wheeler-Zerilli master variables and Lorenz-gauge perturbations \cite{Durkan:2022fvm}.
A limitation in this approach is that it is unlikely that a partial annihilator operator exists for the full second-order calculation.%, and so other alternative approaches are needed.

The second alternative approach leaves the source intact but directly solves the field equation in each domain using a spectral method rather than through convolution with homogeneous solutions. Like our scheme, this approach is naturally suited to multi-domain techniques. Because it does not involve a basis of homogeneous solutions, it is better adapted to hyperboloidal slicing: rather than having to resolve oscillatory homogeneous solutions near $\mathscr{H}^+$ and $\mathscr{I}^+$, one only deals with slowly varying functions, allowing one to compactify the numerical domain; and rather than having to construct high-order asymptotic expansions to impose boundary conditions on the homogeneous solutions, the field equation itself imposes boundary conditions on the retarded solution at $\mathscr{H}^+$ and $\mathscr{I}^+$. We expand on the latter point below. % to avoiding the variation of parameters method makes use of multi-domain techniques.
%This method divides the spacetime into domains and within each domain the sourced field equation is solved.
%The true global solution is obtained by enforcing conditions on the field at its derivatives at the domain boundaries (continuity, jump discontinuities, etc).
%The construction of asymptotic boundary conditions is also simplified by combining pseudo-spectral techniques with compactified hyperboloidal coordinates.
This multi-domain, spectral, compactified hyperboloidal approach was implemented for a scalar-field toy-model in Ref.~\cite{PanossoMacedo:2022fdi}.
In that work the method was shown to be very efficient for distributional sources, extended sources, and sources with unbounded support, where for the latter the calculation of a parametric $r_0$ derivative  was used as an example.

Despite variation of parameters' disadvantages, it does have some clear advantages. One benefit is that it is a straightforward way of solving problems with complicated distributional sources. For example, in Ref.~\cite{Upton:2021oxf}, the source for the second-order retarded metric perturbation was shown to have the form of a highly nontrivial limit of a sequence of distributions. Dealing with such sources is simplest if one can write the solution immediately as an integral against a Green's function, as in variation of parameters, and then manipulate the integral (e.g., through integration by parts) before any numerical evaluation. Hence, a valuable approach might be to combine methods, using variation of parameters within a domain containing the particle and alternative methods outside that domain.

We also wish to stress that most obstacles encountered in our complete second-order calculations are independent of our use of variation of parameters. As mentioned above, one advantage of compactified hyperboloidal slicing is that it can avoid the need to calculate boundary conditions. More precisely, if the numerical variable is regular at the compactified boundaries, then the field equations themselves reduce to regularity conditions at the boundaries, and there is no need to construct boundary conditions of the form described in Appendix~\ref{sec_basis}. However, this does not mean boundary conditions never need to be calculated, nor does it mean that punctures are unnecessary. At second order, we do not generically have regular fields at the boundaries: as analyzed in Ref.~\cite{Pound:2015wva} and alluded to in Sec.~\ref{sec_BCs}, the second-order solution in the multiscale expansion is irregular at the boundaries. The correct physical boundary conditions for the multiscale field equations can be derived from a post-Minkowski expansion near $\mathscr{I}^+$ and an analogous expansion near the horizon. These physical boundary conditions contain hereditary terms, integrals over the system's entire past history, which are impossible to determine from the field equations in the numerical domain, regardless of one's choice of slicing or compactification. 

The framework in this paper readily incorporates such boundary conditions into punctures at the boundaries. Our analysis in Sec.~\ref{boundary punctures} also provides a diagnostic for when a puncture is required and the conditions it must satisfy. That type of analysis should continue to serve a key purpose even when the method of variation of parameters is not used.

We also note that other aspects of our scheme are independent of the use of variation of parameters. One obvious example is the overarching multiscale method, which we have presented in a more geometrical way than in previous literature. Derivatives of the numerical fields with respect to orbital parameters are an essential ingredient in that method~\cite{Pound:2021qin} and in closely associated ones~\cite{vandeMeent-Warburton:18,Lynch:2021ogr,Lynch:2023gpu}. Our analysis has highlighted how calculations of such parametric derivatives depend crucially on the choice of slicing. If standard, constant-$t$ slicing is used, infrared divergences arise. Such divergences can be treated by introducing punctures at the horizon and infinity to enforce physical boundary conditions. However, hyperboloidal slicing entirely evades these divergences (at least for broad classes of fields).

Followup papers will explain how the second-order self-force results in Refs.~\cite{Pound-etal:19,Warburton:2021kwk,Wardell:2021fyy} were obtained by combining (i) the puncture scheme in this paper, (ii) the punctures in Ref.~\cite{Pound:2014xva}, (iii) the coupling formulas presented in Ref.~\cite{coupling-paper}, (iv) the multiscale expansion of the Lorenz-gauge field equations in Ref.~\cite{Miller:2020bft} (reviewed in this paper), and (v) the strategies developed in Refs.~\cite{Pound:2015wva,Miller:2016hjv} to overcome infrared divergences and poor convergence of mode sums. 

In the longer term, our scheme can be applied to eccentric orbits~\cite{Leather:2023dzj}. As we emphasised throughout the body of this paper, the bulk of our analysis applies equally well for eccentric as for quasicircular orbits.

%We have calculated $h^1$ and $\partial_{r_0}h^1$
%using three different time slices: ordinary Schwarzschild-time  slices, smooth hyperboloidal-time slices
%and discontinuous hyperboloidal-time slices.
%We found that using hyperboloidal-time slicing, the source term to the second-order equation behaves in a way that the %Green's-function integral
%against the source converges and we do not need punctures at the boundaries. On $\tilde t$ slices the source
%diverges at the boundaries such that the integral does not converge and the integral has to be regularised using a puncture at the horizon and infinity.
%-------------------------------------------------------------------------------------------------------------------------------------------------------%

\acknowledgments

We are grateful to Leor Barack for suggesting that we formulate a field equation for $\partial_{r_0}\bar h^{1}_{ilm}$ and to both him and Barry Wardell for countless helpful discussions. AP acknowledges support from a Royal Society University Research Fellowship and a UKRI Frontier Research Grant under the Horizon Europe Guarantee scheme [grant number EP/Y008251/1]. AP and JM's early work on this paper was supported by the European Research Council under the European Union's Seventh Framework Programme (FP7/2007-2013)/ERC Grant No.~304978.
NW acknowledges support from a Royal Society~-~Science Foundation Ireland University Research Fellowship.
This publication has emanated from research conducted with the financial support of Science Foundation Ireland under Grant numbers 16/RS-URF/3428, 17/RS-URF-RG/3490 and 22/RS-URF-R/3825.

%-------------------------------------------------------------------------------------------------------------------------------------------------------%
\appendix

\begin{widetext}
\section{Coupling matrices and operators in the Lorenz-gauge field equations}\label{sec_Mij0}

In this appendix we give explicit expressions for the quantities ${\cal M}^{(n)}_{ij}$, ${\cal M}_h$, ${\cal M}_{\partial h}$, $\delta{\cal M}_h$, and $Z^{(n)}_{kj}$ appearing in the Lorenz-gauge equations~\eqref{decomposed EFE} [via \eqref{Eijlm} and \eqref{E1ijlm}]; \eqref{fieldEquations2}  [via \eqref{AB_gravity}]; \eqref{dr0_field_equation} [via \eqref{dr0_AB_gravity}]; and the gauge conditions \eqref{eq:gauges-w0} and \eqref{eq:gauges-w1}. %, (\ref{fieldEquations3}).
For brevity, we omit $\ell m$ labels on the fields $\bar h_{i\ell m}$ and frequency $\omega_m$, and we follow Ref.~\cite{Miller:2020bft} by adopting the shorthand 
\beq\label{lambda defs}
\lambda := (\ell+ 2)(\ell -1) \quad \text{and} \quad \lambda_1:=\ell(\ell+1). 
\eeq

\subsection{Coupling matrices}

The quantities ${\cal M}^{(0)}_{ij}\bar h_j$ in Eq.~\eqref{Eijlm}  are given by
\begingroup
\allowdisplaybreaks
\begin{align}
{\cal M}^{(0)}_{1j}\bar h_j &= \frac{ f^2 f'}{2}\left( \partial_r \bar h_{3}+\frac{i\omega   H}{f}\bar h_3\right)
										+\frac{f \left(1-\frac{4M}{r}\right)}{2r^2}\left(\bar h_1-\bar h_5 -f \bar h_3 \right)
										-\frac{f^2}{2r^2}\left(1-\frac{6M}{r}\right)\bar h_6, \label{M0-1}\\
{\cal M}^{(0)}_{2j}\bar h_j &= \frac{f^2 f'}{2}\left(\partial_r\bar h_3+\frac{i\omega   H}{f}\bar h_3\right)
										+\frac{ff'}{2}\partial_r \left(  \bar h_2 - \bar h_1\right)-\frac{i\omega  }{2}\left(1-H\right)f'\left(  \bar h_2 - \bar h_1\right)\nonumber\\
									&\quad +\frac{ f^2}{2 r^2} \left(\bar h_2-\bar h_4\right) 
										-\frac{ff'}{2r}\left(\bar h_1-\bar h_5-f\bar h_3-2 f \bar h_6\right), \label{M0-2}\\
{\cal M}^{(0)}_{3j}\bar h_j &= -\frac{f}{2r^2}\left[\bar h_1-\bar h_5-\left(1-\frac{4M}{r}\right)\left(\bar h_3 + \bar h_6\right)\right],\label{M0-3}\\
{\cal M}^{(0)}_{4j}\bar h_j &= \frac{ff'}{4}\partial_r\left( \bar h_4- \bar h_5\right)-\frac{i \omega   \left(1-H\right)f'}{4}\left( \bar h_4- \bar h_5\right)
										-\frac{\lambda_1}{2}\,\frac{f}{r^2}\bar h_2
										-\frac{ ff'}{4 r}\left(3\bar h_4+2\bar h_5-\bar h_7+\lambda_1\bar h_6\right),\label{M0-4}\\
{\cal M}^{(0)}_{5j}\bar h_j &= \frac{f}{r^2}\left[\left(1-\frac{9M}{2r}\right)\bar h_5 -\frac{\lambda_1}{2}\left(\bar h_1-f\bar h_3\right)
										+\frac{1}{2}\left(1-\frac{3M}{r}\right)\left(\lambda_1\bar h_6-\bar h_7\right)\right],\label{M0-5}\\
{\cal M}^{(0)}_{6j}\bar h_j &= -\frac{f}{2r^2}\left[\bar h_1-\bar h_5-\left(1-\frac{4M}{r}\right)\left(\bar h_3+\bar h_6\right)\right],\label{M0-6}\\
{\cal M}^{(0)}_{7j}\bar h_j &= -\frac{f}{2r^2}\left(\bar h_7+\lambda\,\bar h_5\right),\label{M0-7}\\
{\cal M}^{(0)}_{8j}\bar h_j &= \frac{ff'}{4}\partial_r\left( \bar h_8- \bar h_9  \right)-\frac{i\omega  \left(1-H\right)f'}{4}\left( \bar h_8- \bar h_9  \right)
										-\frac{ff'}{4r} \left(3\bar h_8+2\bar h_9-\bar h_{10}\right),\label{M0-8}\\
{\cal M}^{(0)}_{9j}\bar h_j &=\frac{f}{r^2}\left(1-\frac{9 M}{2r}\right)\bar h_9 - \frac{f}{2r^2}\left(1-\frac{3M}{r}\right)\,\bar h_{10},\label{M0-9}\\
{\cal M}^{(0)}_{10j}\bar h_j &= -\frac{f}{2r^2}\left(\bar h_{10} +\lambda \bar h_9\right).\label{M0-10}
\end{align}
\endgroup

In the matrix representation~\eqref{fieldEquations2} of the field equations, these coupling terms appear in the form $-\frac{4}{f^2}{\cal M}^{(0)}_{ij}\bar h_j$, which we write explicitly in terms of (algebraic) matrices ${\cal M}_{h}$ and ${\cal M}_{\partial h}$ acting on the vector $\psi$ defined in Eq.~\eqref{psiColumnVector} and its radial derivative $\partial_r\psi$. The matrix ${\cal M}_{h}$ has the explicit form 
\begin{equation}
{\cal M}_{h}=\frac{2}{r^2 f}\times\begin{pmatrix}
														- 2\left(1-\frac{9M}{2r}\right) & 1-\frac{3M}{r}\\[.5em]
  														\lambda & 1
														\end{pmatrix}
\end{equation}
for $\ell>0,\,\,m>0$ and $\ell+m\,\,\text{odd}$;
\begin{equation}\label{Mh_w_even}
{\cal M}_{h}=\frac{2}{r^2 f}\times
	\begin{pmatrix}
	-\left(1-4M/r\right) &f\left(1-4M/r\right)-2 iM\omega H& \left(1-4M/r\right)& f\left(1-6M/r\right)&0\\
	1& -\left(1-4M/r\right)&-1& -\left(1-4M/r\right)&0\\
	\lambda_1 & -\lambda_1f&-2(1-9M/(2r))& -\lambda_1\left(1-3M/r\right)&\left(1-3M/r\right)\\
	 1 & -\left(1-4M/r\right)&-1& -\left(1-4M/r\right)&0\\
	0& 0&\lambda&0&1
	\end{pmatrix}
\end{equation}
\end{widetext}
for $\ell>0,\,\,m>0$ and $\ell+m$ even; and the same matrix~\eqref{Mh_w_even}
%\begin{equation}{\cal M}_{h}=\frac{2}{r^2 f}\times
%%\arraycolsep=7pt%\def\arraystretch{2.2}
%\begin{pmatrix}
%-\left(1-4M/r\right) &f\left(1-4M/r\right)-2 iM\omega H& \left(1-4M/r\right)& f\left(1-6M/r\right)\\
%1& -\left(1-4M/r\right)&-1& -\left(1-4M/r\right)\\
%\lambda_1\left(1-3M/r\right)& -\lambda_1f&-2f& -\lambda_1\left(1-3M/r\right)\\
%1 & \left(1-4M/r\right)&-1& -\left(1-4M/r\right)
%\end{pmatrix}\label{Mh_w_dipole}
%\end{equation}
for $\ell=1,\,\,m=1$ but with the bottom row and rightmost column omitted. 

The matrix ${\cal M}_{\partial h}$  in Eq.~(\ref{AB_gravity}) has the form ${\cal M}_{\partial h }=\mathbf{0}_{2\times 2}$
for $\ell>0,\,\,m>0$ and $\ell+m\,\,\text{odd}$,
\begin{equation}\arraycolsep=6pt\def\arraystretch{0.6}
{\cal M}_{\partial h }=
-\frac{4M}{r^2} \times\begin{pmatrix}
0 & 1  &0&0&0\\
0 &0  &0&0&0\\
 0 &0  &0&0&0\\
0 &0  &0&0&0\\
0 &0  &0&0&0
\end{pmatrix}\label{Mdh_w_even}
\end{equation} 
for $\ell>0,\,\,m>0$ and $\ell+m$ even; and the same matrix~\eqref{Mdh_w_even}
%\begin{equation}
%%\arraycolsep=4pt\def\arraystretch{0.7}
%{\cal M}_{\partial h }=
%-\frac{4M}{r^2}\times
%\begin{pmatrix}
%0 & 1 &0&0\\
%0 &0  &0&0\\
% 0 &0  &0&0\\
%0 &0  &0&0
%\end{pmatrix}\label{Mdh_w_dipole}
%\end{equation} 
for $\ell=1,\,\,m=1$ but with the bottom row and rightmost column omitted.

Here we have only provided the explicit matrices for $\omega_m\neq 0$ cases. For $\omega_m=0$, the gauge conditions~\eqref{eq:gauges-w0} and \eqref{eq:gauges-w1} are used to eliminate $\bar h_{6}$ and $\bar h_{7}$, reducing the dimensions of the matrices.

The matrix $\delta {\cal M}_{h}$   in Eqs.~(\ref{dr0_AB_gravity})
 is given explicitly as $\delta {\cal M}_{h}=\mathbf{0}_{2\times 2}$
for $\ell>0,\,\,m>0$ and $\ell+m\,\,\text{odd}$;
\begin{equation}\label{dr0_Mh_w_even}
\delta {\cal M}_{h}=
-\frac{4M}{r^2 f}i\left(\omega   \delta  H+\delta \omega   H\right)\times
\begin{pmatrix}
0 & 1   &0&0&0\\
0 &0  &0&0&0\\
 0 &0  &0&0&0\\
0 &0  &0&0&0\\
0 &0  &0&0&0
\end{pmatrix} 
\end{equation} 
for $\ell>0,\,\,m>0$ and $\ell+m$ even; and
\begin{equation}\label{dr0_Mh_w_dipole}
\delta {\cal M}_{h}=
-\frac{4M}{r^2 f}i\left(\omega  \delta  H+\delta \omega   H\right)\times
\begin{pmatrix}
0 & 1   &0&0\\
0 &0  &0&0\\
 0 &0  &0&0\\
0 &0  &0&0
\end{pmatrix} 
\end{equation} 
for $\ell=1,\,\,m=1$.

Finally, the quantities $\mathcal{M}^{(1)}_{ij}$ appearing in the second-order source via Eq.~\eqref{E1ijlm} are given by
\begin{subequations}\label{Mij1}
\begin{align}
\mathcal{M}^{(1)}_{1j}\bar h_j &= -\frac{1}{2}ff'H\vec\partial_{\cal V}\bar h_3, \label{M1-1}\\
\mathcal{M}^{(1)}_{2j}\bar h_j &= -\frac{f'}{2}\left[ f H \vec\partial_{\cal V}\bar h_3 - \left(1-H\right)\vec\partial_{\cal V}\left(\bar h_2-\bar h_1\right) \right],\label{M1-2}\\
\mathcal{M}^{(1)}_{4j}\bar h_j &= \frac{f'}{4}\left(1-H\right)\vec\partial_{\cal V}\left(\bar h_4-\bar h_5\right),\label{M1-4}\\
\mathcal{M}^{(1)}_{8j}\bar h_j &= \frac{f'}{4}\left(1-H\right)\vec\partial_{\cal V}\left(\bar h_8-\bar h_9\right),\label{M1-8}\\
\mathcal{M}^{(1)}_{ij}\bar h_j &= 0\,\,\,\,\text{for}\,\,\,\,\,i=3,5,6,9,10.
\end{align}
\end{subequations}

\subsection{Gauge conditions}

The operators in the gauge conditions~\eqref{eq:gauges-w0} and \eqref{eq:gauges-w1} are given by
\begin{subequations}\label{eq:gauge1-w0}
\allowdisplaybreaks
\begin{align}
Z^{(0)}_{1j}\bar h_{j} &=i \omega_m\left(\bar h_{1} + f \bar h_{3}+H \bar h_{2}\right)\nonumber\\*
  							&\quad +\frac{f}{r} \left( r \partial_r\bar h_{2} + \bar h_2 - \bar h_4 \right),\label{Z01}\\%[1em]
Z^{(0)}_{2j}\bar h_j &= i \omega_m\left(\bar h_{2}+H \bar h_{1}-H f \bar h_{3}\right) + f\Big(\partial_r\bar h_{1}- f \partial_r\bar h_{3}\nonumber\\*
							&\quad  +\frac{1}{r} \big[\bar h_{1} -\bar h_{5} - f\bar h_{3} - 2f\bar h_{6} \big]\Big),\label{Z02}\\%[1em]
Z^{(0)}_{3j}\bar h_j  &= i \omega_m\left(\bar h_{4}+H \bar h_{5}\right)\nonumber\\*
							&\quad + \frac{f}{r} \left[ r\partial_r\bar h_{5} + 2\bar h_{5} +\ell(\ell+1)\bar h_{6} -\bar h_{7} \right],\label{Z03}\\%[1em]
Z^{(0)}_{4j}\bar h_j &= i \omega_m\left(\bar h_{8}+H \bar h_{9} \right)\nonumber\\*
							&\quad +\frac{f}{r} \left(r\partial_r\bar h_{9} + 2\bar h_{9} -\bar h_{10} \right),\label{Z04}
\end{align}
\end{subequations}
and
\begin{subequations} \label{eq:gauge1-w1}
\begin{align}
Z^{(1)}_{1j}\bar h_j &= -\vec\partial_{\cal V}(\bar h_{1}+f\bar h_{3} + H\bar h_{2}),  \\
Z^{(1)}_{2j}\bar h_j &= -\vec\partial_{\cal V}(\bar h_{2}+H\bar h_{1} - f H\bar h_{3}),  \\
Z^{(1)}_{3j}\bar h_j &= -\vec\partial_{\cal V}(\bar h_{4}+H\bar h_{5}), \\
Z^{(1)}_{4j}\bar h_j &= -\vec\partial_{\cal V}(\bar h_{8}+H\bar h_{9}).
\end{align}
\end{subequations}

%-------------------------------------------------------------------------------------------------------------------------------------------------------% 
\section{Basis of homogeneous solutions}\label{sec_basis}
%
%-------------------------------------------------------------------------------------------------------------------------------------------------------
\subsection{Lorenz Gauge}
\label{sec_lorenz_gauge_basis}
Our method of variation of parameters requires the construction of a basis of homogeneous solutions, as described around Eq.~\eqref{phi2}. We obtain these basis solutions following Ref.~\cite{Akcay:2010dx}, for example. Half the members of the basis are regular at $\mathscr{I}^+$, and half are regular at $\mathscr{H}^+$. We denote the former as $\psi^{k+}_{\ell m}$ and the latter as $\psi^{k-}_{\ell m}$ ($k=1,\dots, d$). For $\omega_m\neq0$ modes, each $\psi^{k+}_{\ell m}$ represents a purely outgoing wave behaving like $\sim e^{-i\omega u }$ for $r\to\infty$, and each $\psi^{k-}_{\ell m}$ represents a purely ingoing wave behaving like $\sim e^{-i\omega v }$ at $r=2M$. There are a total of $2d$ basis solutions, where $d$ is the dimension of the system, equal to the number of elements in the vector $\psi_{\ell m}$; see, e.g., the vectors in Eq.~\eqref{psiColumnVector}. 

We construct this basis by first choosing some inner and outer boundaries $r_{\rm in}$ and $r_{\rm out}$, setting them as close to $r=2M$ and $r=\infty$ as is practicable. Concretely, $r_{\rm in}$ and $r_{\rm out}$ are chosen such that any change making $r_{\rm out}$ larger, or bringing $r_{\rm in}$ closer to $2M$, does not affect the first 16 significant digits of the numerical solution. For each $\ell m$ mode, at each boundary, we construct a set of $d$ boundary conditions, $\psi^{k-}_{\ell m}(r_{\rm in})$ or $\psi^{k+}_{\ell m}(r_{\rm out})$. For the nonstationary modes ($\omega_m\neq0$), we use the expansions
\begin{subequations}\label{BCs_radiative}
\begin{align}
\psi^{k+}_{\ell m}(r_{\rm out})&=e^{i\omega_m [r^\ast_{\rm out}-k(r^*)]}\sum\limits^{n^+_{\rm max}}_{n=0} a_{k,n}r^{-n}_{\rm out}, \label{eq:BCout}\\
\psi^{k-}_{\ell m}(r_{\rm in})&=e^{-i\omega_m [r^\ast_{\rm in}-k(r^*)]}\sum\limits^{n^-_{\rm max}}_{n=0} b_{k,n}\left( r_{\rm in}-2M\right)^n.\label{eq:BCin}\!\!
\end{align}\end{subequations}
For the stationary modes ($\omega_m=0$), we use 
\begin{subequations}\label{BCs_stationary}
\begin{align}
\psi^{k+}_{\ell0}(r_{\rm out})&= \sum\limits^{n^+_{\rm max}}_{n=\ell}  \left(a_{k,n}+\bar a_{k,n}\log r_{\rm out}\right)r^{-n}_{\rm out}, \label{eq:BCoutStatic}\\
\psi^{k-}_{\ell0}(r_{\rm in})&= \sum\limits^{n^-_{\rm max}}_{n=0} b_{k,n}\left( r_{\rm in}-2M\right)^n.\label{eq:BCinStatic}
\end{align}
\end{subequations}
Both of these apply for a generic time function $s=t-k(r^*)$.
%
%\begin{subequations}
%\begin{align}
%A^{\ell m}_{vv} &= A^{\ell m}_{tt} = (A_{1\ell m} + A_{3\ell m})Y_{\ell m},\\
%A^{\ell m}_{vr_{EF}} &= A^{\ell m}_{tt} = (A_{1\ell m} + A_{3\ell m})Y_{\ell m},\\
%A^{\ell m}_{r_{EF}r_{EF}} &= A^{\ell m}_{tt} = (A_{1\ell m} + A_{3\ell m})Y_{\ell m},\\
%A^{\ell m}_{vA} &= A^{\ell m}_{tt} = (A_{1\ell m} + A_{3\ell m})Y_{\ell m},\\
%A^{\ell m}_{rA} &= A^{\ell m}_{tt} = (A_{1\ell m} + A_{3\ell m})Y_{\ell m},\\
%A^{\ell m}_{AB} &= (A_{1\ell m} + A_{3\ell m})Y_{\ell m},\\
%\end{align}
%\end{subequations}

%Here by regularity at the horizon we mean regularity of each component in ingoing Eddington-Finkelstein coordinates $(v,r,\theta^A)$. By expressing the Eddington-Finkelstein components of a generic tensor $A_{\mu\nu}$ in terms of $A_{i\ell m}$,  we find the following: $A_{\mu\nu}$ is smooth at the future horizon if and only if 
%\begin{subequations}\label{horizon regularity}
%\begin{align}
%\text{All }A_{i\ell m}& \text{ are smooth at } r=2M,\\
%A_{2\ell m} &= A_{1\ell m}+\O(f^2),\label{i=2 horizon regularity}\\
%A_{i\ell m} &= A_{i+1,\ell m}+\O(f) \text{ for } i=4,8. \label{i=4,8 horizon regularity}
%\end{align}
%\end{subequations}

The  coefficients here are $d$-dimensional column vectors. They are different for each $\ell m$ and
are determined from recurrence relations derived by substituting the  ansatzes~\eqref{BCs_radiative} and~\eqref{BCs_stationary} 
into the field equations.
Recurrence relations for the Lorenz-gauge boundary conditions can be found in 
Appendix A of \cite{Akcay:2010dx}.
 $n^\pm_{\text{max}} $ is fixed
by an accuracy requirement.% the next term in the sum to have a relative magnitude less than $10^{-14} $,
%compared to the partial sum.

Once the boundary conditions~$\psi^{k+}_{\ell m}(r_{\rm out})$ and $\psi^{k-}_{\ell m}(r_{\rm in})$ are determined, we find the basis solutions $\psi^{k\pm}_{\ell m}$ everywhere in the spacetime by integrating the homogeneous field equations inward from $r_{\rm out}$ or outward from $r_{\rm in}$, as appropriate. We note that we need the inner and outer homogeneous solutions over the entire domain, not just at the particle, for the retarded integrals in the calculation of the $r_0$ derivative. 

For our Lorenz-gauge calculations, we integrated the homogeneous equations using an $8$th-order Runge-Kutta Prince-Dormand (RKPD) routine from the GNU Scientific Library (GSL) repositories~\cite{GSL}. This is an adaptive routine. In it we set the absolute accuracy goal ($\epsilon_{\rm abs}$) to $10^{-16}$ and the relative accuracy goal ($\epsilon_{\rm rel}$) to $10^{-14}$. $\epsilon_{\rm abs}$ and $\epsilon_{\rm rel}$ were determined such that reducing them made no difference to our numerical results  up to the $16$th significant figure. We set the outer boundary  to be $r_{\rm out}=10^4 M$, taking into account that moving the boundary further out did not change our results for the homogeneous solutions up to the $16$th significant figure. From similar considerations  the inner boundary needs to be $r_{\rm in}=(2+10^{-8})M$ or closer to the horizon. The GSL routine cannot take us closer than $r_{\rm in}=(2+10^{-5})M$ without severe computational burdens setting in, due to factors of $1/f$ in the differential equation. To obtain accurate data closer to the horizon we used a greater-than-machine-precision (GMP) routine for solving coupled differential equations, based on the C++ library of GMP variables and functions~\cite{GMP}.

\subsection{Teukolsky}
\label{sec_teukolsky_basis}
To construct an appropriate basis of homogeneous solutions for the Teukolsky equation, one can follow the same procedure as 
Appendix~\ref{sec_lorenz_gauge_basis} by pescribing boundary conditions at some finite radii $r_{\rm in}$ and $r_{\rm out}$
for the radiative modes.  The boundary conditions can take the form of Eq.~(\Ref{BCs_radiative}) as an asymptotic series solution:
\begin{subequations}\label{BCs_teukolsky}
\allowdisplaybreaks
	\begin{align}
	\psi^{+}_{\ell m}(r_{\rm out})&= e^{i\omega_m [r^\ast_{\rm out}-k(r^*)]}\sum\limits^{n^+_{\rm max}}_{n=0} a_{n}\frac{f(r_{\rm out})^{\sf s}}{(\omega r_{\rm out})^{n}}, \label{eq:BCoutTeukolsky}\\
	\psi^{-}_{\ell m}(r_{\rm in})&= e^{-i\omega_m [r^\ast_{\rm in}-k(r^*)]}\sum\limits^{n^-_{\rm max}}_{n=0} b_{n}\, r_{\rm in} \left( r_{\rm in}-2M\right)^n.\label{eq:BCinTeukolsky}
\end{align}
\end{subequations}
Substituting the ansatzes in Eqs.~(\ref{eq:BCoutTeukolsky}) and (\ref{eq:BCinTeukolsky}) into the field equation yields the following recursion relations for the coefficients $a_{n}$ and $b_{n}$:
\begin{subequations}\label{recurrenceRelations_teukolsky}
	\begin{align}
	a_{n}&= \frac{i}{2 (k-2 {\sf s}-1)} \big[ (\ell+n-{\sf s}-1) (\ell-n+{\sf s}+2)a_{n-1} \nonumber\\
	&+2 M (n-2) \omega_{m} (n-{\sf s}-2)a_{n-2} \big], \label{eq:aoutTeukolsky}\\
	b_{n}&= \frac{1}{2 M n (n - {\sf s} - 4 i M \omega_{m})} \big[ \big( \ell (\ell+1) - {\sf s}({\sf s}+1)\nonumber\\
	&+ 4 i M \omega_{m}  (2 n-2 {\sf s}-1)+2 n {\sf s} - n(n-1) \big) b_{n-1} \nonumber\\
	&+2 i \omega_{m} (n-2 {\sf s}-1) b_{n-2} \big],\label{eq:ainTeukolsky}
\end{align}
\end{subequations}
where $a_{2{\sf s} + 1} = b_{0} = 0$, and all of the remaining terms in the series expansion are determined by imposing
$a_{n < 2{\sf s} + 1} = 0$ and $b_{n < 0} = 0$ respectively.  Other similar asymptotic expansions for the hyperboloidal
Teukolsky equation we have presented here have been derived in \cite{Piovano:2021iwv, Nasipak:2021qfu}.

We have validated these boundary conditions by comparing our solutions to the homogeneous solutions
produced by the \texttt{Teukolsky} package of the BHPToolkit.  Furthermore, we have compared numerical values of the expansions 
with boundary conditions used within the \texttt{Numerical Integration} module, which utilises the Mano, Suzuki, and Tagkasugi 
 method \cite{Mano:1996vt, Mano:1996mf} of solving the Teukolsky equation. % This is done by evaluating a semi-analytic 
%series of hypergeometric functions evaluated with high precision at $r_{\rm in}$ and $r_{\rm out}$.

The boundary conditions for the homogeneous solutions $\psi^{\pm}_{\ell m}$ are then used to construct homogeneous solutions over the entire domain
from $r_{\rm in}$ to $r_{\rm out}$.  Similar to the Lorenz gauge homogeneous solutions, $\psi^{+}_{\ell m}$ is obtained by integrating inwards 
from $r_{\rm out}$ with the boundary condition Eq.~\eqref{eq:BCoutTeukolsky} while $\psi^{-}_{\ell m}$ is found by integrating outwards 
from $r_{\rm in}$ using Eq.~\eqref{eq:BCinTeukolsky}.  In contrast to the Lorenz gauge homogeneous equations, we integrate the homogeneous 
Teukolsky equation with \emph{Mathematica's} \texttt{NDSolve} routine.  The use of \emph{Mathematica} allows us to solve the homogeneous equation to beyond-machine precision, when such accuracy is necessary.  We find setting $r_{\rm out} = 10^{4}M$ and $r_{\rm in} = (2 + 10^{-5})M$ to be sufficient boundaries
to obtain a similar absolute accuracy goal as the Lorenz gauge case we discussed previously.
\section{Teukolsky source term}\label{sec_Tlmw}

In this appendix we explicitly give expressions for the Teukolsky source utilised in our calculations in Sec.~\ref{sec_demonstration_Teukolsky}
as well as a brief summary of their derivation.

\subsection{Kinnersley tetrad}

Our calculations in the Teukolsky formalism use the Kinnersley tetrad~\cite{Kinnersley:1969zza}.  
In Schwarzschild coordinates, the components of the Kinnersley tetrad are given by 
\begin{align}
	l^{\alpha} &= \frac{1}{f} \bigg[ 1, f, 0, 0 \bigg], \\
	n^{\alpha} &= \frac{1}{2} \bigg[ 1, -f, 0, 0 \bigg], \\
	m^{\alpha} &= \frac{1}{\sqrt{2} r} \bigg[0, 0, 1, \frac{i}{\sin\theta} \bigg], \\
	\bar{m}^{\alpha} &= \frac{1}{\sqrt{2} r} \bigg[0,0, 1, -\frac{i}{\sin\theta} \bigg].
\end{align}
The null vectors are constrained by the normalisation
\begin{equation}
	l^{\alpha}n_{\alpha} = - m^{\alpha}\bar{m}_{\alpha} = -1,
	\label{eq:null_vector_normalization}
\end{equation}
with all other contractions between tetrad legs vanishing.

\subsection{GHP operators}
\label{sec_ghp}
Our presentation uses the GHP formalism~\cite{Geroch:1973am}, which we briefly summarize here; we refer the reader to Sec.~$4.1$ of Ref.~\cite{Pound:2021qin} for a detailed review.
The central idea of the GHP formalism, in its modification of the Newman-Penrose formalism, is the
introduction of the concepts of spin and boost weights.  Under
spin and boost transformations, the null vectors transform as
\begin{alignat}{3}
	l^{\alpha} &\longrightarrow \zeta \bar{\zeta} l^{\alpha}, &\qquad
	n^{\alpha} &\longrightarrow \zeta^{-1} \bar{\zeta}^{-1} n^{\alpha},\nn \\
	m^{\alpha} &\longrightarrow \zeta \bar{\zeta}^{-1} m^{\alpha}, &\qquad
	\bar{m}^{\alpha} &\longrightarrow \zeta^{-1}\bar{\zeta} \bar{m}^{\alpha},
	\label{eq:tetrad_transformations}
\end{alignat}
where $\zeta$ is an arbitrary complex number.  A GHP quantity, $\chi$, is then labelled
as type $\{p,q\}$ if under the transformation~(\ref{eq:tetrad_transformations}), the quantity transforms as
$\chi \longrightarrow \zeta^{p}\bar{\zeta}^{q} \chi$. The GHP weight of a quantity is denoted $\chi \circeq \{p,q\}$.  
One can relate the GHP weights $p$ and $q$ to the spin-weight ${\sf s}$ and 
boost-weight ${\sf b}$ via ${\sf s} = (p-q)/2$ and ${\sf b} = (p+q)/2$.
For the tetrad legs, one can read off the following GHP weights:
\begin{align}
	l^{\alpha} \circeq \{1,1\}, &\qquad
	n^{\alpha} \circeq \{-1,-1\}, \nn \\
	m^{\alpha} \circeq \{1, -1\}, &\qquad
	\bar{m}^{\alpha} \circeq \{-1,1\}.
	\label{eq:tetrad_ghp_weights}
\end{align}
From the definition~\eqref{psi4 def}, one can read off $\psifour \circeq \{-4,0\}$.

The GHP derivative operators $\th$, $\th^{\prime}$, $\edth$, and $\edth^\prime$ that appear in Eqs.~\eqref{O_operator} and \eqref{eq:S4_operator} act on spin- and boost-weighted objects. In the Kinnersley tetrad, they are given by
\begin{align}
    \th &= \frac{1}{f} \left(\frac{\partial}{\partial t} 
    + f\frac{\partial}{\partial r}\right),\\
    \th^{\prime} &= \frac{1}{2} \left(\frac{\partial}{\partial t} 
	- f\frac{\partial}{\partial r} - 
    \frac{2 {\sf b} M}{r^2}\right),
\end{align}
and
\begin{align}
    \edth &= \frac{1}{\sqrt{2}r} \left(\frac{\partial}{\partial \theta} 
    + i\csc\theta \frac{\partial}{\partial \phi} - {\sf s}\cot\theta \right),\\
    \edth^{\prime} &= \frac{1}{\sqrt{2}r} \left(\frac{\partial}{\partial \theta} 
    - i\csc\theta \frac{\partial}{\partial \phi} + {\sf s}\cot\theta \right).
\end{align}
When acting on a generic object of spin weight ${\sf s}$ and boost weight ${\sf b}$, $\th$ raises ${\sf b}$ by 1, and $\th'$ lowers it by 1; $\edth$ raises ${\sf s}$ by 1, and $\th'$ lowers it by 1. When acting on spin-weighted spherical harmonics in particular, $\edth$ and $\edth^\prime$ act as spin-raising 
and lowering operators such that
\begin{align}
	\sqrt{2}r\, \edth ( _{\sf s}Y_{\ell m}) &=  
	- [ \ell(\ell + 1) - {\sf s}({\sf s} + 1) ]^{1/2}\,
	_{{\sf s} + 1}Y_{\ell m}, 	\label{eq:spin_raising_spin_raising}\\
	\sqrt{2}r\, \edth^{\prime}( _{\sf s}Y_{\ell m}) &=  
	[ \ell(\ell + 1) - {\sf s}({\sf s} - 1) ]^{1/2}\,
	_{{\sf s} - 1}Y_{\ell m}.
	\label{eq:spin_raising_spin_lowering}
\end{align}

\subsection{Point particle source}

At leading order in our multiscale expansion, the stress-energy tensor~\eqref{Detweiler T} for a particle on a quasiciruclar orbit reduces to
\begin{equation}
	\e T^{\mu\nu}_{(1,0)} = \frac{\mu}{r^2} \frac{u^{\mu}_{(0)} u^{\nu}_{(0)}}{u^{t}_{(0)}}
	\delta(r - r_0) \delta(\theta - \pi/2) \delta(\phi - \phi_{p}).\label{T1munu}
\end{equation} 
%We shall now specialise to a point particle in a circular orbit since this is the main
%focus of this work.  
We focus on the quasicircular case, but the computation of the Teukolsky source proceeds in a similar manner for more generic orbital configurations. We write the four-velocity of the particle in terms
of the leading-order orbital energy and angular momentum, ${\cal E}_{0}$ and ${\cal L}_{0}$, such that 
$u_{\mu} = (-{\cal E}_{0}, 0, 0, {\cal L}_{0})$, with
\begin{equation}
	{\cal E}_{0} = \frac{f_{0}}{\sqrt{1 - 3M / r_0}}, \quad
	{\cal L}_{0} = \frac{r_0\sqrt{M}}{\sqrt{r_0 - 3M}},
\end{equation}
where $f_{0} := 1-2M/r_0$.

To construct the source for the Teukolsky equation given in Eq.~(\ref{eq:S4_mode_source}), we act on the
stress-energy tensor with the operator in Eq.~(\ref{eq:S4_operator}) before decomposing the resulting expression
into the basis of spin-weighted spherical harmonic and Fourier modes such that
\begin{equation}
	_{\sf{s}}S_{\ell m} = -\frac{1}{2\pi}\int_0^{2\pi}\!\!d\phi_p\; e^{im\phi_p} \int\! d\Omega
	\;_{\sf s}\bar{Y}_{\ell m} \,_{\sf s}S,
	\label{eq:stress_energy_decomposition}
\end{equation}
where $d\Omega=\sin\theta\, d\theta\, d\phi$.
Here we have given the expression for generic spin weight~${\sf s}$.  We see from the expression in 
Eq.~(\ref{eq:S4_operator}) that, for ${\sf s=-2}$, angular derivatives appear in the form of $\edth^{\prime}$ derivatives. These can be moved onto the spin-weighted harmonic in Eq.~\eqref{eq:stress_energy_decomposition} using $\int d\Omega\,{}_{\sf s}\bar Y_{\ell m} \edth' {}_{\sf s+1}\Psi = - \int d\Omega\,\edth'{}_{\sf s}\bar Y_{\ell m}\,{}_{\sf s+1}\Psi$ for any ${\sf s}$ and any spin-weighted object ${}_{\sf s+1}\Psi$. We can then exploit the spin-raising and lowering properties of $\edth$ using $\edth'{}_{\sf s}\bar Y_{\ell m}=\overline{\edth{}_{\sf s} Y_{\ell m}}$ followed by Eq.~(\ref{eq:spin_raising_spin_raising}), reducing the angular integral to an integral against ${}_{\sf s+1}\bar Y_{\ell m}$.

In Eq.~(\ref{eq:S4_operator}), the point particle stress-energy tensor enters through its tetrad components. In the Kinnersley tetrad, the relevant projections of Eq.~\eqref{T1munu} are
\begingroup
\allowdisplaybreaks
\begin{align}
	T^{(1,0)}_{nn} &= \frac{Mf^{2}_{0}\, \delta(r - r_0)}{4 r^{2}_{0}\sqrt{1 - 3M / r_0}}\delta(\theta - \pi/2) \delta(\phi - \phi_{p}),\label{T1nn}\\
	T^{(1,0)}_{n\bar{m}} &= \frac{i M^{3/2} f_{0} \,\delta(r - r_0)}{2\sqrt{2}\, r^{2}_{0}\sqrt{r_0-3M}}\delta(\theta - \pi/2) \delta(\phi - \phi_{p}),\\
	T^{(1,0)}_{\bar{m}\bar{m}} &= -\frac{M^2 \,\delta(r - r_0)}{2 r^{3}_{0}\sqrt{1-3M/r_0}}\delta(\theta - \pi/2) \delta(\phi - \phi_{p}).\label{T1mbmb}
\end{align}
\endgroup
Here we have used the distributional identity
\begin{equation}
	X(r)\delta(r - r_0) = X(r_0)\delta(r - r_0)
	\label{eq:first_dirac_delta_identity}
\end{equation}
for smooth $X(r)$. %The angular delta functions can be expanded using the completeness relation
%\beq
%\delta(\theta - \pi/2) \delta(\phi - \phi_{p}) = \sum_{\ell=|\sf s|}^\infty\sum_{m=-\ell}^\ell {}_{\sf s}\bar Y_{\ell m}(\pi/2,\phi_p){}_{\sf s} Y_{\ell m}(\theta,\phi),
%\eeq
%with the appropriate spin weight chosen for each tetrad component (i.e., ${\sf s}=0$ for $T^{(1)}_{nn}$, ${\sf s}=-1$ for $T^{(1)}_{n\bar{m}}$, and ${\sf s}=-2$ for $T^{(1)}_{\bar m\bar{m}}$).  
%It is important to note the GHP weights of the different tetrad components,
%\begin{align}
%	T_{nn} \circeq \{0, -2\},\quad
%	T_{n\bar{m}} \circeq \{-1, -1\},\quad
%	T_{\bar{m}\bar{m}} \circeq \{-2, 0\},
%\end{align}
%as this will determine the spin weights of the spherical harmonics onto which each component is decomposed.  
The full first-order Teukolsky source term then has the form
\begin{equation}
	\,_{\sf s}S^{(1,0)}_{\lmw} =  \,_{\sf s}S^{(A)}_{\lmw} + \,_{\sf s}S^{(B)}_{\lmw} 
	+ \,_{\sf s}S^{(C)}_{\lmw},
\end{equation}
where for ${\sf s} = -2$,
\begin{widetext}
\begin{align}
	&\,_{-2}S^{(A)}_{\lmw} = \pi r^{4} \frac{M \sqrt{\lambda\lambda_{1}} f^{2}_{0} \,_{0}\bar{Y}_{\ell m}(\pi/2, 0)\,}{r^{3/2}_{0} \sqrt{r_0 - 3M}}
	\,\delta(r - r_0),\\
	&\,_{-2}S^{(B)}_{\lmw} = - i\pi r^{3}\frac{\sqrt{M^3\lambda} f_{0}\,_{-1}\bar{Y}_{\ell m}(\pi/2, 0)\,}{r^{2}_{0}\sqrt{r_0 - 3M}}
	\Big[ (2M + r - 2ir^{2}\omega_{m} -7r^{2}f)\,\delta(r - r_0) - 2r^{2}f\,\delta^{\prime}(r - r_0) \Big],\\
	&\begin{aligned}
	\,_{-2}S^{(C)}_{\lmw} = \pi r^{4} 
	\frac{M^2 \,_{-2}\bar{Y}_{\ell m}(\pi/2, 0)}{r^{5/2}_{0}\sqrt{r_0 - 3M}}
	\Big[ \big( \omega_{m}(2 i M + r^{2}\omega_{m}) &- r^{2}f(5f - 6i\omega_{m}) \big)\delta(r - r_0)\\
	&- r^{2}f \big( (2i\omega_{m} + 6f)\delta^{\prime}(r - r_0) + f \delta^{\prime\prime}(r - r_0) \big) \Big].
	\end{aligned}
\end{align}
\end{widetext}
Here we have used the angular delta functions in Eqs.~\eqref{T1nn}--\eqref{T1mbmb} to evaluate the angular integral in Eq.~\eqref{eq:stress_energy_decomposition} (after integrating by parts as explained below that equation). We have also used $\,_{\sf s}\bar{Y}_{\ell m}(\pi/2, \phi_p) = \,_{\sf s}\bar{Y}_{\ell m}(\pi/2, 0)e^{-im\phi_p}$ (for all $\sf s$)  to evaluate the integral over~$\phi_p$.

Finally, to express the source terms in the form we use in our numerical worldtube calculations, we need
to express $_{-2}S^{(1,0)}_{\lmw}$ in the canonical form~\eqref{J1 Teukolsky}, in which every term takes the form $X(r_0)\delta^{(n)}(r-r_0)$, with all coefficients of radial delta functions evaluated at $r_0$ rather than $r$.
To achieve this, we use the relation Eq.~(\ref{eq:first_dirac_delta_identity}) along with similar identities for higher derivatives of
Dirac delta functions:
\begin{align}
	X(r)\delta^{\prime}(r - r_0) &= X(r_0)\delta^{\prime}(r - r_0)
	- X^{\prime}(r_0)\delta(r - r_0), \nonumber \\
	X(r)\delta^{\prime\prime}(r - r_0) &= 
	X(r_0)\delta^{\prime\prime}(r - r_0)
	- 2X^{\prime}(r_0)\delta^{\prime}(r - r_0) \nonumber \\
	&\quad+ X^{\prime\prime}(r_0)\delta(r - r_0).
	\label{eq:further_dirac_delta_identities}
\end{align} 
After applying those identities, we find the coefficients in Eq.~\eqref{J1 Teukolsky} are given by
\begin{widetext}
\begingroup
\allowdisplaybreaks
\begin{align}
	&\begin{aligned}
	J^{(A)}_{\rm pp} &= -\frac{4\pi r^{2}_{0}}{f^{3}_{0}\sqrt{r_0 - 3M}} \Bigg[ f_{0} \bigg( 
	\sqrt{\lambda\lambda_{1}} r^{3/2}_{0} f_{0} \,_{0}\bar{Y}_{\ell m}(\pi/2, 0)\, - i\sqrt{M\lambda} r^{2}_{0} \,_{-1}\bar{Y}_{\ell m}(\pi/2, 0)
	\Big( r_0 f_{0} \big( r_0(7 + 2i\omega_{m})- 13 \big) \\*
	&- 2M(7r_0 - 15) \Big) \bigg)
	- M\sqrt{r_0}\,_{-2}\bar{Y}_{\ell m}(\pi/2, 0) \Big( 
	M \big(4 M^2 (r_0 (5 r_0-48)+56) - 2 M r_0 (i r_0\omega_{m}(6 r_0-17)\\* 
	&- 10 r_0(r_0-9) + 96) - 
	r_0^{3}(r_0\,\omega_{m} (\omega_{m} - 6i) - 5r_0 + 14 i \omega_{m} + 42) - 42r_0^{2} \big)
	 \Big) \Bigg],
	\end{aligned}\\
	&J^{(B)}_{\rm pp} = \frac{8i\pi M^{3/2}r^{2}_{0}}{f^{2}_{0}\sqrt{r_0 - 3M}} \Big[ \sqrt{\lambda}\, r^{2}_{0} f_{0} \,_{-1}\bar{Y}_{\ell m}(\pi/2, 0)
	- i\sqrt{M r_0} \,_{-2}\bar{Y}_{\ell m}(\pi/2, 0) \big( 2M + r_0f_{0}(3r_0 - 7) + i r_0\omega_{m} \big) \Big],\\
	&J^{(C)}_{\rm pp} = \frac{4M^2\pi r^{9/2}_{0}\,_{-2}\bar{Y}_{\ell m}(\pi/2, 0)}{f_{0}\sqrt{r_0 - 3M}}.
\end{align}
\endgroup
The parametric derivatives with respect to $r_0$ are then given by
\begin{align}
	&\begin{aligned}
	\delta J^{(A)}_{\rm pp} &= \frac{2\pi}{\sqrt{M} r_0^{3/2}(r_0 - 3M)^{3/2}f^{4}_{0}} \Bigg[ 
	r_0^{2}f_{0} \bigg(
	\sqrt{\lambda\lambda_{1}} r_0^{3/2} f_{0} \,_{0}\bar{Y}_{\ell m}(\pi/2, 0) \big(4M(r_0 - 3M) + 3r_0f_{0}(7M - 2r_0)\big) \\
	&- iM\sqrt{\lambda} \,_{-1}\bar{Y}_{\ell m}(\pi/2, 0) \Big( 
	\big(24 M r_0^{2} f_{0}^{2}(9-7 r_0) + r_0^{3}f_{0} (49 r_0-60) +12 M^2 (7 r_0-18)\\
	&+ 4 M r_0(21 - 12 i r_0 \omega_{m} -7 r_0)+r_0^2 (14 i r_0 \omega_{m} - 5)\big) + 8 M (3 M-r_0) (6 M+r_0 
	\big(2 i r_0\omega_{m} - 1)\big) \Big) \\
	& + M \sqrt{Mr_0} \,_{-2}\bar{Y}_{\ell m}(\pi/2, 0) \Big( 12M(3M - r_0)\big( 8M^{2} + 6iMr_0^{2} - r_0^{4}\big)
	+ r_0^{4} f_{0} \big( 24M^{2}(M(12r_0 - 25) - 4r_0(r_0 - 2)) \\
	&+ 2i M r_0^{2} \omega_{m}( 3M(24r_0 - 71) + 4r_0(17 - 6r_0) ) + r_0^{4}\omega_{m}^{2}(27M - 8r_0)
	+ r_0^{2} f_{0}^{2} (4 r_0 (r_0 (10 r_0-63)+42) \\
	&-9 M (r_0 (15 r_0-98)+70))+6 i r_0^2 \omega_{m}(M (49-27 r_0)+2 r_0 
	(4 r_0-7))+4 M (3 M (r_0 (5 r_0-57)+60) \\
	&+ r_0^{2}(54-5 r_0) - 54r_0) \big) \Big) \bigg)
	\Bigg],
	\end{aligned}\\
	&\begin{aligned}
	\delta J^{(B)}_{\rm pp} &= -\frac{4iM^{3/2}\pi}{r_0^{3}(r_0 - 3M)^{3/2}f_{0}^{3}} \Bigg[
	\sqrt{\lambda}r_0^{5} \,_{-1}\bar{Y}_{\ell m}(\pi/2, 0) \big(60M^{2} - 42M r_0 + 7r_0^{2}\big) \\
	&- i\sqrt{M} r_0^{7/2} \,_{-2}\bar{Y}_{\ell m}(\pi/2, 0) \Big(M^{3}(864 - 396r_0) + 2r_0^{3}\big( 4r_0(3 + i\omega_{m})\big)
	+ 2M^{2}r_0\big( 3r_0(80 + 13 i\omega_{m})\big) \\
	&+ 3Mr_0^{2}\big(117 - r_0(63 + 17i\omega_{m})\big) \Big)
	\Bigg],
	\end{aligned}\\
	&\delta J^{(C)}_{\rm pp} = \frac{2M^2\pi r_0^{5/2}\,_{-2}\bar{Y}_{\ell m}(\pi/2, 0)}{(r_0 - 3M)^{3/2}f^{2}_{0}} 
	(66M^{2}-47Mr_0 + 8r_0^{2}).
\end{align}
\end{widetext}

\bibliography{refs}

\end{document}